\documentclass[aps,reprint,twocolumn,pra,groupedaddress,floatfix]{revtex4-1}
\usepackage{amsmath,amssymb,amsfonts}
\usepackage{mathrsfs}
\usepackage{stackengine}
\usepackage{soul}
\usepackage{graphicx}
\usepackage{color}
\graphicspath{{./Figs/}}
\begin{document}
\title{Multifaceted dynamics and gap solitons in $\mathcal{PT}$-symmetric periodic structures}
\author{S. Vignesh Raja}
\email{vickyneeshraja@gmail.com}
\author{A. Govindarajan}
\email{govin.nld@gmail.com}
\author{A. Mahalingam}
\email{drmaha@annauniv.edu}
\author{M. Lakshmanan}
\email{lakshman.cnld@gmail.com}
\affiliation{$^{*,\ddagger}$Department of Physics, Anna University, Chennai - 600 025, India}
\affiliation{$^{\dagger,\mathsection}$Centre for Nonlinear Dynamics, School of Physics, Bharathidasan University, Tiruchirappalli - 620 024, India}
\begin{abstract}
We report the role of $\mathcal{PT}$-symmetry in switching characteristics of a highly nonlinear fiber Bragg grating (FBG) with cubic-quintic-septic nonlinearities. We demonstrate that the device
shows novel bi-(multi-) stable states in the broken regime as
a direct consequence of the shift in the photonic band gap influenced
by both $\mathcal{PT}$-symmetry and higher-order nonlinearities.  We also numerically depict that such FBGs provide a productive test bed where the broken $\mathcal{PT}$-symmetric regime
can be exploited to set up all-optical applications such as binary switches, multi-level signal processing and optical computing. Unlike optical bistability (OB) in the traditional and unbroken $\mathcal{PT}$-symmetric FBG, it exhibits many peculiar
features such as flat-top stable states and ramp like input-output characteristics before the onset of OB phenomenon in the broken regime. The gain/loss parameter plays a dual role in controlling the switching intensities between the stable states  which is  facilitated by reversing the direction of light incidence. We also find that the gain/loss parameter tailors the formation of gap solitons pertaining to transmission resonances which clearly indicates that it can be employed to set up optical storage devices. Moreover, the interplay between gain/loss and higher order nonlinearities brings notable changes in the nonlinear reflection spectra of the system under constant pump powers.
The influence of each control parameters on the switching operation
is also presented in a nutshell to validate that FBG offers more degrees
of freedom in controlling light with light.
\end{abstract}
\maketitle
\section{Introduction}
In the era of high-speed information exchange, all-optical switches and logic devices are versatile components in all-optical communication systems which have lead to  widespread research across different devices such as couplers \cite{jensen, govindaraji2015}, Bragg gratings \cite{zang2012analysis}, ring resonators \cite{sethi2014all} and so on. Among these devices, FBGs have engrossed an ever-mounting attention as they afford a larger degree of freedom and flexibility to engineer any spectral characteristics of interest \cite{erdogan1997fiber} and are potential candidates for provisioning, protection, packet switching, and external modulation applications \cite{ramaswami2009optical} in addition to sensing, dispersion compensation, and filtering functionalities \cite{hill1997fiber}. Since the bandwidth of the device is very narrow, a small change in refractive index introduced into the system via an external signal is sufficient enough to detune the inbuilt photonic band gap from its resonance wavelength. Thus the structure allows light transmission at the wavelengths which were inhibited to transmit previously. This gives the comprehensive picture of the underlying mechanism to realize all these functionalities including optical switching \cite{radic1995theory}.

Optical bistability is a ubiquitous phenomenon in the framework of nonlinear feedback systems such as Bragg gratings \cite{radic1995theory}, nonlinear Fabry--P\'{e}rot cavity \cite{jeong2006all}, ring resonators \cite{li2014bistability}, and even in the case of nonlinear directional couplers with the aid of metamaterials \cite{litchinitser2007optical}. As the name suggests, any nonlinear variation in input intensity results in two or more stable states for the given incident intensity. Apart from its conventional application in the form of optical switches or memory devices where the two stable states can be customized as binary logic, researchers have also exploited the possibility to build all-optical transistors, limiters, inverters \cite{ping2005bistability}, and signal processing devices \cite{yousefi2015all}. In principle, optical bistability emanates itself as a result of light-dependent refractive index changes or absorption changes inside the structure upon which they are categorized as dispersive or absorptive optical bistability \cite{gibbs2012optical}. Ever since the breakthrough work by Winful \emph{et al.} \cite{winful1979theory}, the studies on optical bistability in nonlinear feedback structures primarily focused on understanding the underlying physics behind its operation at various conditions \cite{shi1995optical, radic1995theory} which in turn aided the possibility to realize various nonlinear applications \cite{ping2005bistability}. These studies indicate that the bistable curve or the hysteresis loop can be manipulated at will with the aid of any control parameter originating from the device such as device length, detuning parameter, the strength of modulation \cite{zang2012analysis} or via an external control in the form of probe and pump pulse parameters \cite{ping2005bistability}. Some of these factors will be discussed in detail in this paper in later sections.

A suitable choice of materials is an essential ingredient to realize an all-optical switch with low threshold and high figure of merit (FOM) \cite{harbold2002highly,chen2006measurement}.  In this regard optical properties of many nonlinear glasses have been reported in the literature from both theoretical \cite{karimi2012all} as well as experimental \cite{harbold2002highly,chen2006measurement} perspectives.  Particularly,  chalcogenide glass that accounts for both cubic- quintic and septic nonlinearities has been subjected to intense investigations \cite{harbold2002highly,chen2006measurement,karimi2012all,porsezian2005modulational,triki2016,triki2017}. The nonlinear coefficient of chalcogenide is quite larger than silica and is of the order of $10^{3}$.  Thus it can reduce the threshold level considerably \cite{broderick1998nonlinear}. Yosia \emph{et al.}  have reported the formation of non-overlapping bistable states influenced by a phase shifted cubic-quintic grating \cite{yosia2007double}. It is noteworthy to mention that unlike Kerr nonlinearity driven bistability, it offered two completely different approaches to switch into the high state \cite{ping2005nonlinear}. Recently, Yosefi \emph{et al.} have demonstrated the soliton switching in nonlinear FBG with higher order nonlinearity \cite{yousefi2015all}. These kinds of studies ensure that higher order nonlinearities when properly exploited can play a remarkable role in the next generation signal processing devices.

Having discussed concisely the general aspects of the device, we now wish to stress the importance of $\mathcal{PT}$-symmetry in the current generation optical devices. In the context of optics, the practicality of inherent loss in any functional device was not considered by the scientific community for many years \cite{lupu2013switching}. But with the advent of parity-time ($\mathcal{PT}$) symmetry concept,  these losses are no longer considered to be detrimental by virtue of the delicate balance between amplification and attenuation \cite{kottos2010optical,el2007theory} in the system as in the case of $\mathcal{PT}$-symmetric couplers \cite{govindarajan2018tailoring, karthi2}, $\mathcal{PT}$-metamaterials \cite{feng2013experimental}, $\mathcal{PT}$-microring laser \cite{longhi2014pt}, $\mathcal{PT}$-gratings \cite{huang2014type,phang2013ultrafast}, $\mathcal{PT}$-laser cavities \cite{feng2014single,longhi2010pt}, etc. Though this notion traces its origin to the field of quantum mechanics way back in 1998 \cite{bender1998real}, the translation of the concept on to photonic platform paved the way for its major theoretical advancements \cite{lin2011unidirectional,razzari2012optics,el2007theory} and the first experimental realization by R\"{u}tter \emph{et al.} on a LiNbO${_3}$ waveguide \cite{ruter2010observation}. $\mathcal{PT}$ symmetric-photonics is regarded as the booming field in the last decade or so thanks to some unconventional features possessed by these devices ranging from broadband unidirectional invisibility \cite{lin2011unidirectional}, coherent perfect absorption \cite{longhi2010pt} to coherent amplification \cite{baum2015parity}, controllable bidirectional laser emission \cite{longhi2014pt} and so forth, which are not viable in the perspective of existing systems. Driven by the luxury that refractive index, gain and loss coefficients can be manipulated at ease \cite{el2007theory},  there is an increasing number of contributions in the literature dedicated to potential applications of $\mathcal{PT}$-symmetric systems namely optical isolators, lower power optical diodes \cite{chang2014parity}, signal processing devices \cite{phang2015versatile}, single mode lasing cavities \cite{feng2014single}, soliton switches \cite{govindarajan2018tailoring}, and logic devices \cite{phang2015versatile}.

To realize a $\mathcal{PT}$-symmetric periodic structure, it is necessary to maintain an equilibrium between generation and annihilation of photons so that it offers no net amplification. In a nutshell, the complex refractive index should satisfy the condition $n(z)=n^*(-z)$. In analogy to its quantum counterpart, optical $\mathcal{PT}$-symmetric media exhibit phase transition at the so called exceptional point.
Hence the operation of the $\mathcal{PT}$-symmetric grating is stable below a critical amount of gain and loss and when it is violated, the grating exhibits exponential energy growth or lasing behavior \cite{phang2015versatile}.

Surprisingly, the inclusion of loss to the traditional structures had an affirmative role in its functionality \cite{lupu2013switching}. For instance, contemporary work by Govindarajan \emph{et al.} on steering dynamics of $\mathcal{PT}$- symmetric coupled waveguides \cite{govindarajan2018tailoring} has marked an immense reduction in the power requirement for switching and huge amplification of pulse power. The interplay between nonlinearity and $\mathcal{PT}$-symmetry has impacted in a fall in intensity of the bistable threshold as reported by  Phang \emph{et al.} \cite{phang2015versatile, phang2013ultrafast, phang2014impact}. They have also reported a switching time of 2.5 ps in one of their works, which hints that these systems are well suited to exploit switching and logic operations \cite{phang2013ultrafast}. It is worthwhile to mention that $\mathcal{PT}$-symmetric optical devices are undoubtedly far more competent than systems exhibiting no loss or systems with only gain \cite{lupu2013switching}. As of now, the switching characteristics in $\mathcal{PT}$-symmetric fiber Bragg gratings is briefly studied only with conventional silica grating with third order nonlinearity \cite{1555-6611-25-1-015102}. Following these works, we here study the role of defocusing quintic nonlinearity combined with self-focusing cubic  and septic nonlinearities on switching of $\mathcal{PT}$-symmetric fiber Bragg gratings. We highlight the role of every individual parameter in dictating the bi- and multi-stable phenomena in both the unbroken as well as broken $\mathcal{PT}$-symmetric regimes. The study also includes the formation of gap solitons corresponding to the resonance peaks of transmission curves in the presence of higher order nonlinearities and $\mathcal{PT}$-symmetry. We have also investigated the effects of nonlinearities and $\mathcal{PT}$-symmetry on the spectra of the system in the presence of constant pump power.

 The plan of the paper is as follows. Section \ref{Sec:2} describes the necessary mathematical formulation for the system of interest.  Sections \ref{Sec:3} and \ref{Sec:4} give brief explanations of the switching characteristics in the unbroken and broken $\mathcal{PT}$-symmetric regimes, respectively. Section \ref{Sec:5} reveals the gap soliton formation at resonance peaks. Section \ref{Sec:6} elucidates the switching characteristics of the system in the presence of constant pump power. Finally in Sec. \ref{Sec:7} we discuss the significance of the results.

\section{Theoretical Model}
\label{Sec:2}
We consider a $\mathcal{PT}$-symmetric fiber Bragg grating of period $\Lambda$ inscribed on the core of a fiber of refractive index $n_{0}$ and length $L$. The nonlinearity of the fiber is not merely restricted to cubic nonlinearity but it also takes account of the quintic and septic nonlinearities. The complex refractive index distribution ($n(z)$) profile that describes such a $\mathcal{PT}$-symmetric system is mathematically written as \cite{sarma2014modulation}
\begin{gather}
n(z)=n_{0}+n_{1R}\cos\left(\frac{2\pi}{\Lambda}z\right)+in_{1I}\sin\left(\frac{2\pi}{\Lambda}z\right)\\\nonumber
+n_{2}|E|^{2}+n_{4}|E|^{4}+n_{6}|E|^{6}.
\label{Eq:Norm1}
\end{gather}

The strength of modulation parameter is defined by ($n_{1})$, which
has both real ($n_{1R}$) and imaginary parts ($n_{1I}$) and the imaginary
term stands for gain ($+n_{1I})$ or loss ($-n_{1I})$ dictated by the
so called $\mathcal{PT}$-symmetric potential, and the last three terms signify self
focusing ($n_{2}$, $n_{6}>0)$ and self defocusing ($n_{4}$$<0$) nonlinearities.
The transverse electric field $E(z,t)$ inside the FBG is the superposition of two counter propagating modulated modes which can be written mathematically as
\begin{gather}
E(z,t)=E_{f}(z,t)\exp[i(\beta_{0}z-\omega_{0}t)]\\\nonumber
+E_{b}(z,t)\exp[-i(\beta_{0}z-\omega_{0}t)],
\label{Eq:Norm_2}
\end{gather}
where the envelope functions $E_{f}(z,t)$ and $E_{b}(z,t)$ that
are used to describe the electric fields in the forward and backward directions obey the slowly varying envelope (paraxial) approximation (SVEA). The propagation constant of the fiber without grating is given by $\beta_{0}$$=2\pi n_{0}$$/\lambda_{0}$, where $\lambda_{0}$ is the free space wavelength. The Bragg wavelength of the grating is expressed as $\lambda_b = 2 n_0 \Lambda$.  Practically, the Bragg wavelength is taken in the telecommunication regime, which lies at  $1.55$ $ \mathrm{\mu m}$. But it can be chosen anywhere from visible region to infrared by suitably altering the grating period ($\Lambda$). Note that the typical values of $\Lambda$ for a short FBG can vary from 200 nm to 800 nm.

The coupled mode equations, which describe the light propagation in
the proposed system, are given by  \cite{yosia2007double, miri2012bragg}
\begin{gather}
\nonumber+i\left(\frac{\partial E_{f}}{\partial z}+\frac{n_{0}}{c}\frac{\partial E_{f}}{\partial t}\right)+\left(k_{0}+g_{0}\right)\exp^{-i2\delta_{0} z}E_{b}\\\nonumber+\gamma_{0}\left(|E_{f}|^{2}+2|E_{b}|^{2}\right)E_{f}\\\nonumber-\varGamma_{0}\left(|E_{f}|^{4}+6|E_{f}|^{2}|E_{b}|^{2}+3|E_{b}|^{4}\right)E_{f}\nonumber\\
+\sigma_{0}\left(|E_{f}|^{6}+12|E_{f}|^{4}|E_{b}|^{2}+18|E_{b}|^{4}|E_{f}|^{2}+4|E_{b}|^{6}\right)E_{f}=0, \\
\nonumber-i\left(\frac{\partial E_{b}}{\partial z}-\frac{n_{0}}{c}\frac{\partial E_{b}}{\partial t}\right)+\left(k_{0}-g_{0}\right)\exp^{+i2\delta_{0} z}E_{f}\\\nonumber+\gamma_{0}\left(|E_{b}|^{2}+2|E_{f}|^{2}\right)E_{b}\\\nonumber-\varGamma_{0}\left(|E_{b}|^{4}+6|E_{f}|^{2}|E_{b}|^{2}+3|E_{f}|^{4}\right)E_{b}\nonumber\\
+\sigma_{0}\left(|E_{b}|^{6}+12|E_{b}|^{4}|E_{f}|^{2}+18|E_{f}|^{4}|E_{b}|^{2}+4|E_{f}|^{6}\right)E_{b}=0.
\end{gather}
We adopt the well-known transformation $E_{f, b}$ = $A_{f, b}$ $\exp(\mp\delta_0 z)$ and further consider the synchronous approximation  (SVEA) to obtain the time independent equations \cite{winful1979theory}. Hence, the governing equations for the propagation of continuous waves (CW) become
\begin{gather}
+i\frac{d A_{f}}{dz}+\delta_{0} A_{f}+\left(k_{0}+g_{0}\right)A_{b}+\gamma_{0}\left(|A_{f}|^{2}+2|A_{b}|^{2}\right)A_{f}\nonumber\\-\varGamma_{0}\left(|A_{f}|^{4}+6|A_{f}|^{2}|A_{b}|^{2}+3|A_{b}|^{4}\right)A_{f}\nonumber\\
+\sigma_{0}\left(|A_{f}|^{6}+12|A_{f}|^{4}|A_{b}|^{2}+18|A_{b}|^{4}|A_{f}|^{2}+4|A_{b}|^{6}\right)A_{f}=0,\\
-i\frac{d A_{b}}{d z}+\delta_{0}A_{b}+\left(k_{0}-g_{0}\right)A_{b}+\gamma_{0}\left(|A_{b}|^{2}+2|A_{f}|^{2}\right)A_{b}\nonumber\\-\varGamma_{0}\left(|A_{b}|^{4}+6|A_{b}|^{2}|A_{f}|^{2}+3|A_{f}|^{4}\right)A_{b}\nonumber\\
+\sigma_{0}\left(|A_{b}|^{6}+12|A_{b}|^{4}|A_{f}|^{2}+18|A_{f}|^{4}|A_{b}|^{2}+4|A_{f}|^{6}\right)A_{b}=0.
\end{gather}

The detuning parameter in the coupled
equation is given by $\delta_{0}=\left(2\pi n_{0}\right)\left(\frac{1}{\lambda_{o}}-\frac{1}{\lambda_{b}}\right)$. The stop band of the grating is expressed as $|\delta_{0}|\leq k_{0}$,  where $k_{0}$ is the strength of coupling between
the oppositely traveling fields. Within this band, no propagating modes are supported by the grating and so the light transmission is prohibited \cite{erdogan1997fiber}. The coefficients of coupling, gain/loss, cubic, quintic, and septic nonlinearities are given by \cite{miri2012bragg,yosia2007double}
\begin{gather}
\nonumber \kappa_0=\pi n_{1 R} / \lambda_{0}, \quad g_0=\pi n_{1 I} / \lambda_0, \quad \gamma_0 = 2 \pi n_2 / \lambda_0, \\  \Gamma_0 = 2 \pi n_4 / \lambda_0, \quad\sigma_0 = 2 \pi n_6 / \lambda_0. 
\end{gather}
From the fundamentals of $\mathcal{PT}$-symmetry, it is well known that the system is said to be in the unbroken regime if $k_{0}>g_{0}$. On the other hand, if $g_{0}>k_{0}$ the system is set to operate in the broken regime. The condition in which $k_{0}=g_{0}$ is called \emph{singularity} or the \emph{exceptional point.}  The following transformation  is adapted  \cite{sarma2014modulation} to obtain normalized coupled mode equations 
\begin{gather}
{\zeta}=\cfrac{z}{z_{0}},\quad  u=\cfrac{A_{f}}{\sqrt{P(0)}},\quad v=\cfrac{A_{b}}{\sqrt{P(0)}},
\end{gather}
where $P(0)=\left|u_{0}\right|^2$ is the intensity of the input laser pulse. The normalized coupled mode equations are given by
\begin{gather}
+i\frac{d u}{d\zeta}+{\delta}u+\left({k}+{g}\right)v+{\gamma}\left(|u|^{2}+2|v|^{2}\right)u-\nonumber\\{\varGamma}\left(|u|^{4}+6|u|^{2}|v|^{2}+3|v|^{4}\right)u\nonumber\\
+{\sigma}\left(|u|^{6}+12|u|^{4}|v|^{2}+18|v|^{4}|u|^{2}+4|v|^{6}\right)u=0, \label{Eq:norm1}\\
-i\frac{d v}{d\zeta}+{\delta}v+\left({k}-{g}\right)u+{\gamma}\left(|v|^{2}+2|u|^{2}\right)v-\nonumber\\{\varGamma}\left(|v|^{4}+6|u|^{2}|v|^{2}+3|u|^{4}\right)v\nonumber\\
+{\sigma}\left(|v|^{6}+12|u|^{2}|v|^{4}+18|v|^2|u|^4+4|u|^{6}\right)v=0. \label{Eq:norm2}
\end{gather}
The normalized parameters are given by
\begin{gather}
\delta={\delta_{0}z_{0}},\qquad k=k_{0}z_{0},\qquad
g={g_{0}} z_{0}, \nonumber\\
\gamma={\gamma_{0}} P(0) z_{0},\quad \varGamma={\varGamma_{0}} P(0) z_{0},\quad \sigma={\sigma_{0}} P(0) z_{0}.
\end{gather}
Further  output intensity can be defined as  $P_{1}(L)=|u(L)|^2$.  
These coupled mode Eqs. (\ref{Eq:norm1}) and (\ref{Eq:norm2}) are solved by implicit Runge-Kutta fourth order method with the following boundary conditions \cite{broderick1998nonlinear}
\begin{gather}
u(0)=u_{0}, \nonumber\\
  v(L)=0.
\end{gather}

Before proceeding to analyze the switching characteristics in detail, we now investigate the role of nonlinearity and the gain/loss coefficient on the photonic band gap of the device. The existence of such a gap is a typical characteristic of any feedback structure and it is already discussed in the context of conventional FBG and $\mathcal{PT}$-symmetric cubic FBG in the literature \cite{sarma2014modulation,porsezian2005modulational,miri2012bragg}.  The mathematical expressions for the dispersion relation of a highly nonlinear $\mathcal{PT}$-symmetric FBG can be found by substituting the continuous wave solution into the normalized coupled equations given by (\ref{Eq:norm1}) and (\ref{Eq:norm2}). The wave solutions represent $u$ and $v$ in terms of forward and backward wave amplitudes $\psi_{1,2}$ along the length of propagation of the incident field inside the grating and they are given by
\begin{gather}
u=\psi_{1}\exp^{iq\zeta}, \qquad v=\psi_{2}\exp^{iq\zeta}.
\label{Eq:Norm13}
\end{gather}
Here $\psi_{1}$ and $\psi_{2}$ are assumed to be real constants.  The ratio of $\psi_{2}/\psi_{1}$ is assumed as a new parameter $f$. The sum of $\psi_{1}^2$ and $\psi_{2}^2$ gives the total power (J) of the propagating wave. The relation between $\psi_{1,2}$ and $J$ is written as
\begin{gather}
\psi_{1}=\sqrt{\cfrac{J}{1+f^{2}}}, \qquad \psi_{2}=\sqrt{\cfrac{J}{1+f^{2}}}\ f.
\label{Eq:Norm14}
\end{gather}
Hence one obtains the nonlinear dispersion relation between $q$ and $\delta$ by using Eqs. (\ref{Eq:Norm13}) and (\ref{Eq:Norm14})  in (\ref{Eq:norm1}) and (\ref{Eq:norm2}) as
\begin{gather}
q=-\cfrac{k}{2f}\left(1-f^{2}\right)+\cfrac{g}{2f}\left(1+f^{2}\right)-\cfrac{\gamma J}{2}\left(\cfrac{1-f^{2}}{1+f^{2}}\right)+\nonumber\\\Gamma J^{2}\left(\cfrac{1-f^{2}}{1+f^{2}}\right)-\cfrac{3\sigma J^{3}}{2}\left(\cfrac{1-f^{6}+2f^{2}-2f^{4}}{\left(1+f^{2}\right)^{3}}\right), \label{Eq:dr1}\\
\delta=-\cfrac{k}{2f}\left(1+f^{2}\right)+\cfrac{g}{2f}\left(1-f^{2}\right)-\cfrac{3\gamma J}{2}+\nonumber\\2 \Gamma J^{2}\left(1+\cfrac{f^{2}}{\left(1+f^{2}\right)^{2}}\right)-\cfrac{5\sigma J^{3}}{2}\left(1+\cfrac{3f^{2}}{\left(1+f^{2}\right)^{2}}\right). \label{Eq:dr2}
\end{gather}

\begin{figure*}[t]
	\centering
\includegraphics[width=0.25\linewidth]{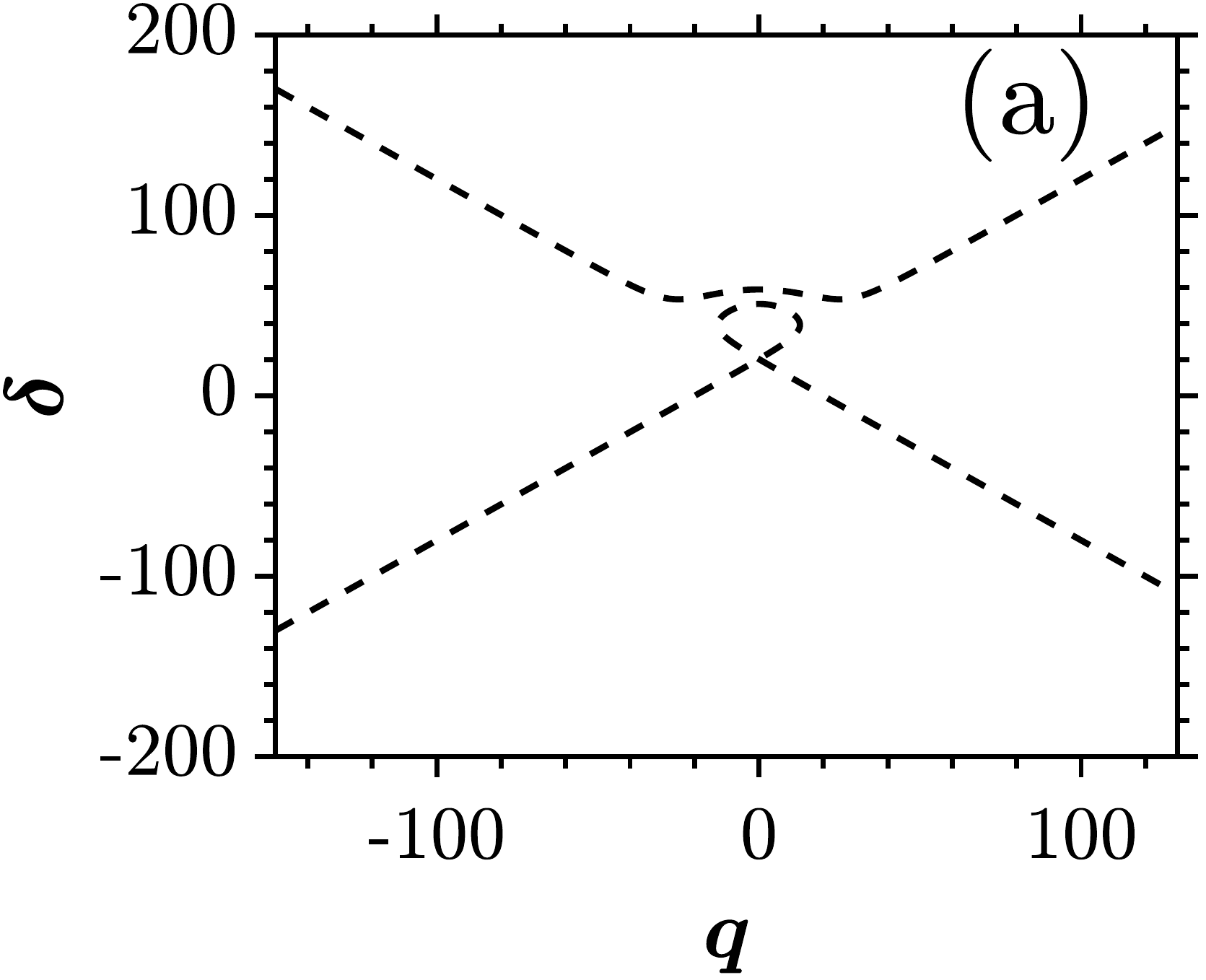}\includegraphics[width=0.252\linewidth]{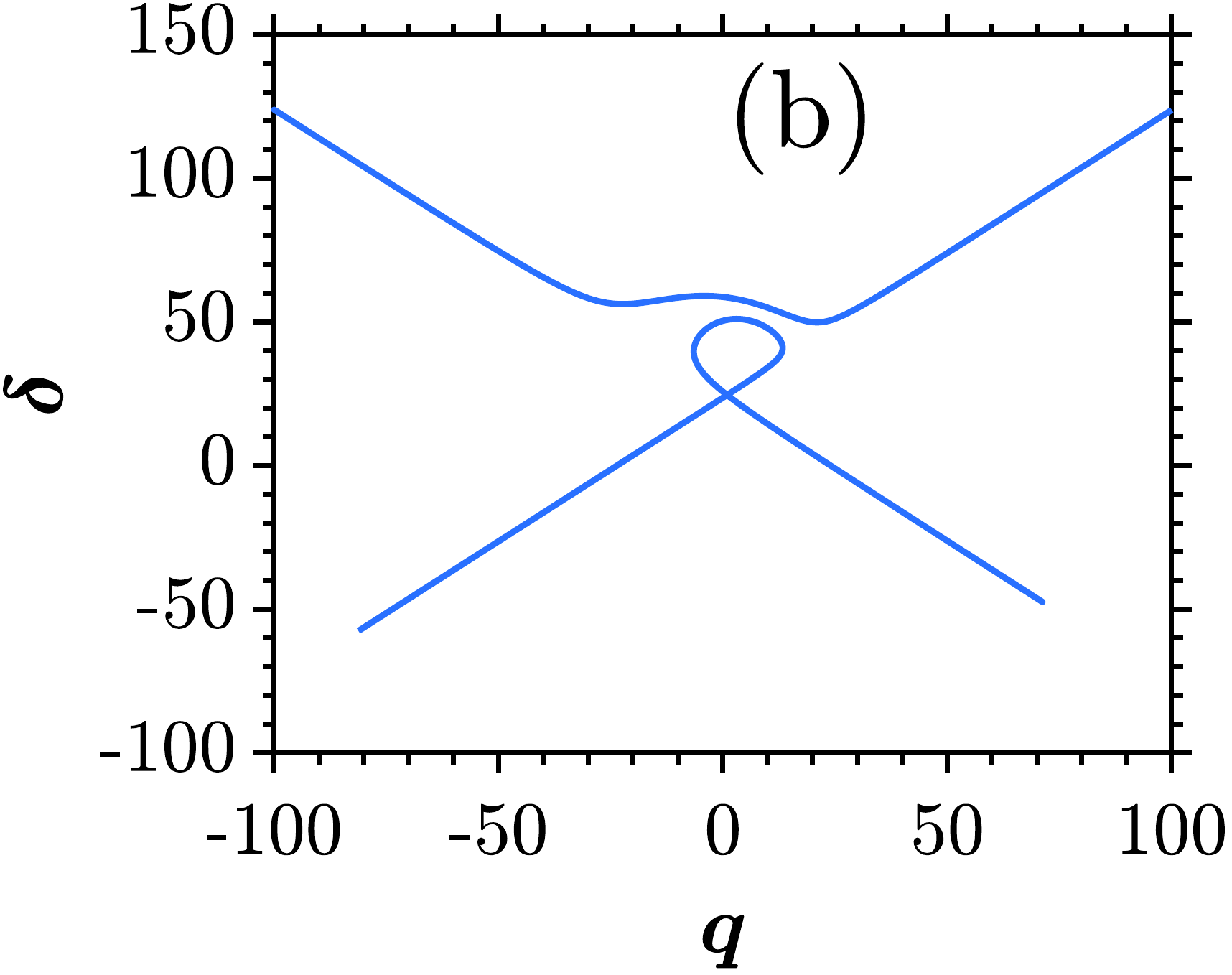}\includegraphics[width=0.252\linewidth]{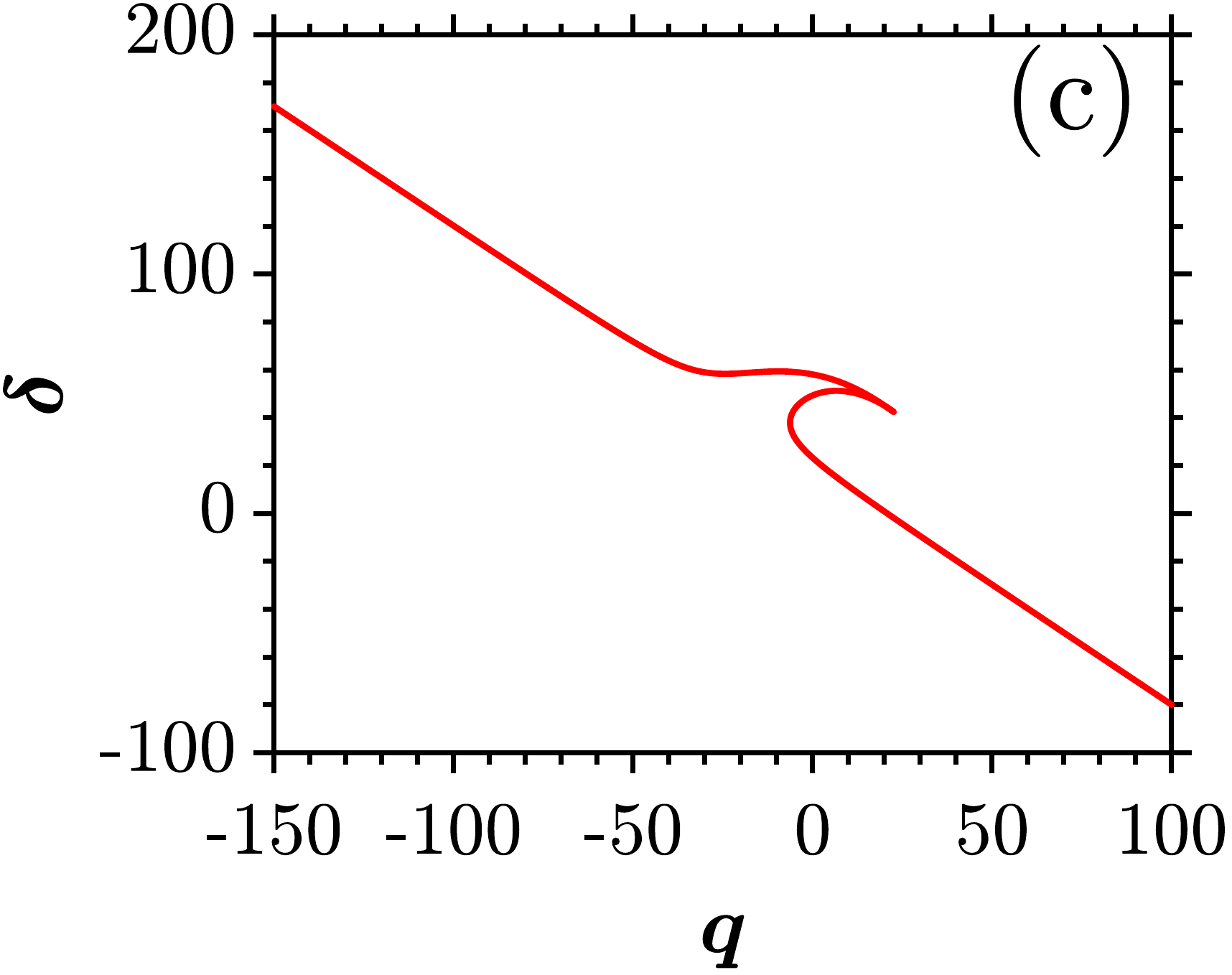}\includegraphics[width=0.25\linewidth]{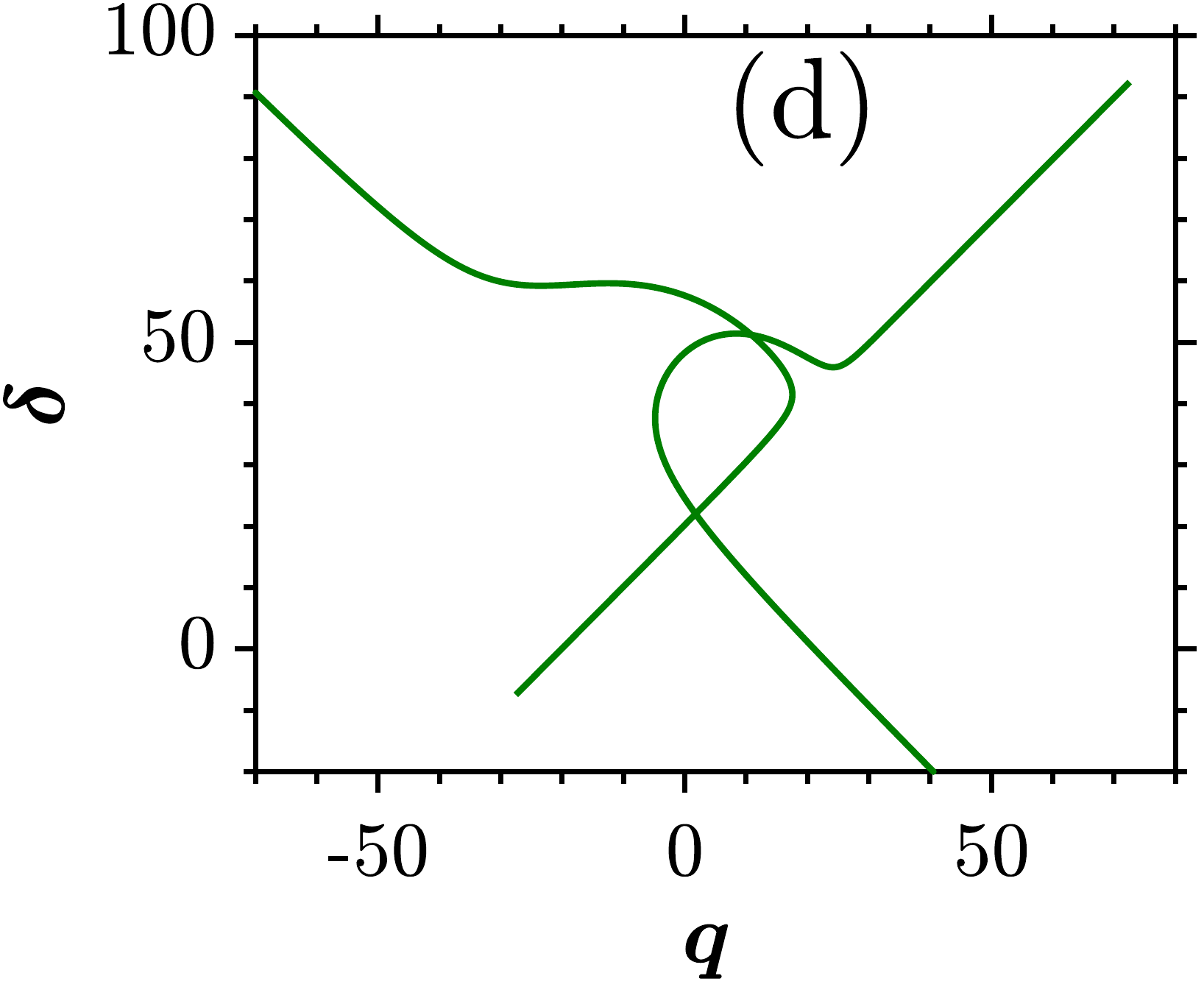}\\
\includegraphics[width=0.25\linewidth]{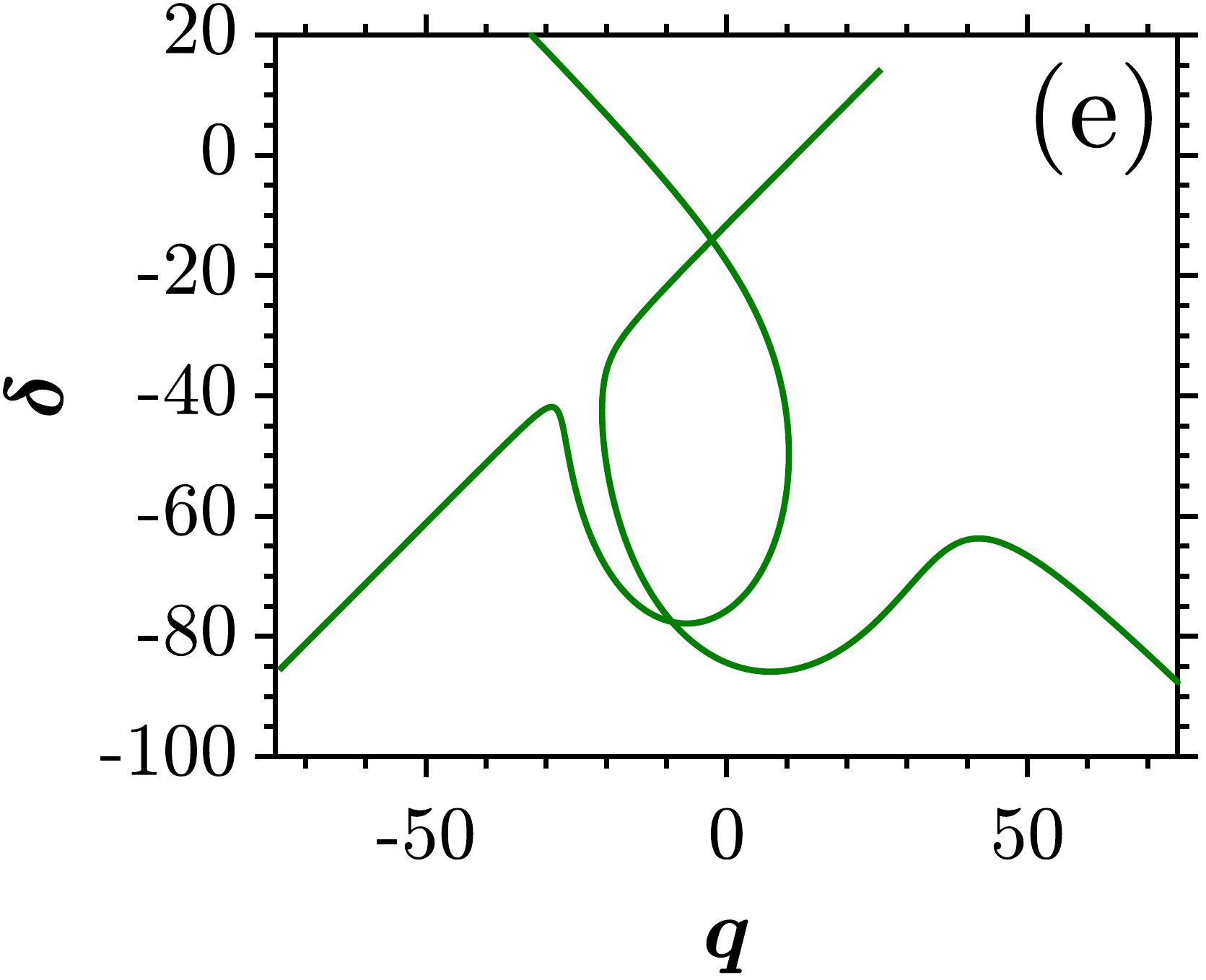}\includegraphics[width=0.25\linewidth]{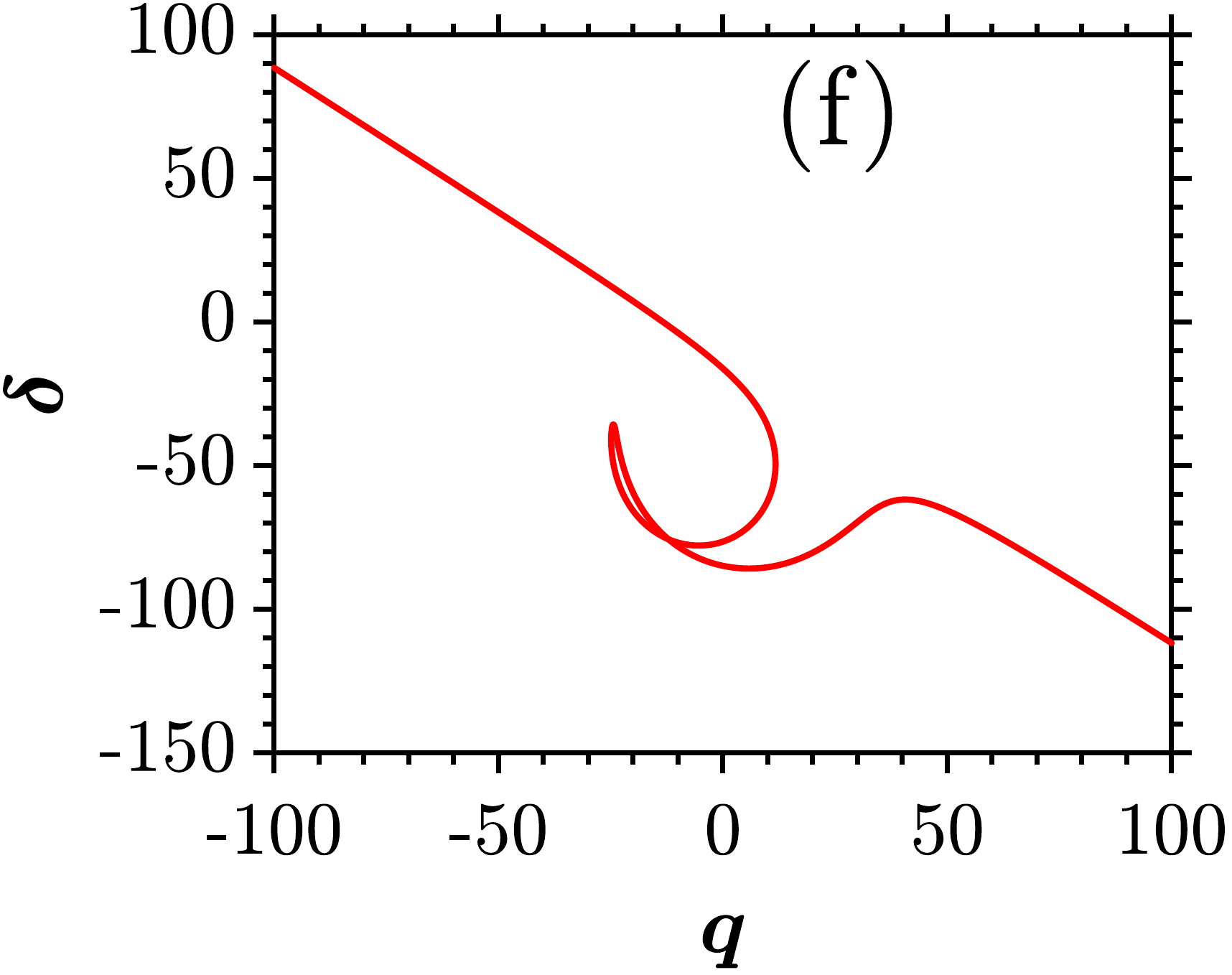}\includegraphics[width=0.25\linewidth]{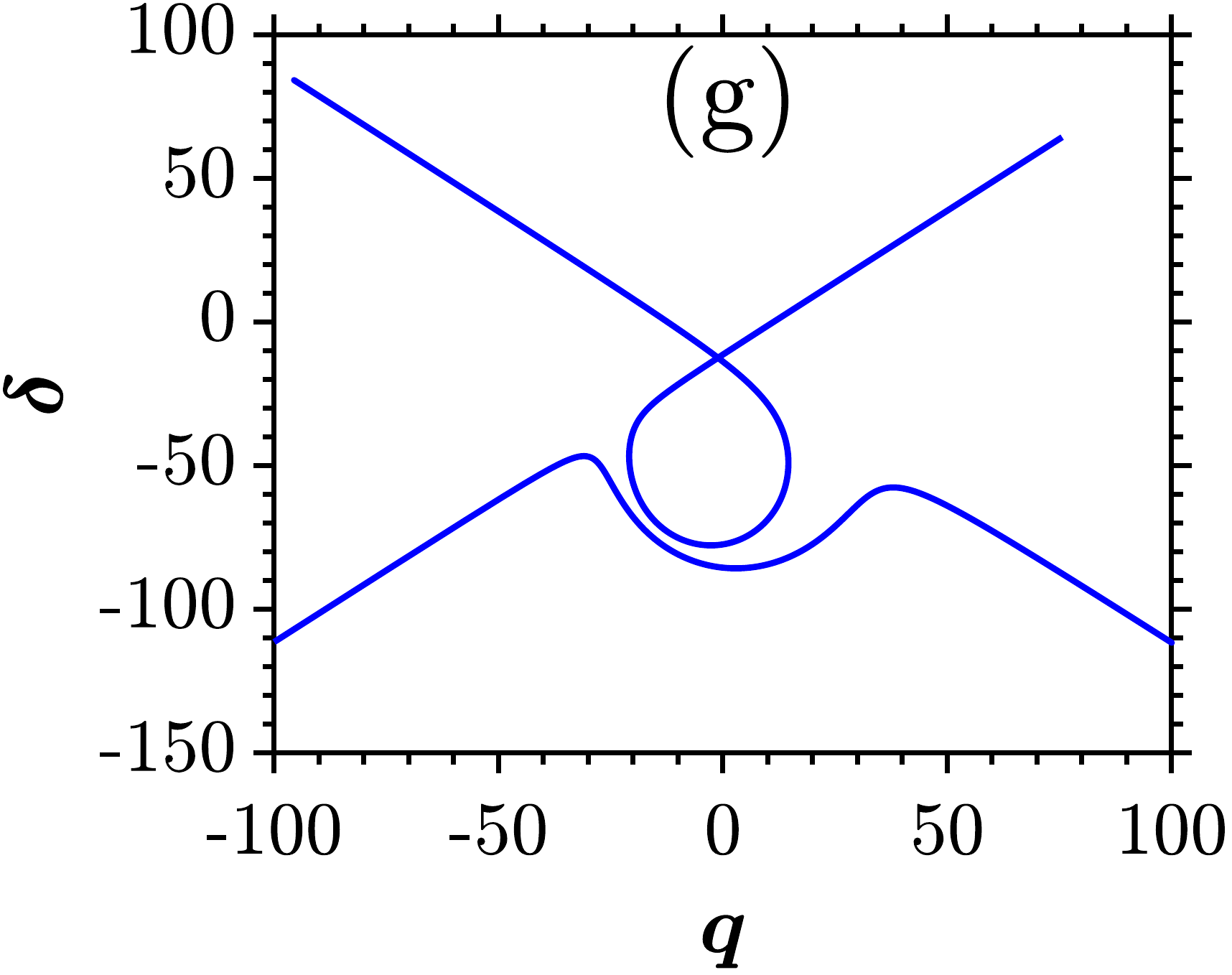}\includegraphics[width=0.25\linewidth]{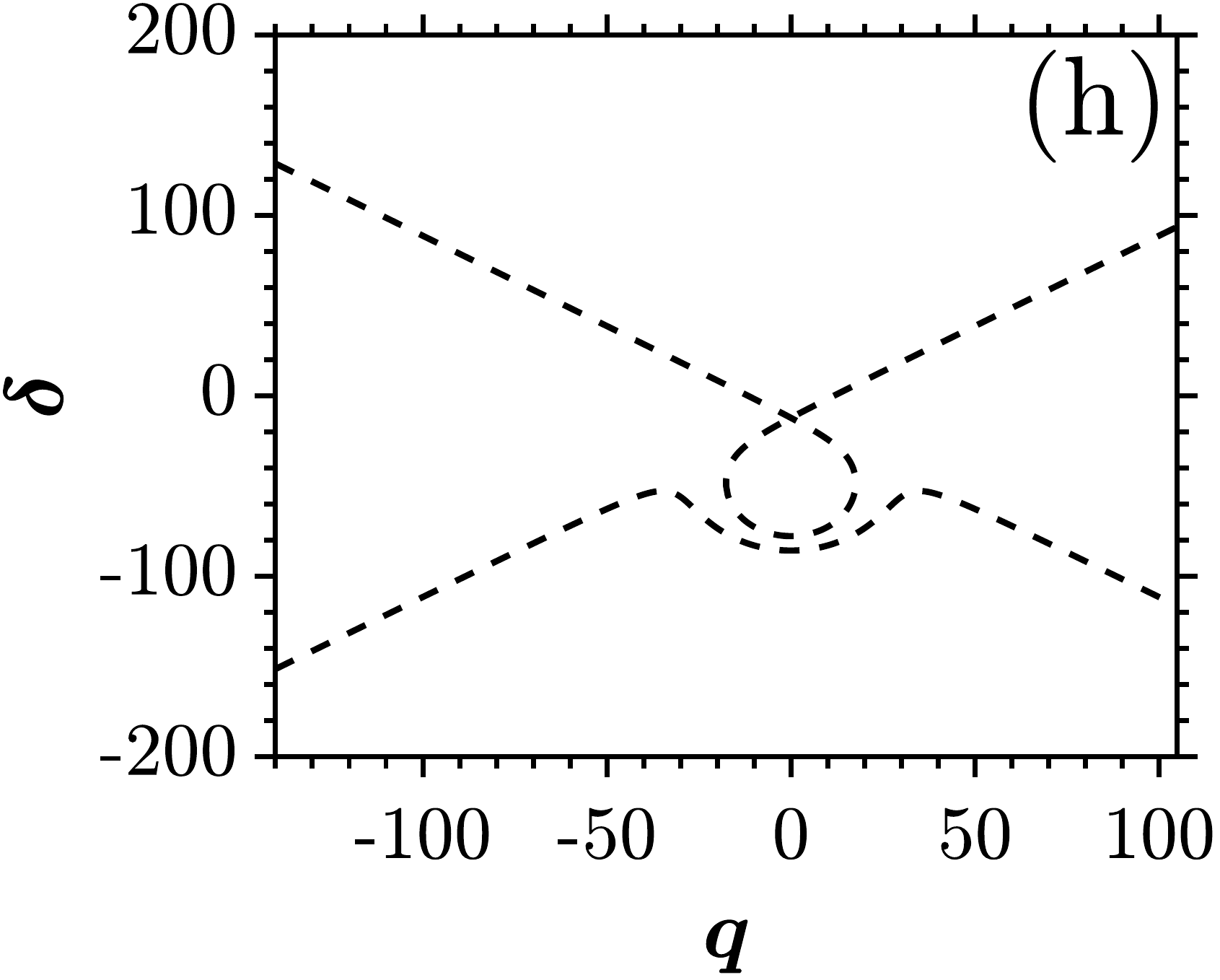}
	\caption{(Color online) Nonlinear dispersion curves plotted at $J = 2.5$ for a FBG with $k=4$. The top panels represent the relation between $q$ and $\delta$ in the presence of cubic-quintic nonlinearities ($\gamma=2$, $\Gamma=4$, $\sigma=0$) and the bottom panels correspond to $q$ vs $\delta$ curve in the presence of cubic-quintic-septic nonlinearities ($\gamma=2$, $\Gamma=4$, $\sigma=2$). Figures (a) and (h) illustrate the existence of photonic band gap in the absence of $\mathcal{PT}$-symmetry. Figures (b) and (g) depict the narrowing of band gap in the unbroken $\mathcal{PT}$-symmetric regime ($g=2$). Figures (c) and (f) are plotted at the exceptional point ($g=4$). Figures (d) and (e) show the novel dispersion curves in the broken $\mathcal{PT}$-symmetric regime ($g=5$).}
	\label{dr_cur}
\end{figure*}
From Figs. \ref{dr_cur}(a) to (h), one can visualize two branches in the dispersion curve admitted by Eqs. (\ref{Eq:dr1}) and (\ref{Eq:dr2}). These two branches correspond to the normal dispersion regime $(f>0)$ and the anomalous dispersion regime $(f<0)$. It should be noted that the dispersion curve pertaining to linear and cubic nonlinearity has already been discussed in Refs. \cite{miri2012bragg,sarma2014modulation} and so we here focus only on quintic and septic nonlinearities. In the presence of quintic nonlinearity, a loop is formed at the lower branch of the $q$ vs $\delta$ curve. Any increase in the value of $\Gamma$ increases the size of the loop and vice-versa. The loop disappears at lower values of $\Gamma$ as a result of perfect balance between self-focusing cubic nonlinearity and self-defocusing quintic nonlinearity. In the unbroken $\mathcal{PT}$- regime, with increase in $g$, the shape remains the same as in the case of conventional FBG (cf. Figs. \ref{dr_cur}(a) and (b)).  As we increase the value of $g$, the loop in the lower branch is shifted towards higher value of $q$ as shown in Fig. \ref{dr_cur}(b). At the exceptional point ($g = k$), the band gap vanishes and the formation of loop in the lower branch still persists. Hence one can control the formation of loop in the dispersion curves by dictating the strength of the nonlinearity. In the broken $\mathcal{PT}$-symmetric regime, we get two curves on the left and right of the center wavelength instead of distinguishable curves on the upper and lower branches \cite{miri2012bragg, komissarova2019pt}. The two curves merge and form a closed dispersion curve as seen in Fig. \ref{dr_cur}(d). The overlap region grows as the value of $\Gamma$ gets higher. In other words, instead of forming a loop in the lower branch, the intersection of two curves occurs at the center and it expands with increase in the value of $\Gamma$ in the $\mathcal{PT}$-symmetric broken regime. 

In the presence of septic nonlinearity, a loop is formed at the upper branch of the curve in contrast to the quintic nonlinearity case as seen in Fig. \ref{dr_cur}(h). Also, the entire dispersion curve is shifted towards lower values of $\delta$. This is because of the additional focusing effect introduced by the septic nonlinearity. Thus the formation of loops in the upper or lower branch is decided by the nature of nonlinearity whether it is self-focusing or self-defocusing, respectively even in the presence of $\mathcal{PT}$-symmetry.  Similar to the previous case, any increase in the value of septic nonlinearity increases the size of the loop in the upper branch. Likewise, any increase in the value of $g$ in the unbroken $\mathcal{PT}$-symmetric regime shifts the loop towards lower value of $q$ as observed in Fig. \ref{dr_cur}(g). The plot in Fig. \ref{dr_cur}(f) also confirms that, at the exceptional point, the band gap closes in the septic nonlinear regime  quite similar to the quintic case, with the difference is being that the loop is now formed in the upper branch.  The variations illustrated in the Fig. \ref{dr_cur}(d) hold true for Fig. \ref{dr_cur}(e) also, except the shift in the curves towards negative values of $\delta$ and the increase in the area of the intersection between the two curves.
It is worthwhile to mention that all these variations occurring in the photonic band gap of the device are the direct consequence of change in refractive index of the medium induced by interplay between $\mathcal{PT}$-symmetric potential and higher order nonlinearities. Having elucidated the necessary mathematical description of the system,
we now look into the switching characteristics of the proposed system.

\section{Transmission properties in unbroken $\mathcal{PT}$-symmetric  regime}
\label{Sec:3}
\begin{figure}[t]
\centering
\centering
\includegraphics[width=0.5\linewidth]{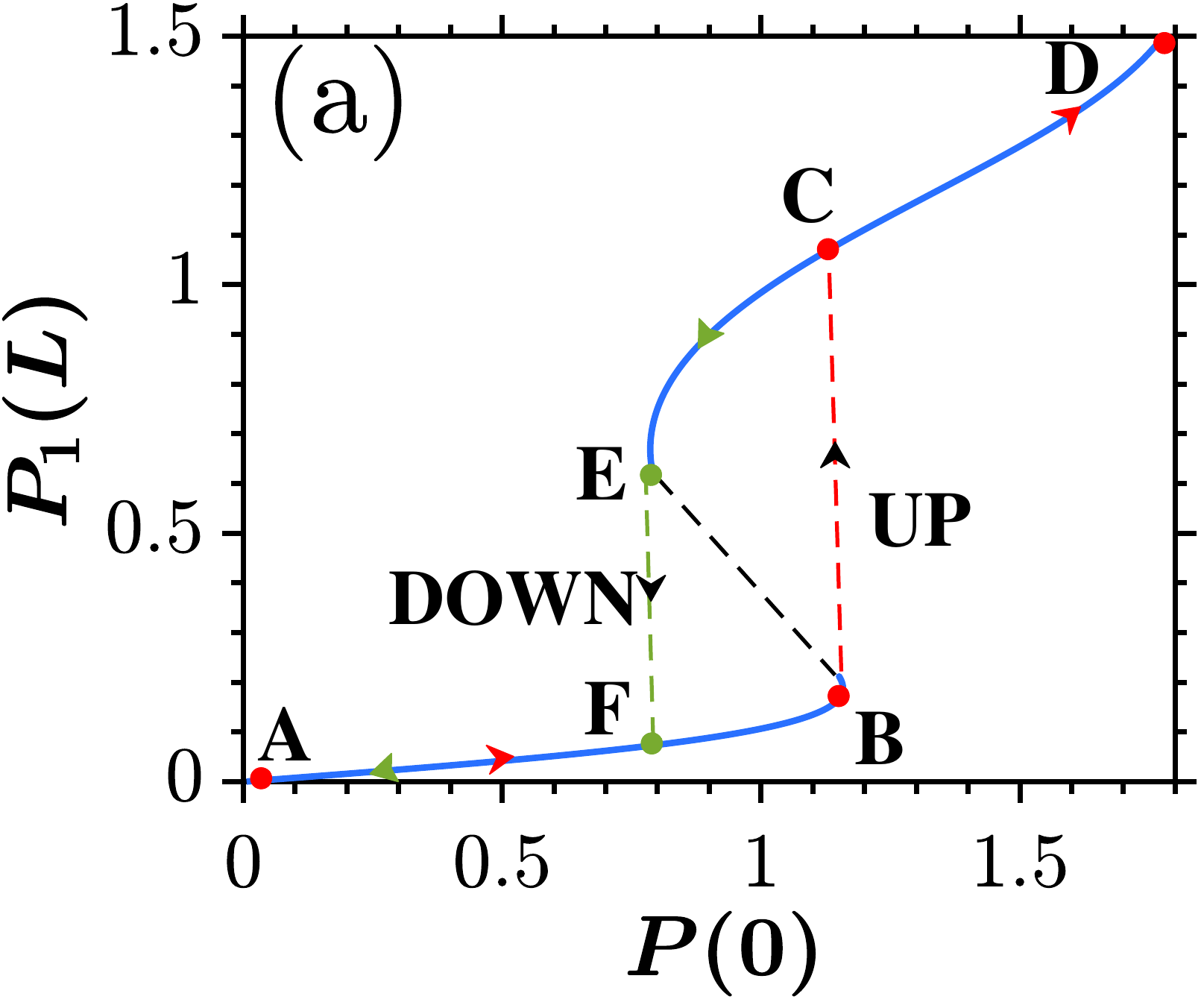}\includegraphics[width=0.5\linewidth]{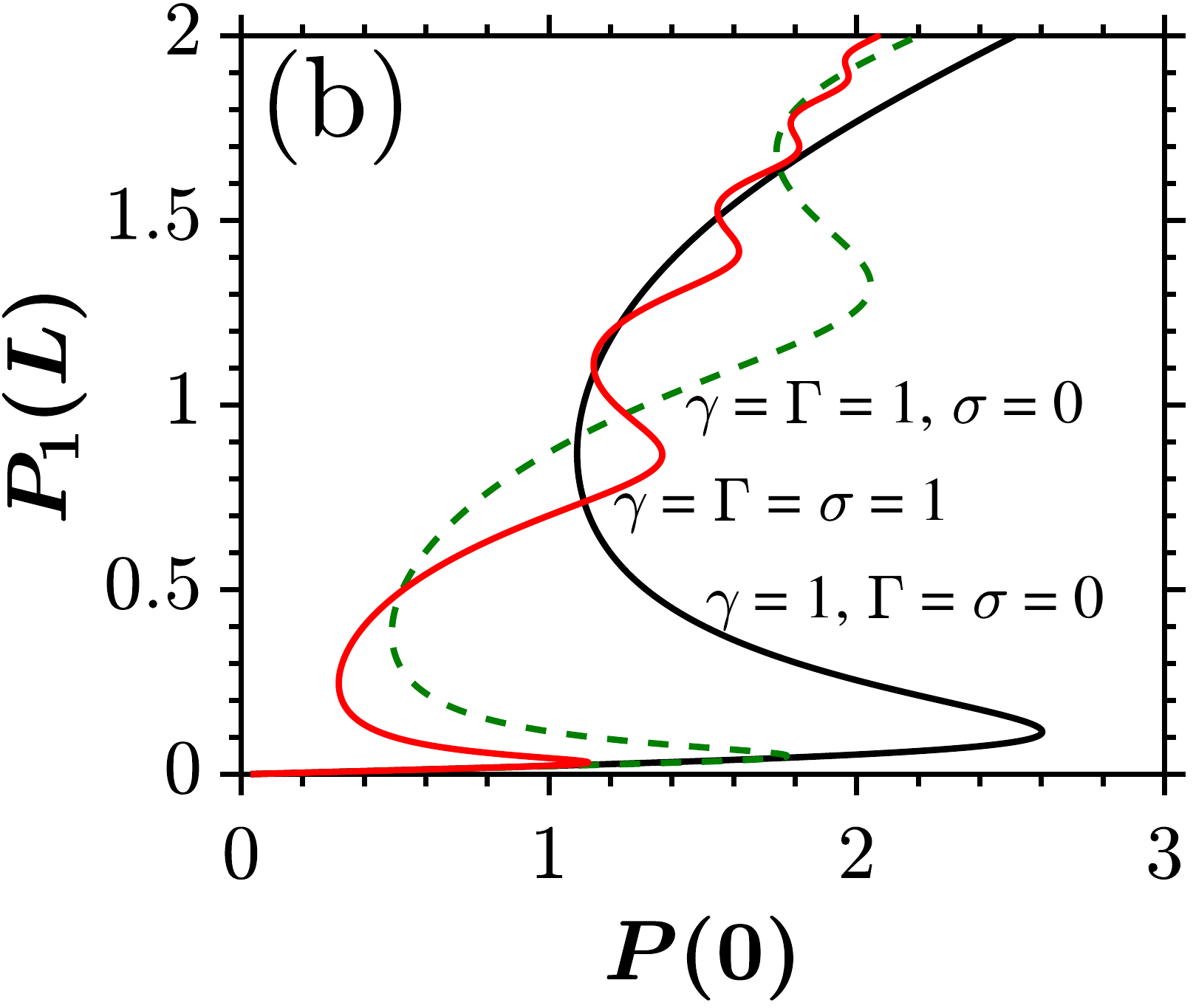}
\caption{(Color online) (a) Schematic sketch showing different branches of a typical bistable curve with their switch-up (marked as B) and switch-down intensities (marked as E). Here the two stable states are indicated by curves AB and DE and the unstable branch is indicated by BE and (b) Comparison of optical bistable (multistable) characteristics of the unbroken $\mathcal{PT}$-symmetric FBG device under different nonlinear regimes for $L=1,k=3$, $g=1.5$, and $\delta=0$. } 
	\label{Fig_01}
\end{figure}

The switching characteristics of the $\mathcal{PT}$
-symmetric FBG has been investigated in the recent work by Liu \emph{et al.} which reveals
that with an increase in gain/loss coefficient the threshold of switching gets higher \cite{1555-6611-25-1-015102}.
Even though this outcome is an undesired one, additional degrees of freedom
offered by the inclusion of $\mathcal{PT}$-symmetry
should not be taken lightly. The additional features of $\mathcal{PT}$-symmetry need to be retained but not at the cost of the threshold. On
the other hand, we find that the inclusion of higher order nonlinearities in the conventional FBG aids in the reduction of the threshold. This is the factor that
drove us to study the switching characteristics of a highly nonlinear
FBG under $\mathcal{PT}$-symmetric notion. Before we proceed
to study the individual effects in detail, we portray the combined
effects of $\mathcal{PT}$-symmetry with cubic-quintic-septic
nonlinearities on the switching threshold. To do so, numerical simulations
were carried out with device parameters $L=1$, $k=3$, $g=1.5$, and $\delta=0$ \cite{lin2011unidirectional}.
The switch-up intensities of cubic, quintic and septic nonlinearities
were found to be descending in the order $2.6$, $1.77$, $1.13$ (see Fig. \ref{Fig_01}(b)) and the corresponding values of switch down intensities are $1.09$, $0.49$, $0.32$ respectively. Such a dramatic decrease in the switch-up intensity and the threshold is provided by the inclusion of higher order nonlinearities alongside the $\mathcal{PT}$-symmetry. Thus our system can uniquely combine the pros of the individual
systems without imposing any impairments. 
\begin{figure}[h!]
	\centering
	\includegraphics[width=0.5\linewidth]{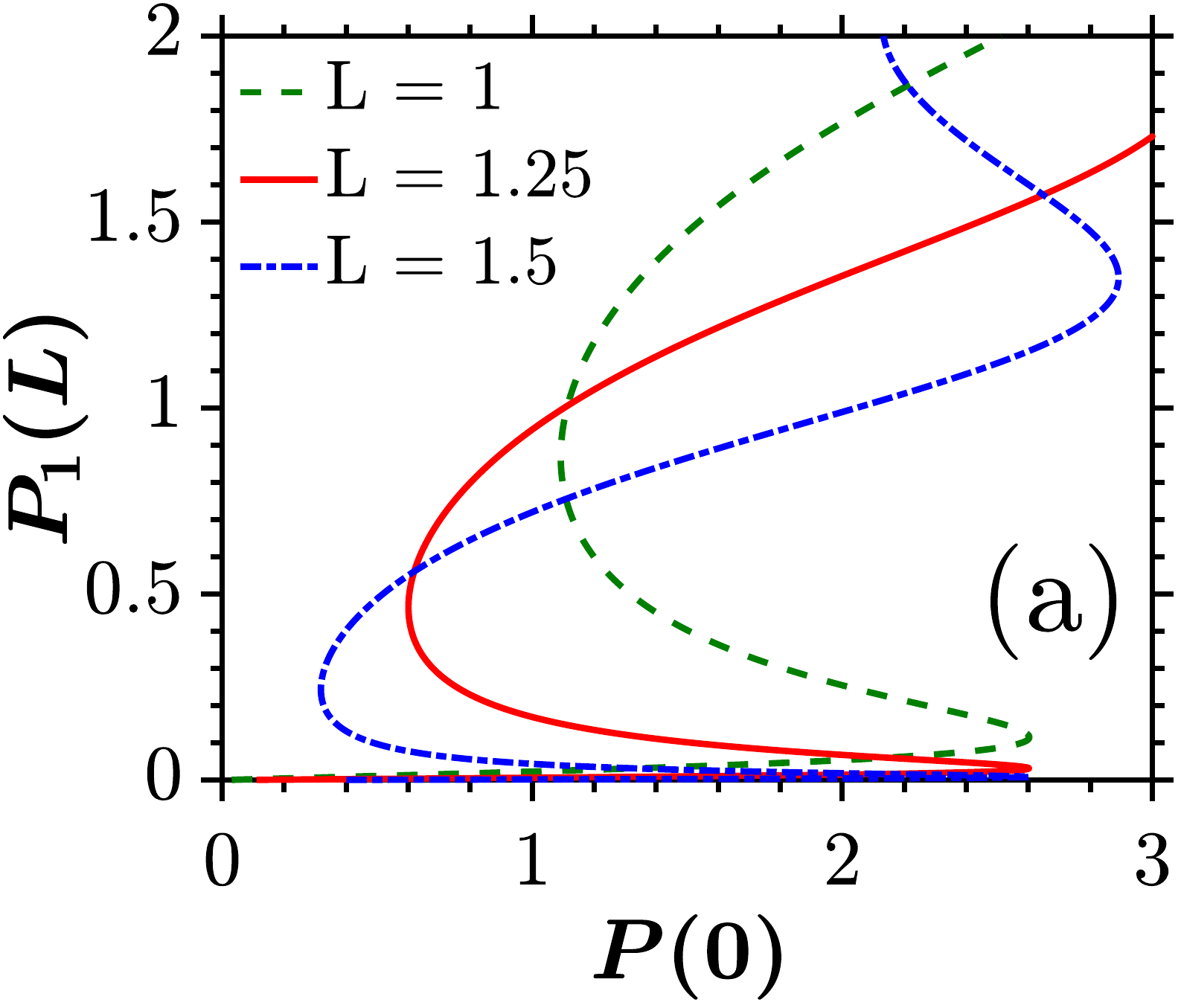}\includegraphics[width=0.5\linewidth]{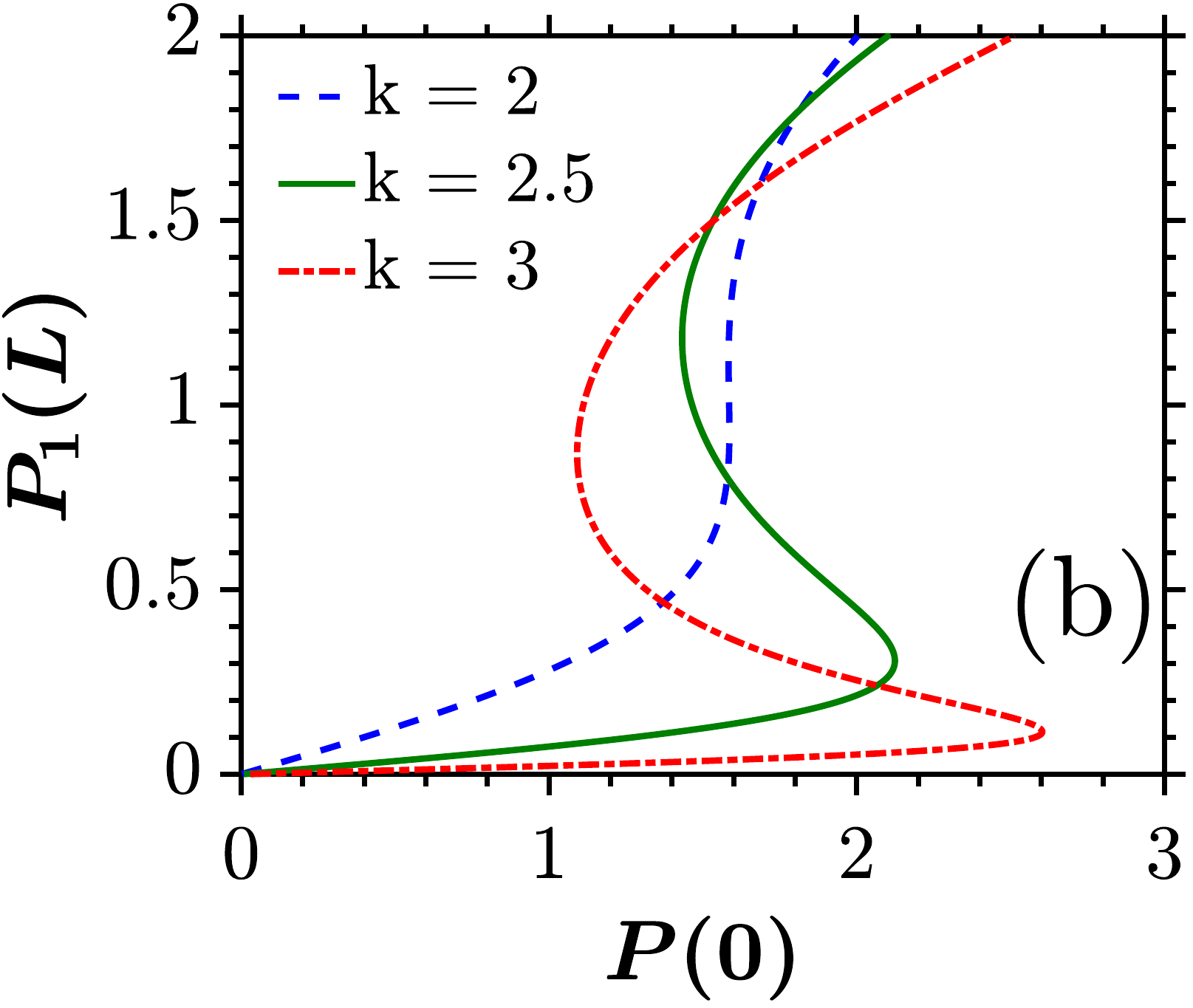}
	\caption{(Color online) Role of (a) device length (b) the strength of coupling coefficient in the optical bistability phenomenon of an unbroken $\mathcal{PT}$-symmetric FBG.}
	\label{Fig_1}
\end{figure}

\subsection{Effect of length and coupling coefficient}
To understand the role of length of the fiber grating and the strength of coupling coefficient, we first consider a simple case in which only the cubic-nonlinearity is present ($\Gamma$= $\sigma$ = 0). Figure \ref{Fig_1}(a) shows the input-output characteristics at $g=1.5$, $k=3$, $\delta=0$
for three different values of the device length L=$1$, $1.25$, and $1.5$.
When the length is shorter ($L=1$), the effective feedback to the system gets reduced
and therefore we observe only two stable states in the output with a switch up intensity of $2.603$ for $L=1$. Below this length, the feedback is
not sufficient to observe bistability when $k=3$. The switching up intensity
between the three curves featured a slim difference but an increase in
$L$ severely influences the hysteresis width. The differences between
the switch-up and switch-down intensities for the three different lengths
are $1.51, 1.99, 2.28$ (approximately). The upper stable
branch gets flatter at higher values of length. The strength of the
coupling also influences the shape and width of the hysteresis curve.
At the given length ($L=1$), the lower values
of $k$ results in insufficient feedback to create a bistable state.
In the simulations, we fix $L=1$, $g=1.5$, $\delta=0$ and for three
different values of $k$ the bistability curves are plotted
in Fig. \ref{Fig_1}(b). When $k=2.5$, the switch up intensity increases to $2.12$ and further
it goes to $2.603$ at $k=3$. From this, it is very clear that $k$ not only
intensifies the switch up intensity value but it has a combined influence alongside the
device length in increasing the feedback to the system. Hereafter, throughout this paper, the length and the strength of the coupling are fixed at $L=2$, $k=4$ as it is practically feasible to tune the values of gain/loss coefficient and the detuning parameter with an external control rather than the device length and the inherent coupling coefficient. Note  that in view of practical realization, the value of coupling parameter $k$ (which can be in the range of 1 to 10 $\mathrm{cm^{-1}}$) goes hand in hand with the length of the grating varying from 1 mm to 10 cm and the optimum value of $kL$ can attain the values from 1 to 10 \cite{agrawal2001applications}.
\begin{figure}[h]
	\centering
	\includegraphics[width=0.5\linewidth]{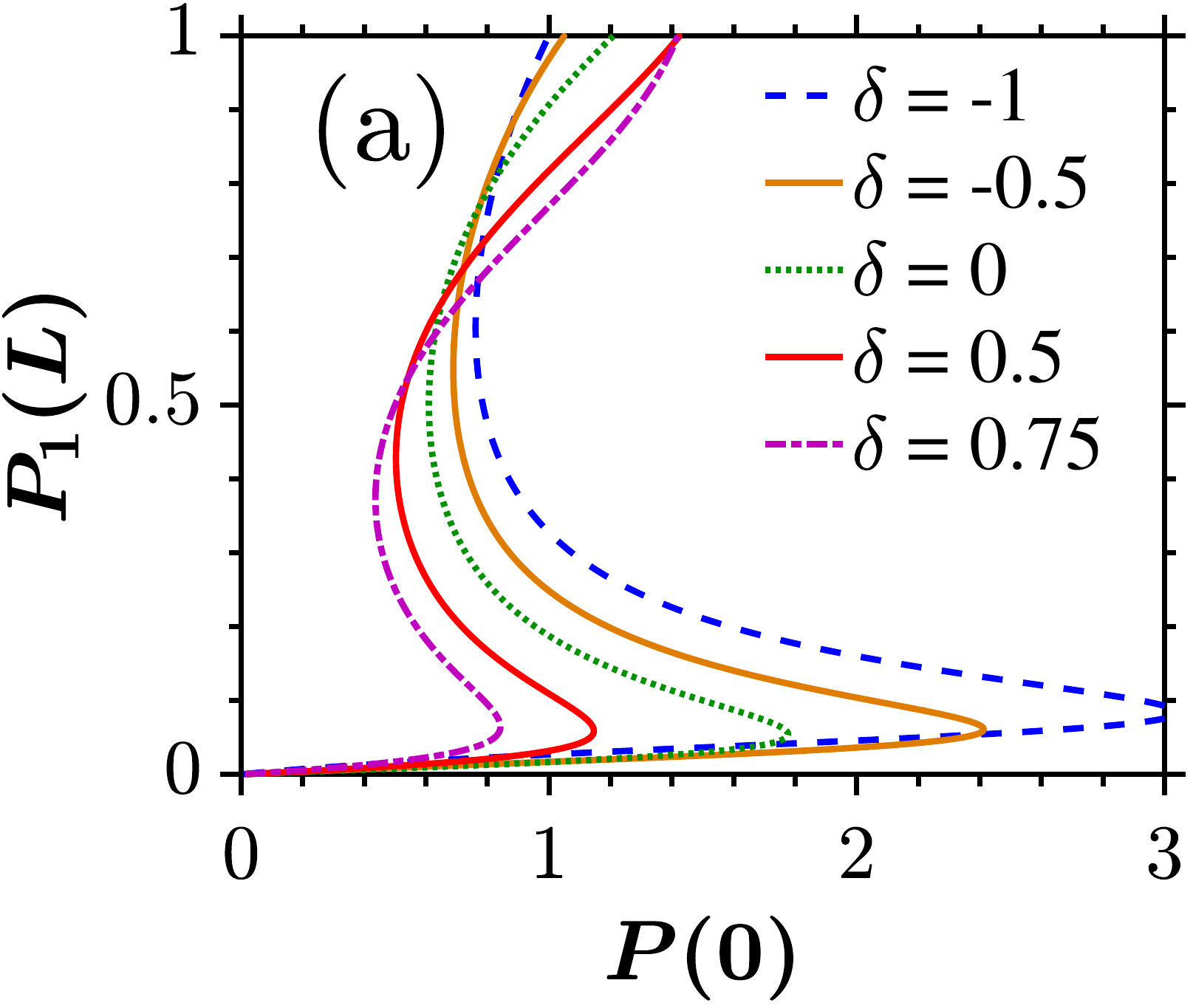}\includegraphics[width=0.5\linewidth]{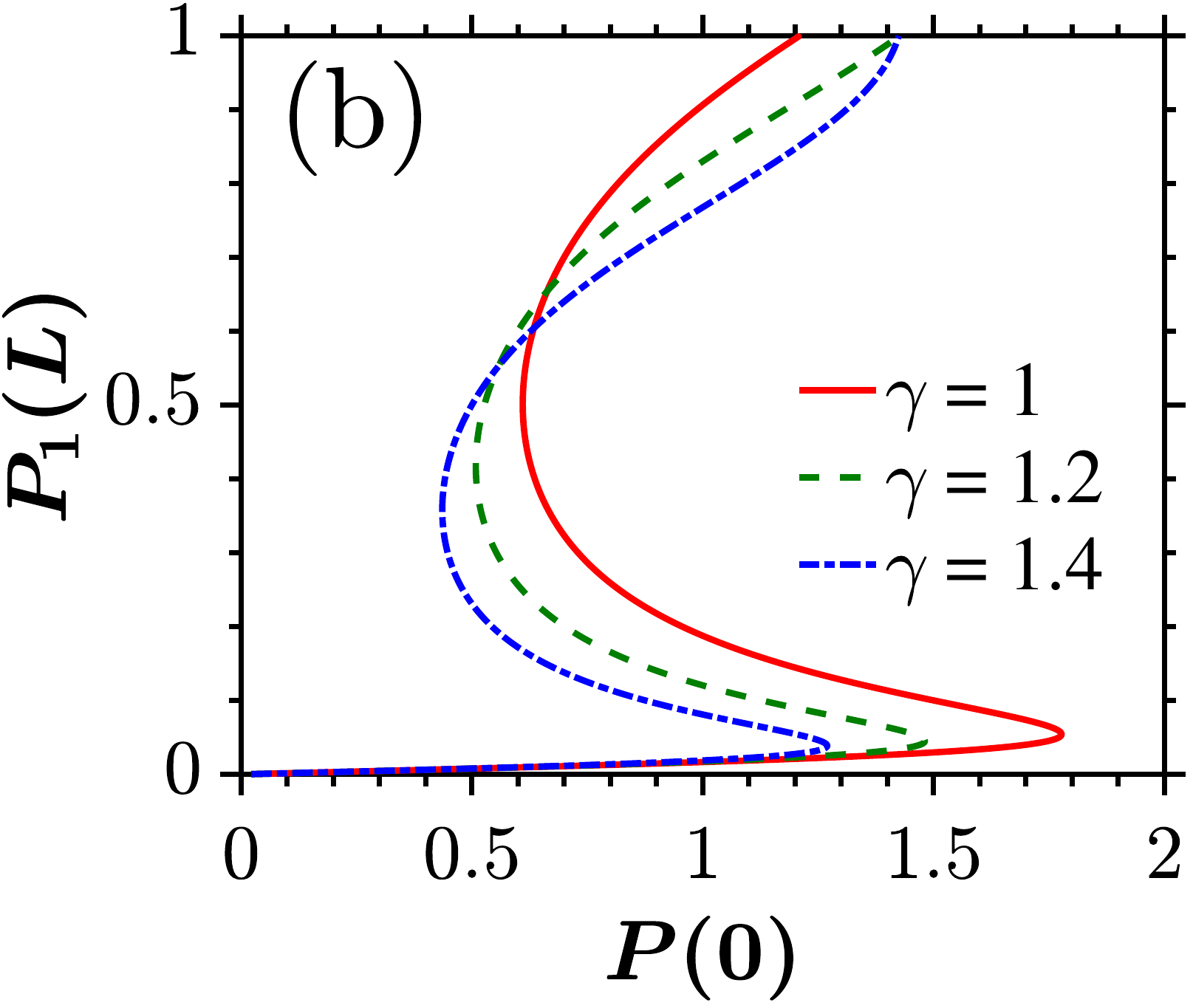}
	\caption{ (Color online) Output intensity $P_{1}$$(L)$ as a function of input intensity $P(0)$ in the unbroken $\mathcal{PT}$-symmetric cubic regime ($\Gamma=\sigma=0$) of the FBG at fixed values of $g=3.75$. Figure (a) is plotted for five different values of detuning parameter at $\gamma=1$. Figure (b) is simulated for three different values of cubic nonlinear coefficient at $\delta=0$.}
	\label{Fig_2}
\end{figure}

\subsection{Role of Kerr (cubic) nonlinearity and detuning parameter}
With a clear-cut idea on the role of length and strength of the
coupling obtained from the previous section, we directly proceed to
study the effect of cubic nonlinearity at a
given value of gain/loss coefficient and other device constraints.
The importance of the detuning parameter ($\delta$) on the bistable phenomenon is pointed out graphically in Fig. \ref{Fig_2}(a). If the signal wavelength is away from the Bragg wavelength, it is said to be detuned and depending on whether it is longer or shorter than the Bragg wavelength, it is designated as negative or positive detuning, respectively. Compared
to the operation at the synchronized wavelength (refer the plot when $\delta=0$), detuning in the shorter
wavelength reduces the threshold and width of the hysteresis whereas the negative
detuning increases the threshold and width of the hysteresis. The switch up intensities for different values of  $\delta=-1$, $-0.5$, $0$,  $0.5$, and $0.75$  at $g=3.75$ are given by $3.041$, $2.413$, $1.777$, $1.145$,
and $0.8422$, respectively. The series of values in descending order representing
the switching intensities look deceiving that one may intend to reduce the threshold
by increasing the detuning further. But this will detune the system
outside the band edges and hence results in insufficient feedback
to create any bistable feature.

The nonlinear parameter of the fiber purely depends on the concentration of the dopant added to the intrinsic
one. Hence we can have a variety of silica fibers possessing different
values of third order nonlinearity. To elucidate the role of nonlinearity
numerically, we set the parameters as $g=3.75$ and $\delta=0$
and vary $\gamma$ from $1$ in steps of $0.2$. As expected,
higher the value of nonlinearity, lesser the intensity required to
switch between the stable states. In Fig. \ref{Fig_2}(b),  the corresponding values of switch up intensity for $\gamma$=1, 1.2, 1.4 are measured as $1.777$, $1.481$, and $1.269$, respectively. Their corresponding switch down intensities are found to be  $0.612$, $0.509$, and $0.436$.
\begin{figure}[t]
	\centering
	\includegraphics[width=0.5\linewidth]{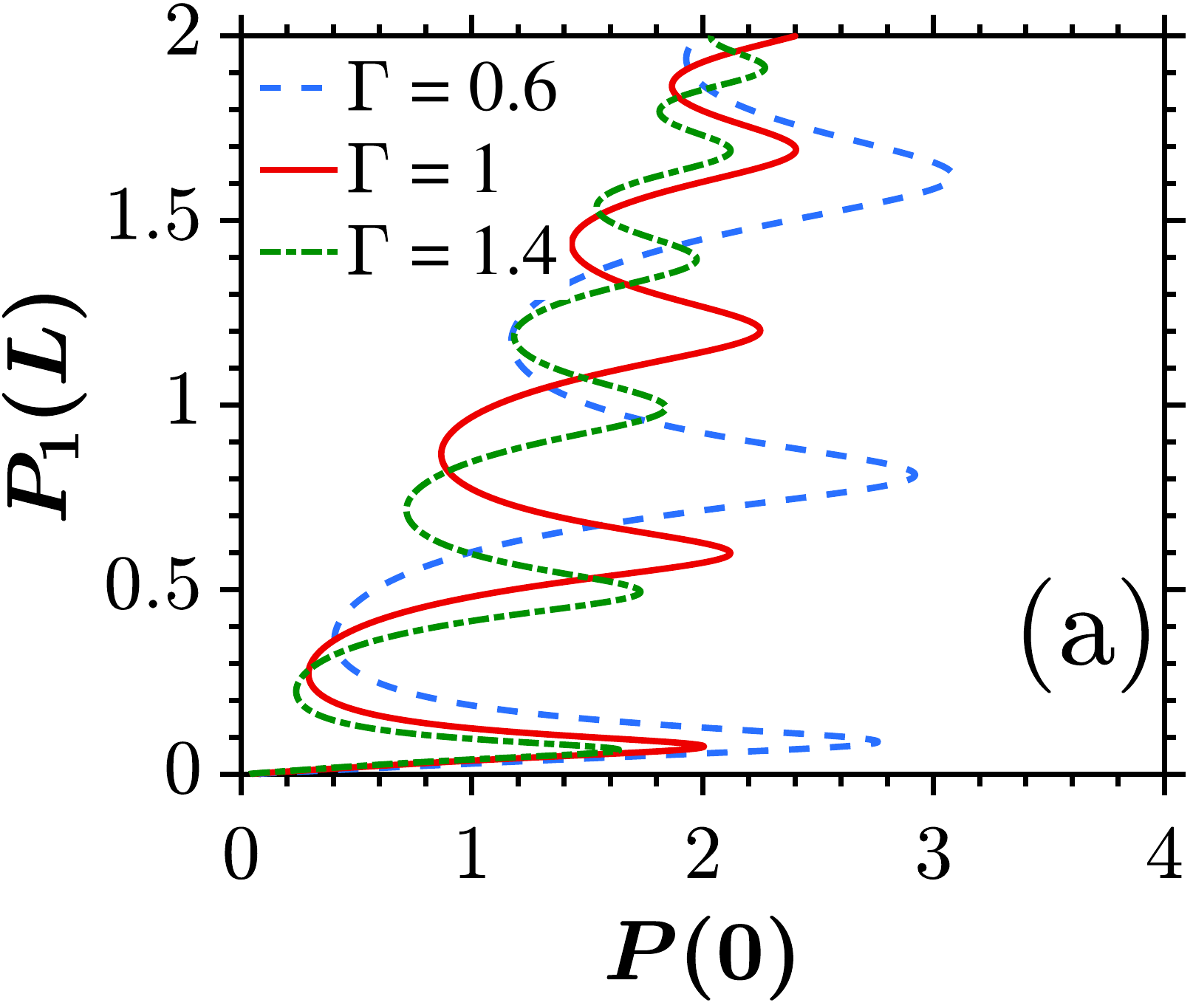}\includegraphics[width=0.5\linewidth]{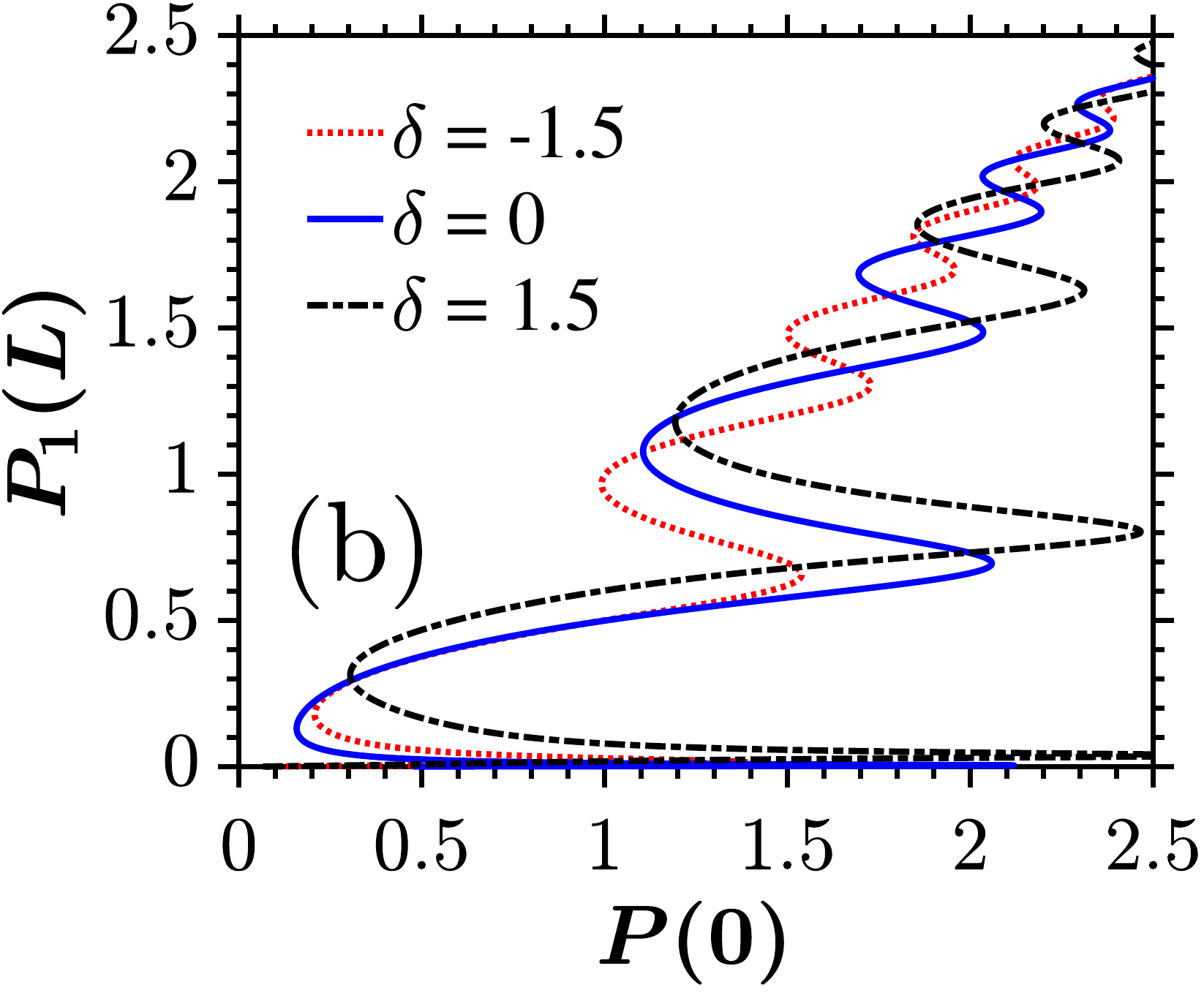}\\
	\includegraphics[width=0.5\linewidth]{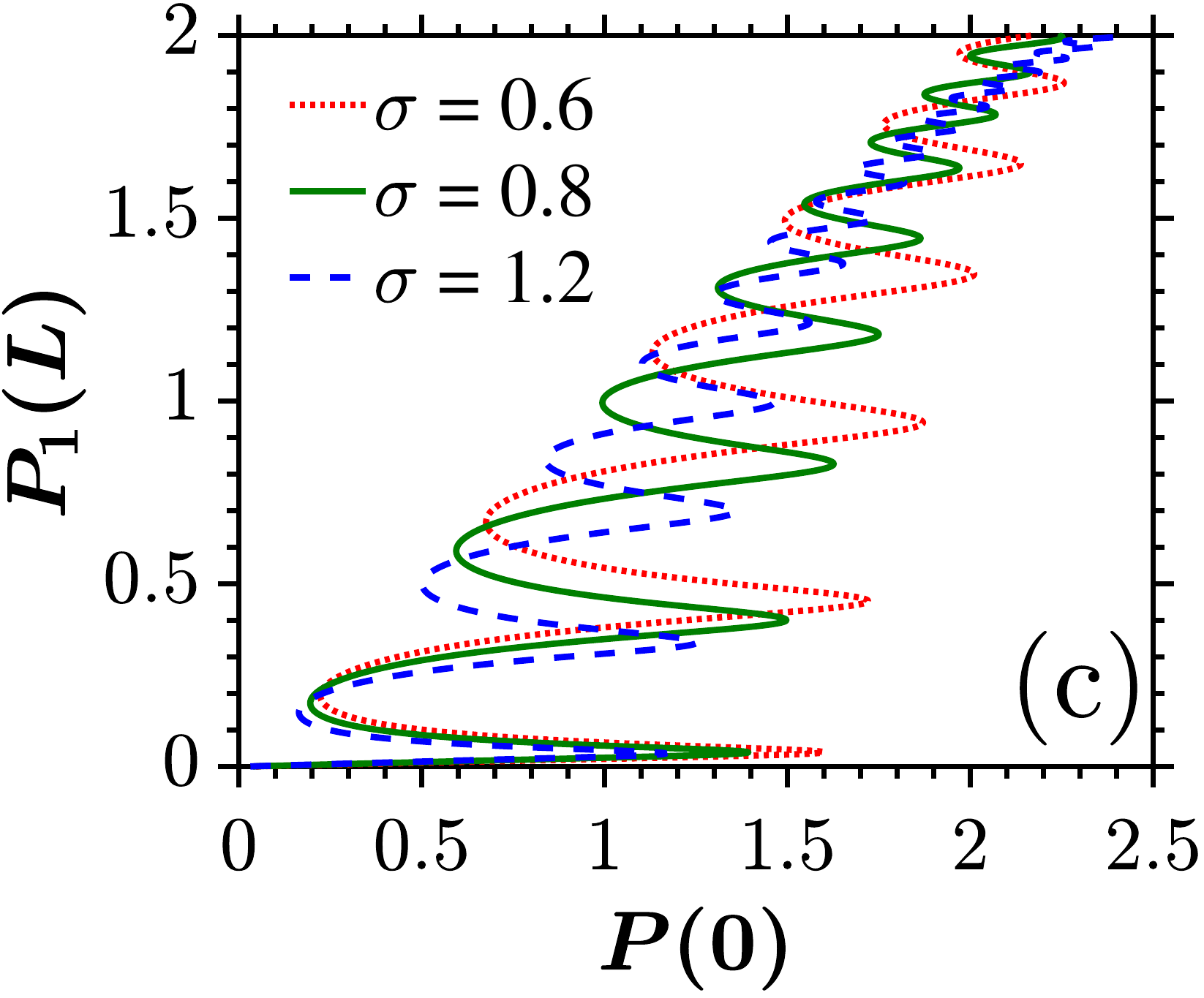}\includegraphics[width=0.5\linewidth]{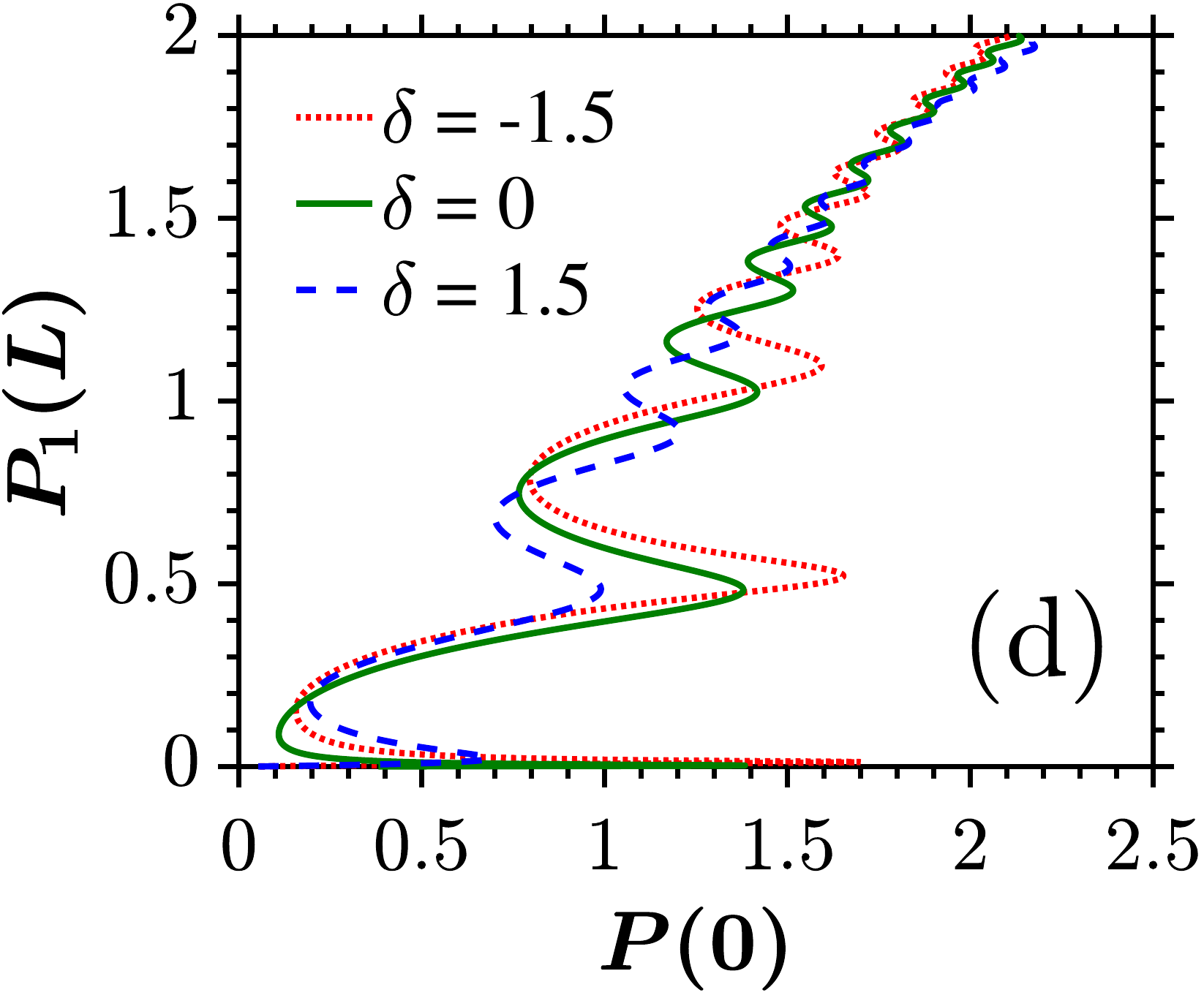}
	\caption{ (Color online) Illustrations of the variations in the output intensity $P_{1}$$(L)$ against input intensity $P(0)$ of an unbroken $\mathcal{PT}$-symmetric FBG. The top panels correspond to the cubic-quintic nonlinear regime  $\gamma=\Gamma=1, \sigma=0$, while the lower panel features FBG with cubic-quintic-septic nonlinearities ($\gamma$= $\Gamma=\sigma=1$). Figures (a) and (b) show the variation with respect to $\Gamma$ and $\delta$, respectively. The bottom panels represent the same for the septic case. The figures on the left and right are plotted at $g=2$ and $g=3.5$, respectively.}
	\label{Fig_4}
\end{figure}

\subsection{Combined effects of cubic-quintic nonlinearities}
To illustrate the effect of cubic-quintic nonlinearity on the switching,
we set $g=2$, $\delta=0$, and $\gamma=1$ (see Fig. \ref{Fig_4}(a)). When $\Gamma=0.6$,
the switch up intensity is very high at $2.76$ and the hysteresis
width is quite large. When $\Gamma$ is increased to unity, the intensity
reduces to $2$ with a slim reduction in switch down intensity from $0.41$
to $0.29$. The other key difference between the two curves is that,
at $\Gamma=1$, there are more stable branches than at $\Gamma=0.6$. The
switching intensities between the adjacent stable branches also get
reduced. More stable states are visible in Fig. \ref{Fig_4}(a) when $\Gamma$ is increased to 1.4. The reduction in the switch up intensity, as well as hysteresis width, is analogous to the cubic nonlinearity
case and so the thumb rule to reduce the switch up intensity is straightforward
to pronounce, that is choose a material with higher nonlinearity regardless
of the regime in which it is operated. But the detuning has a dissimilar
influence on switch-up intensity compared to the cubic effect. In the presence of
cubic nonlinearity alone, the intensity falls off in the positive detuning
regime whereas in the presence of an additional defocusing (quintic) nonlinearity, 
the intensity decreases when operated in the negative detuning regime
as seen from Fig. \ref{Fig_4}(b). The values of switch-up intensities between second and third stable states for $\delta=1.5$, $0$, $-1.5$ are
$2.45$, $2.059$, and $1.539$, respectively. These values are measured at $g=3.5$ and $\gamma=\Gamma=1$.  Interpretation of these outcomes implies that when we include the higher order nonlinearities to the system without imposing any changes to the other parameters, multistable states are observed in its input-output characteristics. These multistable states can be employed in {\it n} level pulse amplitude modulation (PAM) scheme to improve the quality of reconstructed signal provided that the intensity of the regenerated signal is stationed in one of these stable states \cite{zhou2014optical}. Compared to binary modulation scheme, \emph{PAM-4} offers two times larger transmission capacity. Hence FBG with higher order nonlinearities in the $\mathcal{PT}$-symmetric unbroken regime can be used in the all-optical short-haul communication networks.

\subsection{Effect of cubic-quintic-septic nonlinearities}
The number of stable states and the threshold of switching in our system can be controlled with ease by carefully adjusting the system parameters. Theoretically, higher order ($2^{n}$) modulation schemes ($n>3$) can further improve the transmission capacity in short-haul communication network. If such schemes are commercially feasible in the near future, then chalcogenide based FBG with septic nonlinearity can play a key role to setup \emph{n-PAM} signal regenerators, since it admits more number of stable states.
The thumb rule stated in the previous section holds good even in the
presence of septic nonlinearity i.e., higher the septic nonlinear
coefficient lower the switch-up intensity required for switching and the width
of the hysteresis also decreases with increase in $\sigma$ as shown in Fig. \ref{Fig_4}(c). The
switch-up powers between the first stable branch and the second branch
for three different values of $\sigma=0.6$, $0.8$, and $1.2$ are measured as $1.598$, $1.393$, and $1.167$,
respectively. The increase in the number of stable branches at larger
nonlinear coefficients is also similar to the cubic-quintic case.   There
is no big difference between the switch-down power between the first
and second branches when $\sigma$ is varied. But the difference keeps
mounting for the succeeding branches on the top of the first bistable
curve at higher values of $\sigma$ as evident from Fig. \ref{Fig_4}(c).  In Fig. \ref{Fig_4}(d) the switch up intensity at $\delta=0$ is measured as 1.384. Since the septic nonlinearity is a focusing effect, operating the device in the positive detuning regime decreases the switch up intensity (0.6848 at $\delta=1.5$), whereas in the negative detuning regime  the switch up intensity is increased (1.7 at $\delta=-1.5$). So we conclude that the role of detuning in power reduction merely depends on the nature of the higher order nonlinearities whether it is self-focusing or defocusing.

\subsection{Role of gain/loss parameter on various nonlinearities}
\begin{figure}[t]
	\centering
	\includegraphics[width=0.5\linewidth]{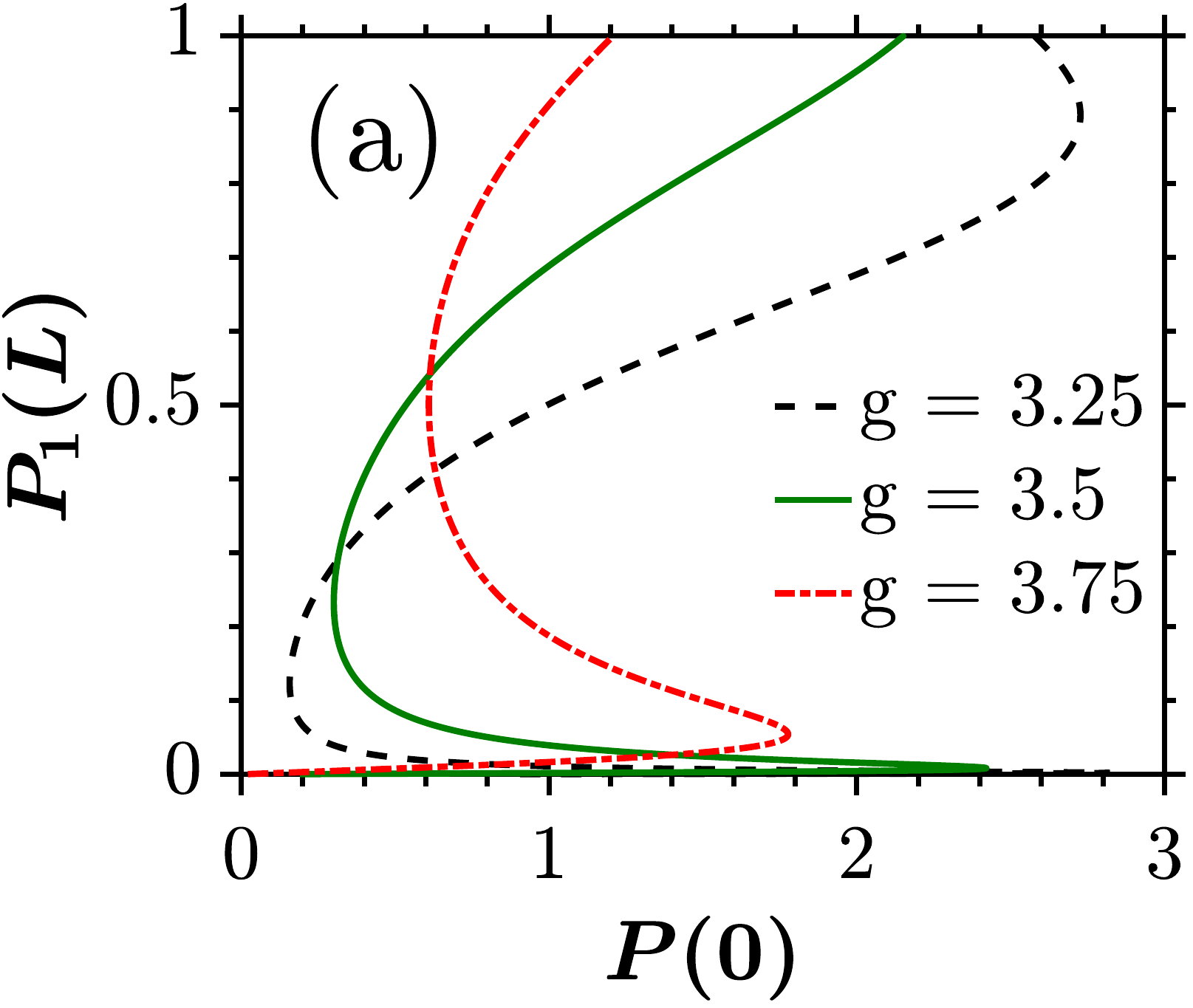}\includegraphics[width=0.5\linewidth]{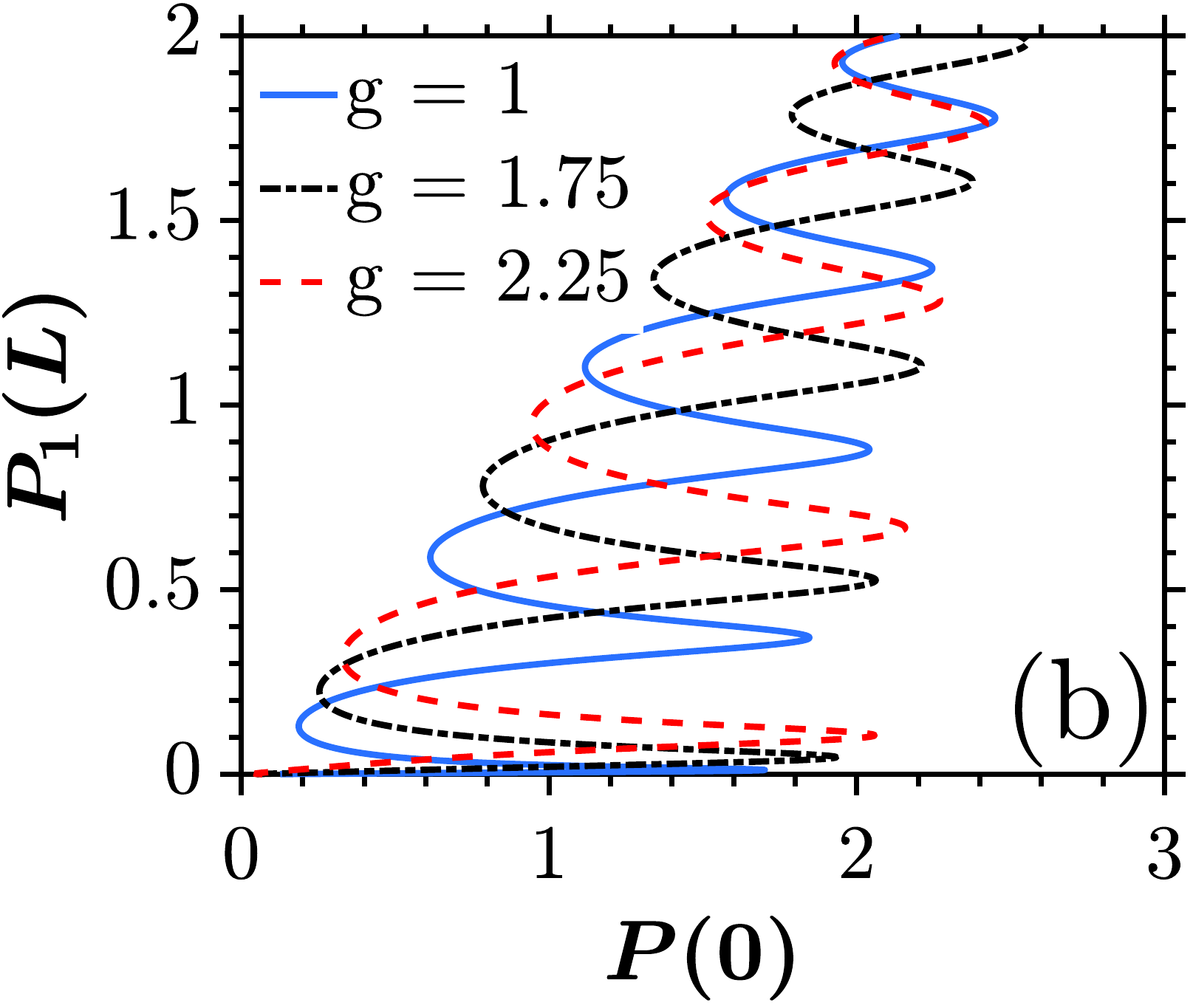}\\\includegraphics[width=0.5\linewidth]{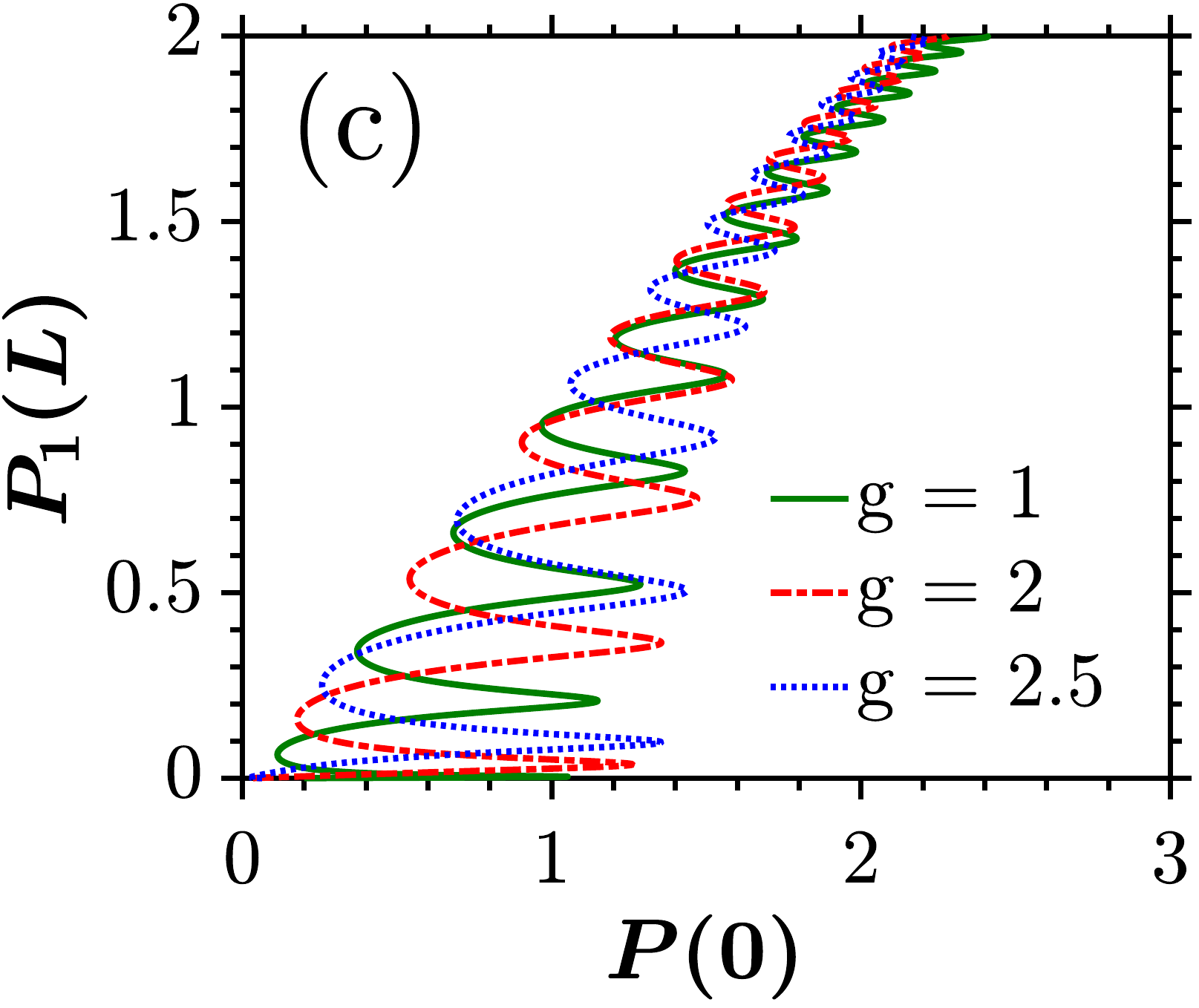}	
	\caption{ (Color online) Plot of the variation in the input-output characteristics of the  unbroken $\mathcal{PT}$-symmetric FBG at different values of $g$ with $\delta=0$ for the left light incidence. The variations are plotted in the presence of  (a) cubic nonlinearity alone ($\gamma=1, \Gamma= \sigma=0$) (b) cubic-quintic nonlinearities ($\gamma=\Gamma=1, \sigma=0$), and (c) combination of cubic-quintic-septic nonlinearities  ($\gamma=\Gamma=\sigma=1$).}
	\label{Fig_3}
\end{figure}

\begin{figure}[t]
	\centering
	\includegraphics[width=0.5\linewidth]{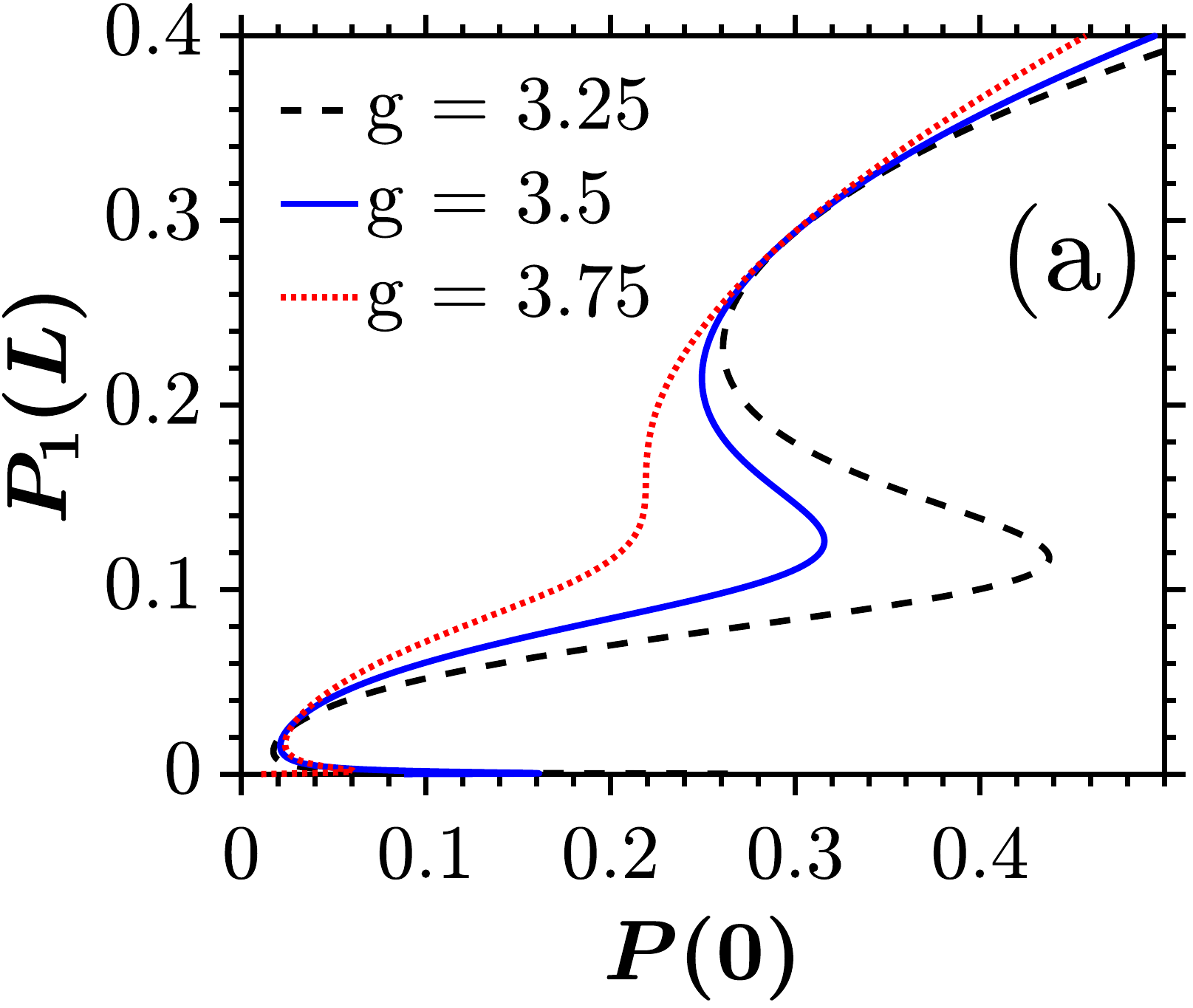}\includegraphics[width=0.5\linewidth]{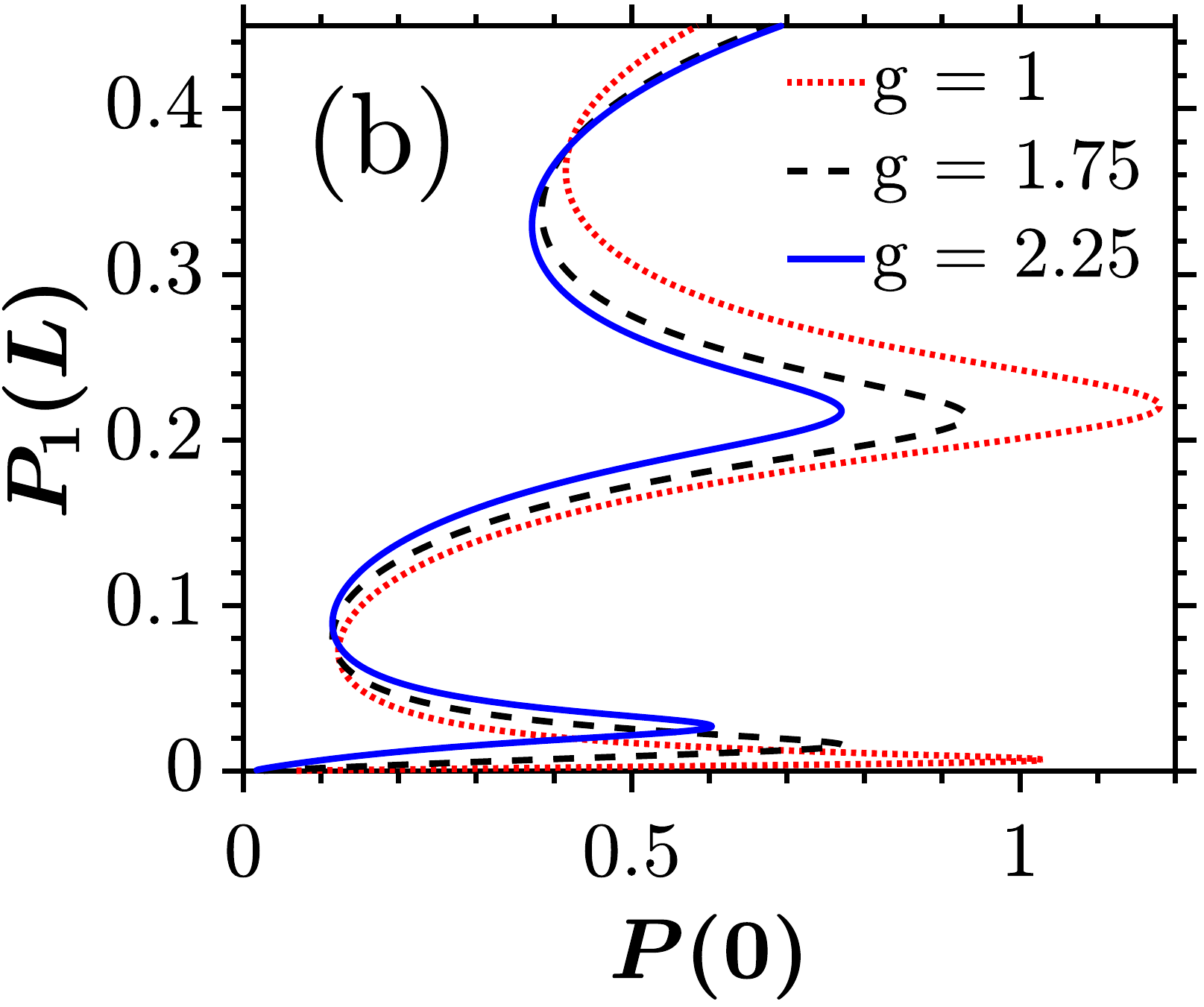}\\
	\includegraphics[width=0.5\linewidth]{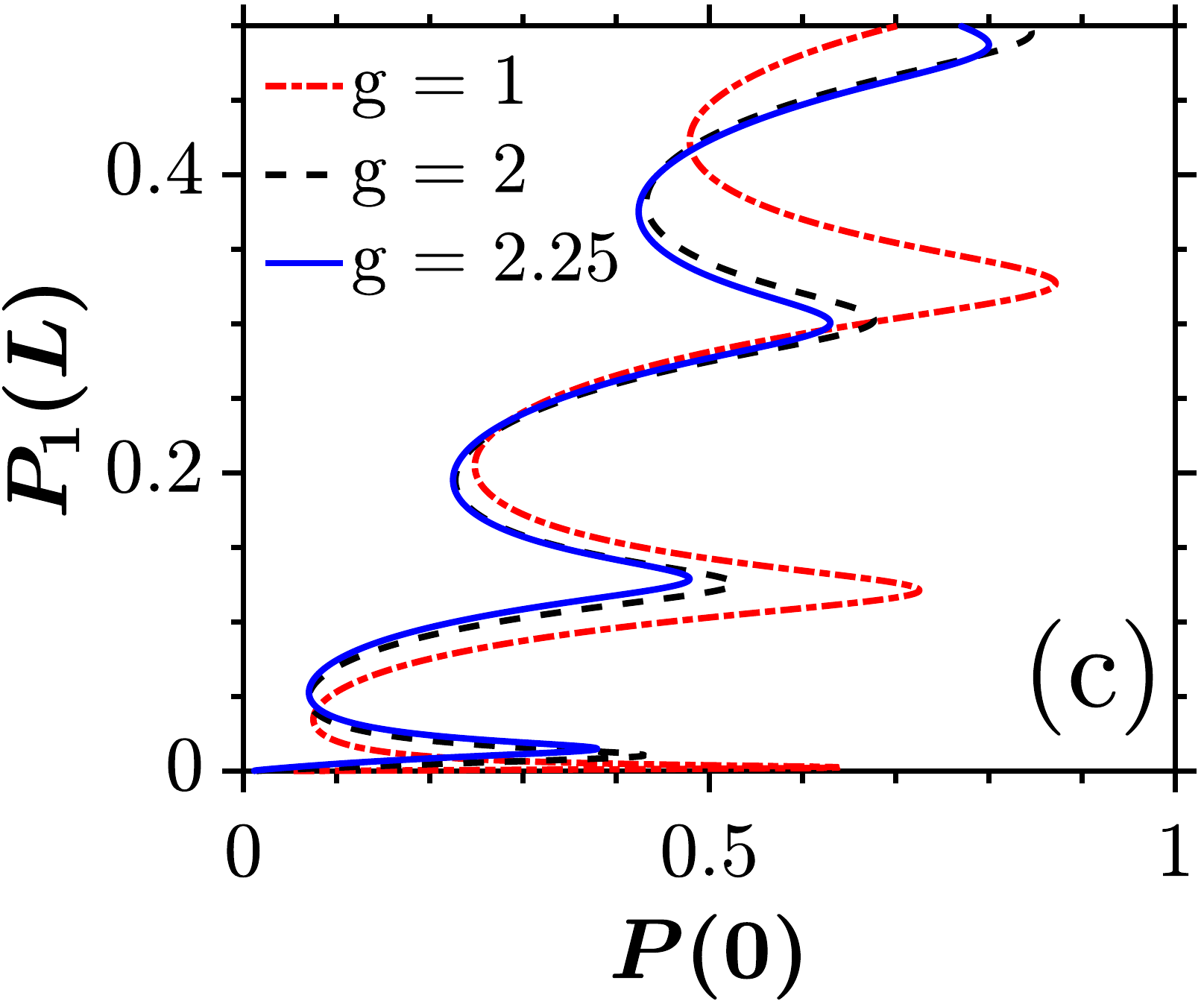}
	\caption{ (Color online)  Plots illustrating the same dynamics with the same system parameter as in Fig. \ref{Fig_3}  when the light is launched from the right input surface of the FBG.}
	\label{Fig_3_rr}
\end{figure}

It has been reported that any increase in the value of gain/loss coefficient ($g$) increases the intensity levels required to switch between the stable states  in the presence of cubic nonlinearity \cite{1555-6611-25-1-015102, lin2011unidirectional, komissarova2019pt}. But this is true only for a certain range of $g$ values. For the values of $g$, closer to the value of $k$ it results in the decrease of the switching intensities as in Fig. \ref{Fig_3}(a). Next, we look into the effect of $g$ in the presence of higher order
nonlinearities. 
When $g$ is increased gradually in the presence of quintic nonlinearities without violating the unbroken $\mathcal{PT}$-symmetric
conditions, the switch-up intensity builds-up to $2.058$ ($g=2.25$) via $1.933$ ($g=1.75$), and $1.702$ ($g=1$)
as seen in Fig. \ref{Fig_3}(b).  It
does not return to the lower branch at the same intensity when the input intensity is
decreased. It sustains in the same branch till the input intensity
is reduced below $0.332$, $0.26$, and $0.19$, respectively, for the above mentioned values of $g$. There is a marginal decrease in the switch-up intensity between the first and second stable branches in the above mentioned values.

Figure \ref{Fig_3}(c) exemplifies the role of gain/loss parameter on the system
that incorporates all the higher order nonlinearities
($\gamma=\Gamma=\sigma=1$) at the Bragg wavelength for three different values
of $g$=$1$, $2$, and $2.25$.  The switch-up intensity starts to ascend in the order
$1.087$, $1.261$, and $1.435$ and the corresponding values of switch down intensities are measured to be $0.1154$, $0.1776$, $0.26$. 
It is very obvious from these studies in the unbroken regime  that the desired bistable or multistable curves can be manipulated at ease by judiciously adjusting the imaginary part of the complex refractive index. 

The phase shifted gratings are well known for exhibiting low intensity switching behaviors compared to other grating structures \cite{radic1994optical,radic1995theory}. The $\mathcal{PT}$-symmetric FBGs can exhibit such a low intensity switching if the direction of incidence of the input pulse is reversed which is not at all feasible in a conventional FBG, since it exhibits same bistable and multistable behaviors in both directions.  Thus the switching phenomenon in a $\mathcal{PT}$-symmetric FBG can be termed as nonreciprocal switching in the sense that it can have an entirely different switching dynamics for left and right incidences. 

It is noteworthy to mention at this juncture that the parameter $g$ has a dual role in controlling the switching intensities of the system. By dual role we mean that the device exhibits  distinguishable OB/OM curves for the same value of $g$ and other system parameters except for left and right light incidence directions. For instance, there is a weak bistable curve formation at $g = 3.75$ in Fig. \ref{Fig_3_rr}(a) for the right incidence. Moreover, the switching intensities decreases with an increase in $g$ for the right incidence in contrast to the left light incidence. This sort of dual nature of parameter $g$ on the switching intensities persists even in the presence of quintic nonlinearity as shown in Fig. \ref{Fig_3_rr}(b) and we can observe more than two stable states for input powers less than unity. Such a formation of low power multistable states was not observed in Fig. \ref{Fig_3}(a). 
This kind of inverse relationship between the parameter $g$ and the switching intensities and the formation of multistable states at low value of input intensities lasts even with the addition of septic nonlinearity to the system as seen in Fig. \ref{Fig_3_rr}(b).  The reason for such a behavior is apparent from the fact that the increase in $g$ 
suppresses the absorption of the forward field intensity for right incidence and vice versa.

\section{Broken $\mathcal{PT}$-symmetric regime}
\label{Sec:4}

\subsection{Influence of gain/loss parameter}

\begin{figure}[h]
	\centering
	\includegraphics[width=0.5\columnwidth]{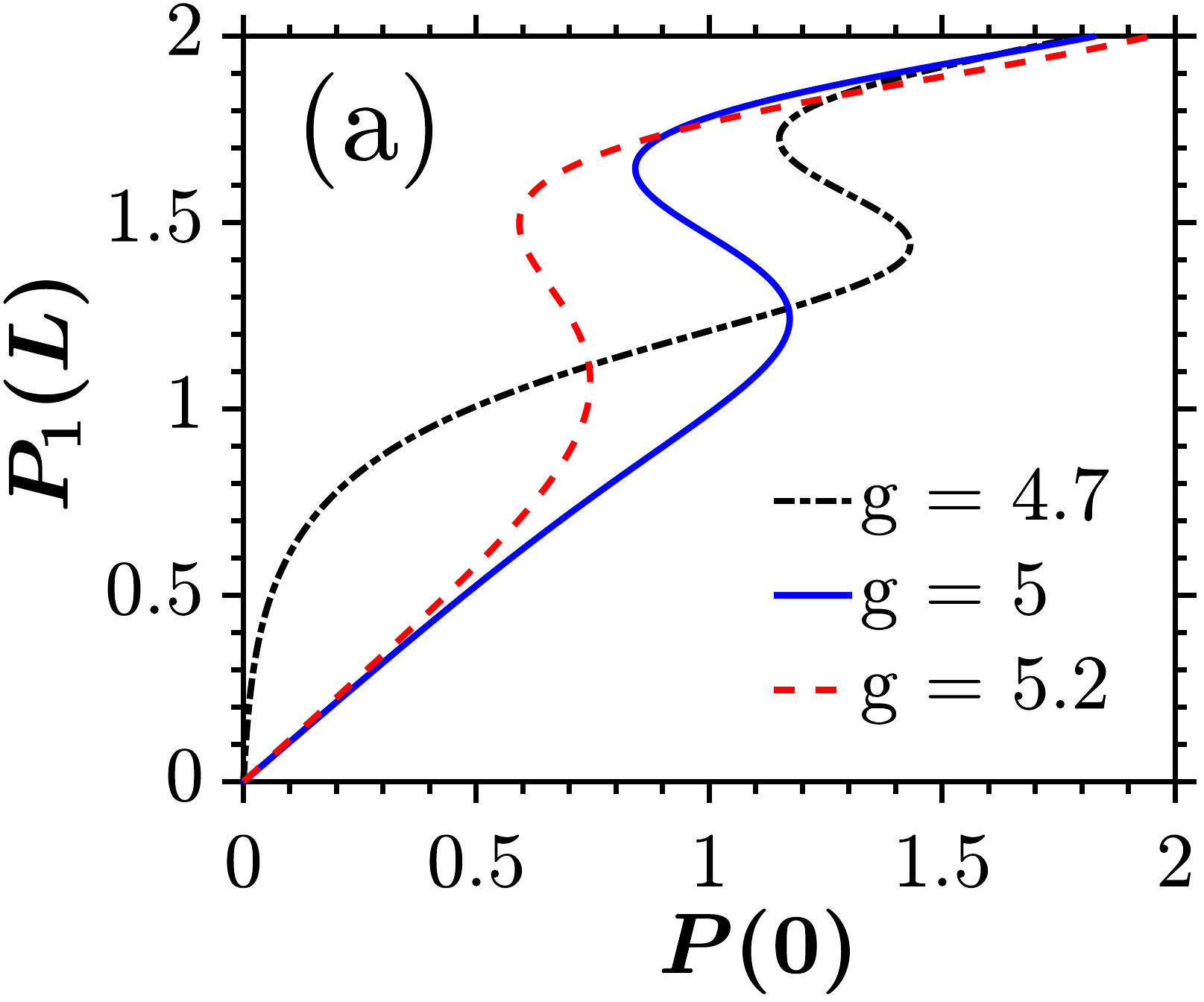}\includegraphics[width=0.5\columnwidth]{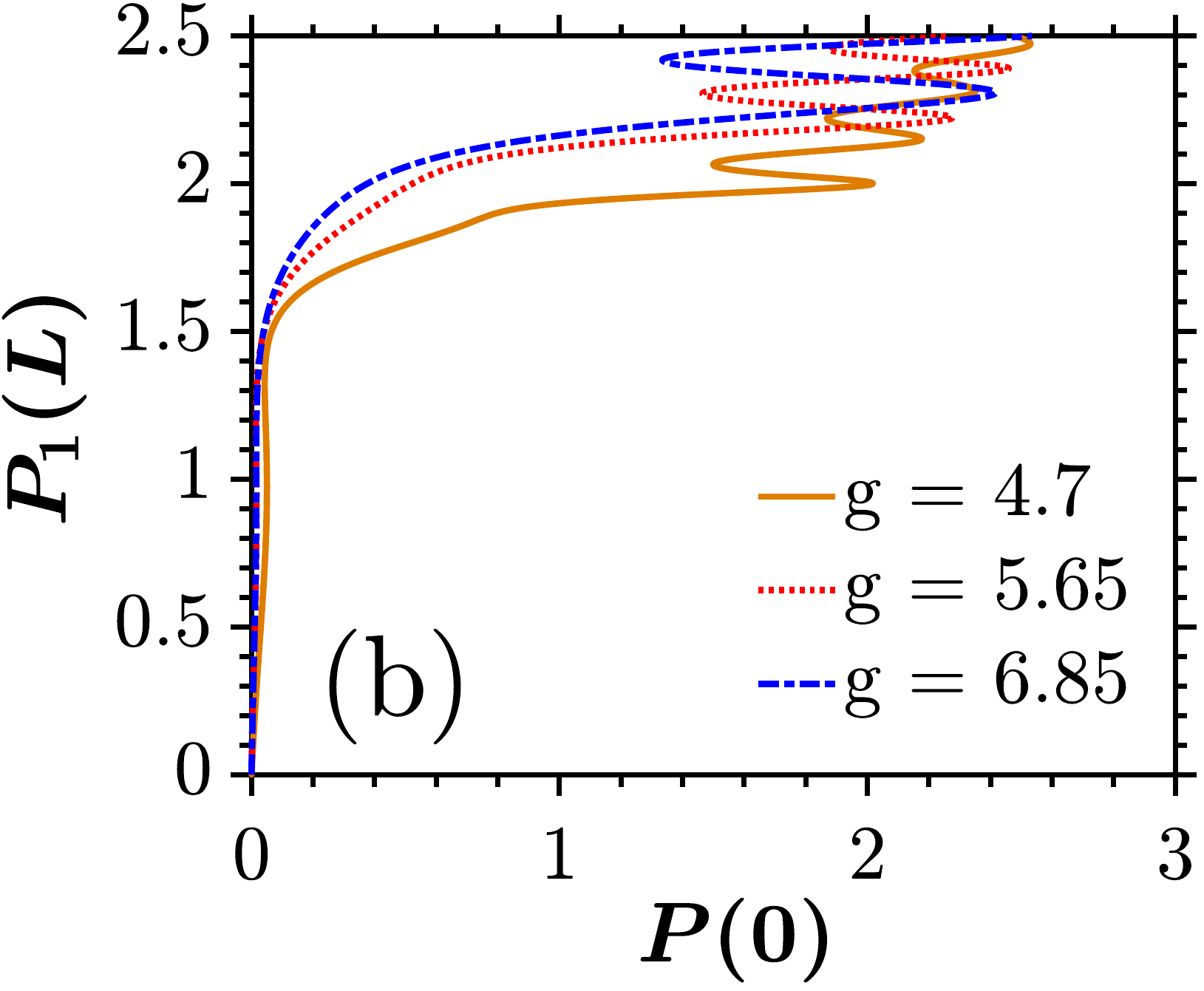}\\\includegraphics[width=0.5\columnwidth]{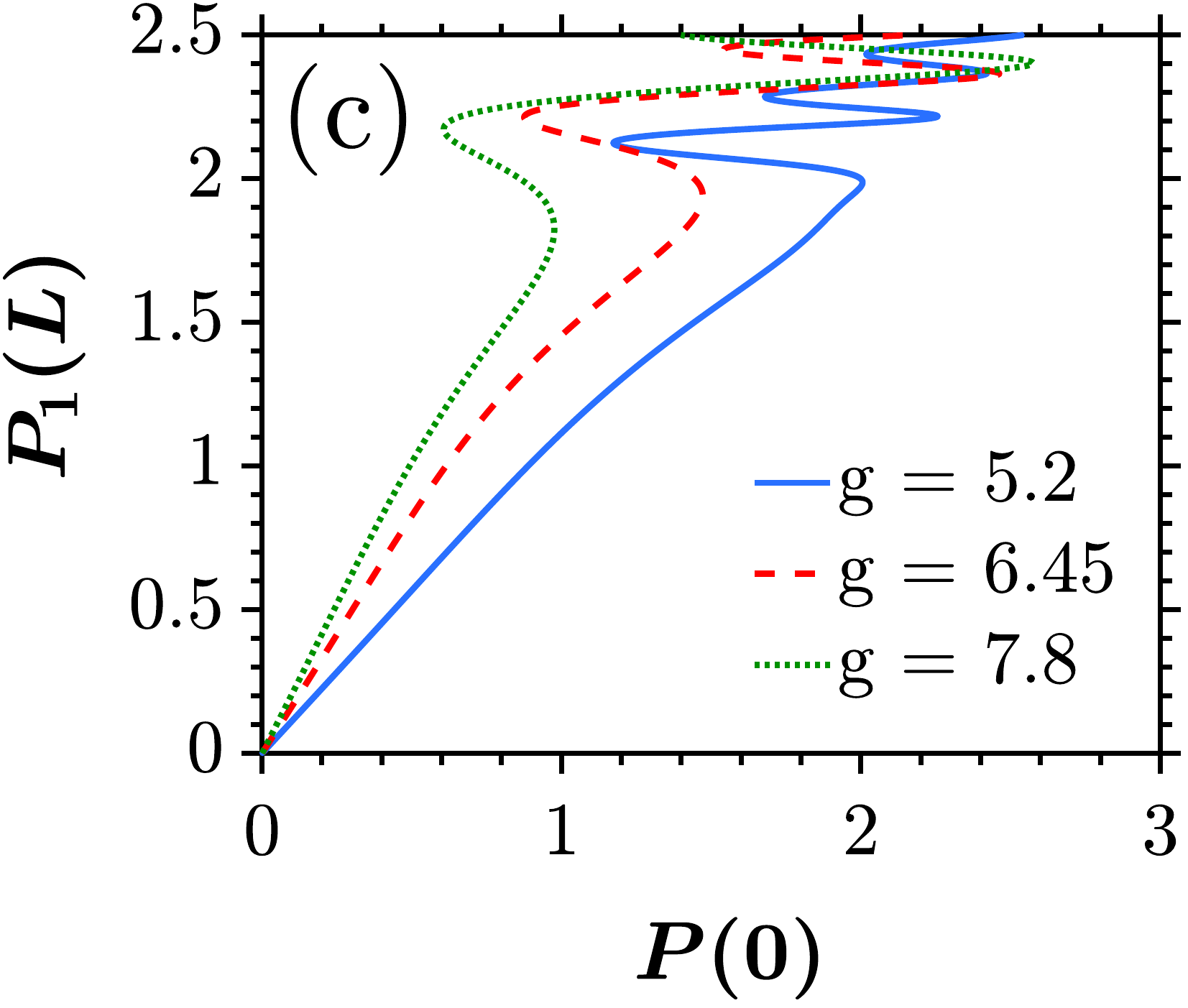}\includegraphics[width=0.5\columnwidth]{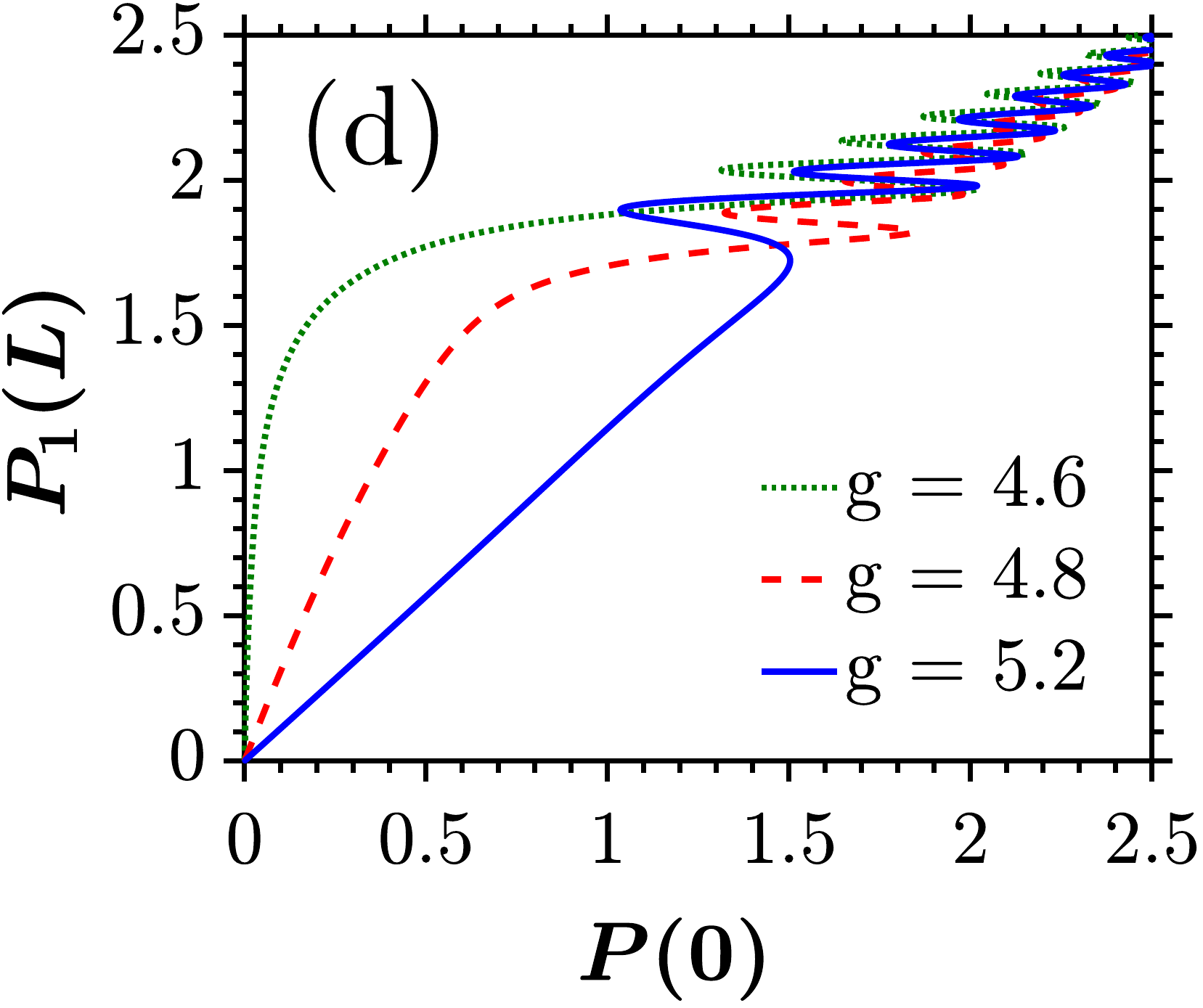}\\
	\caption{ (Color online) Plots depicting the novel optical bistability in a broken $\mathcal{PT}$-symmetric FBG for different values of $g$ at $\delta=0$ for left incidence. Here the output and the input intensities of the system are represented by the variables $P_{1}(L)$ and $P(0)$ respectively. Figure (a) is simulated in the presence of cubic nonlinearity alone ($\gamma=1$, $\Gamma= \sigma= 0$).  Figures (b) and (c) represent the role of $g$ in the presence of cubic-quintic nonlinearities ($\gamma$= $\Gamma=1$, $\sigma=0$). Figure (d) is plotted in the presence of  cubic-quintic-septic nonlinearities ($\gamma$ = $\Gamma=1$, $\sigma=0.6$), respectively. }
	\label{Fig1}
\end{figure}

\begin{figure}[h]
	\centering
	\includegraphics[width=0.5\columnwidth]{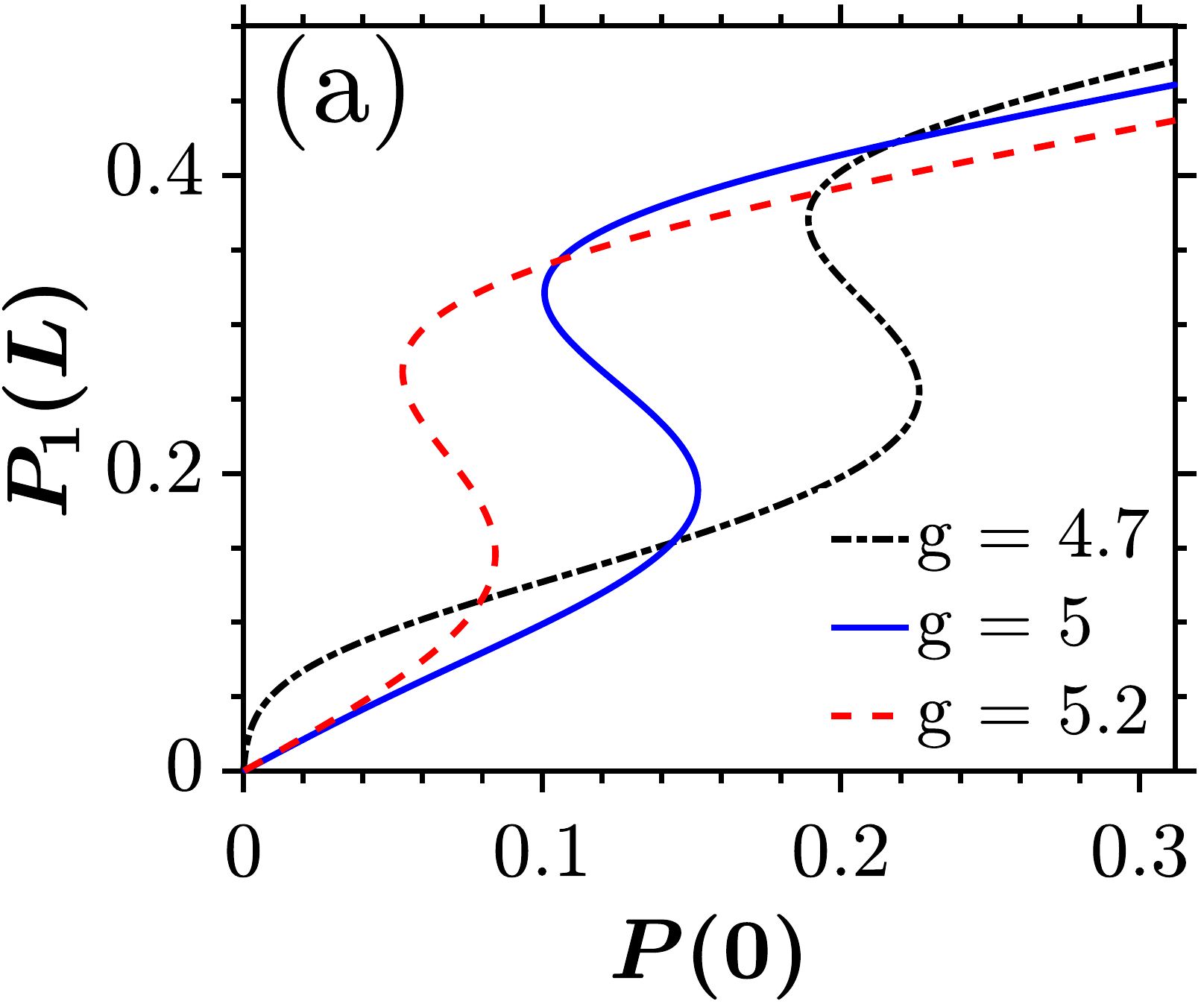}\includegraphics[width=0.5\columnwidth]{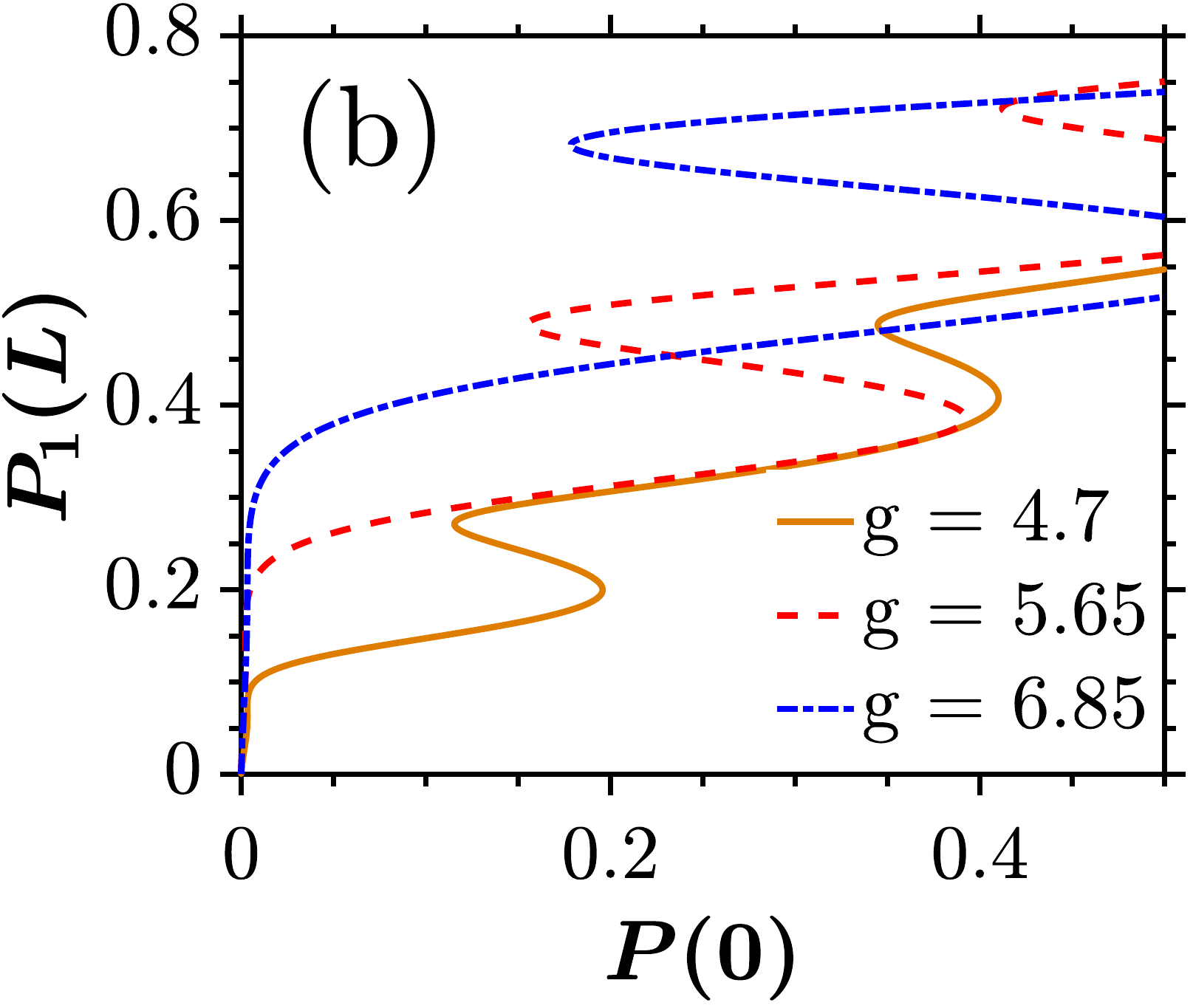}\\\includegraphics[width=0.5\columnwidth]{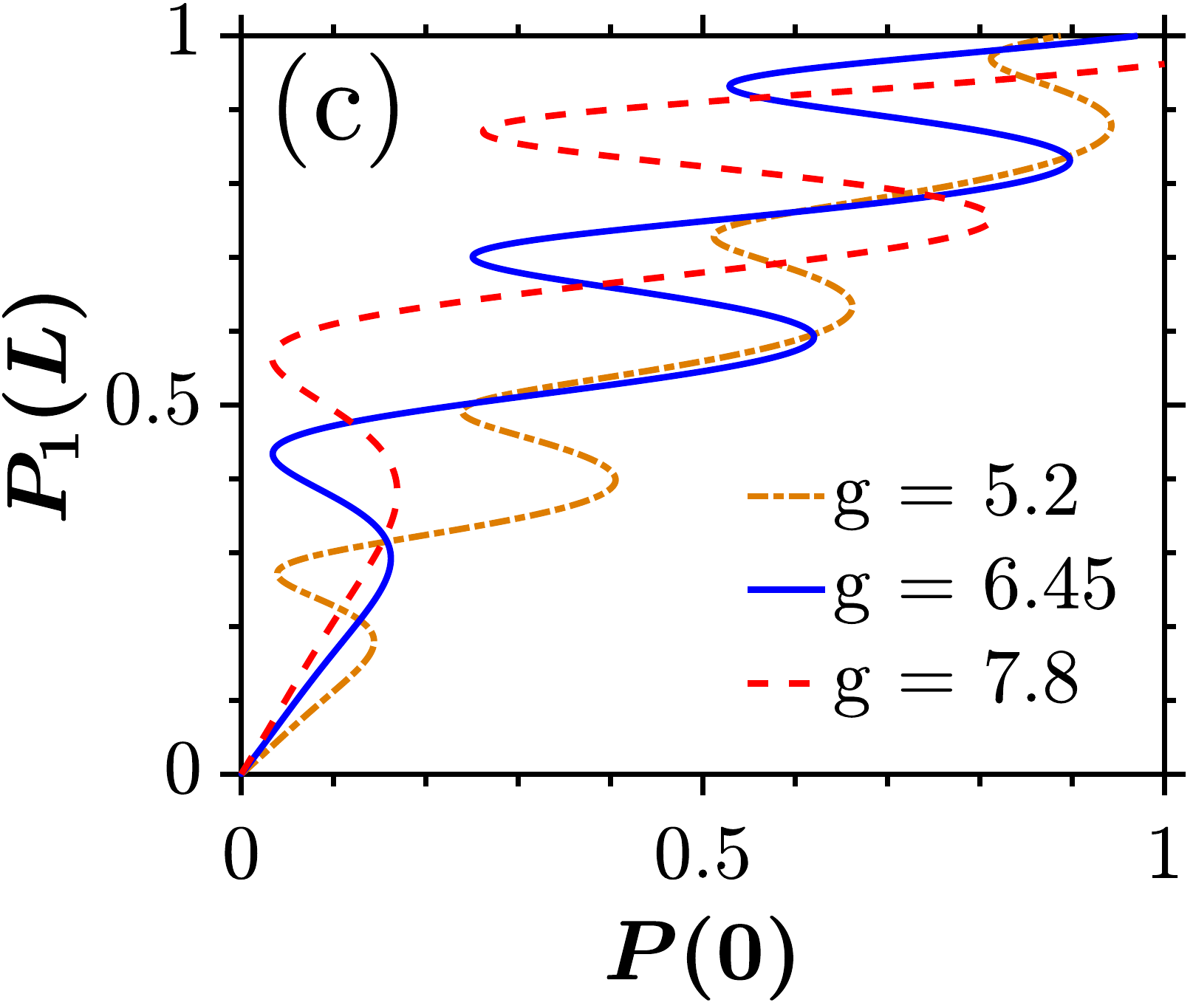}\includegraphics[width=0.5\columnwidth]{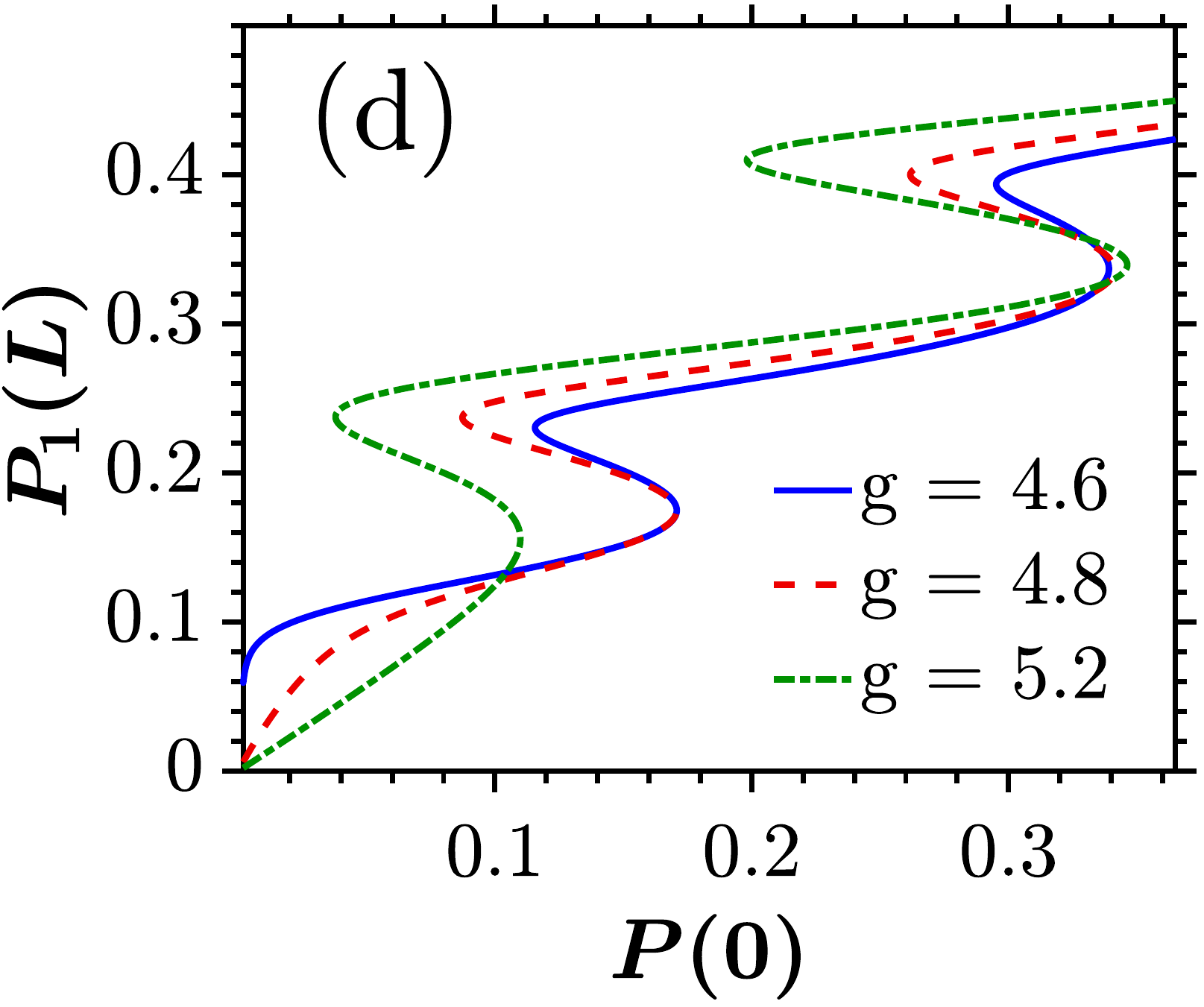}\\
	\caption{ (Color online) Plots depicting the same dynamics with the same system parameters as in Fig. \ref{Fig1} for the right light incidence. }
	\label{Fig1_rr}
\end{figure}

\begin{figure}[h]
	\centering
	\includegraphics[width=0.53\columnwidth]{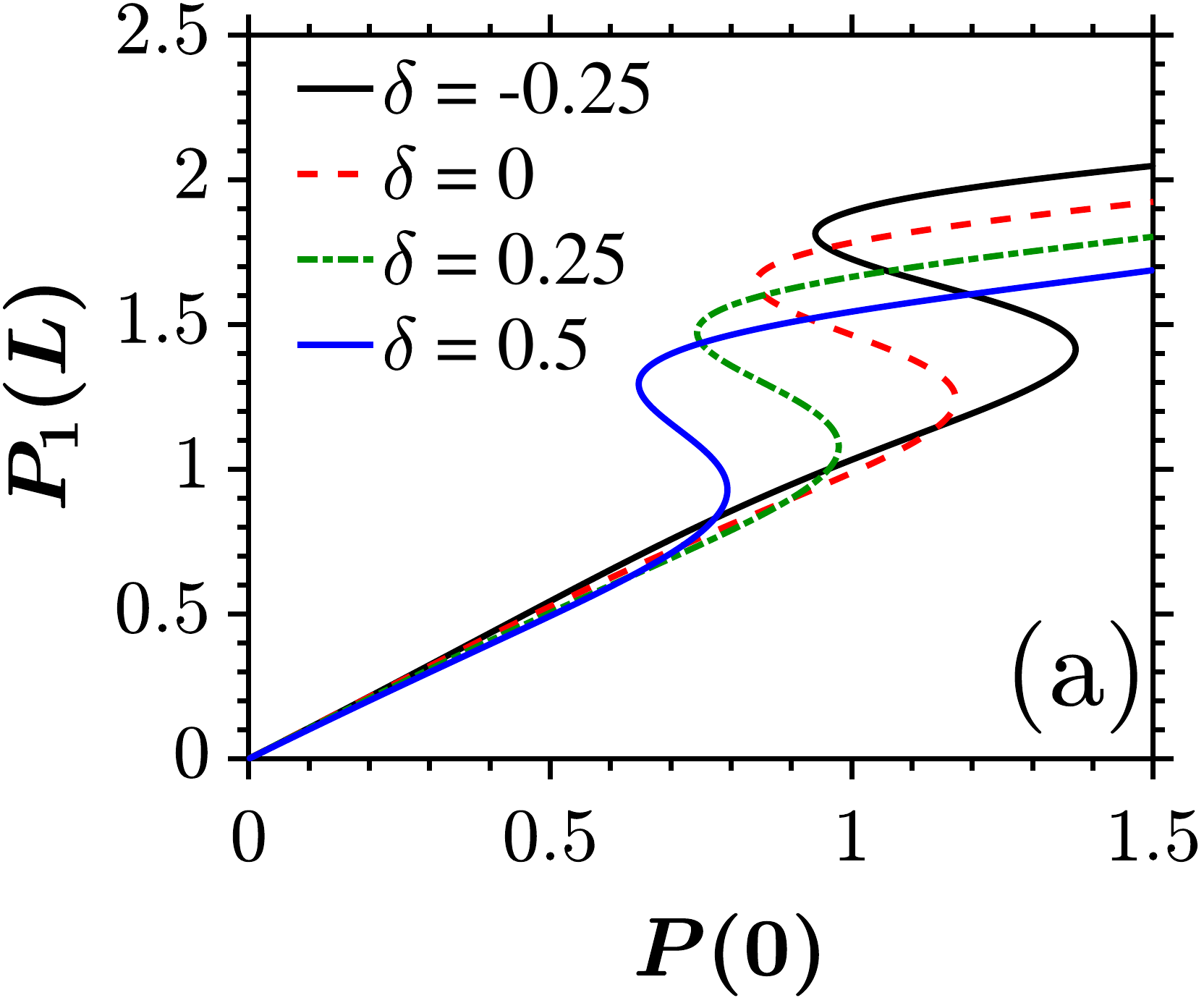}\includegraphics[width=0.5\columnwidth]{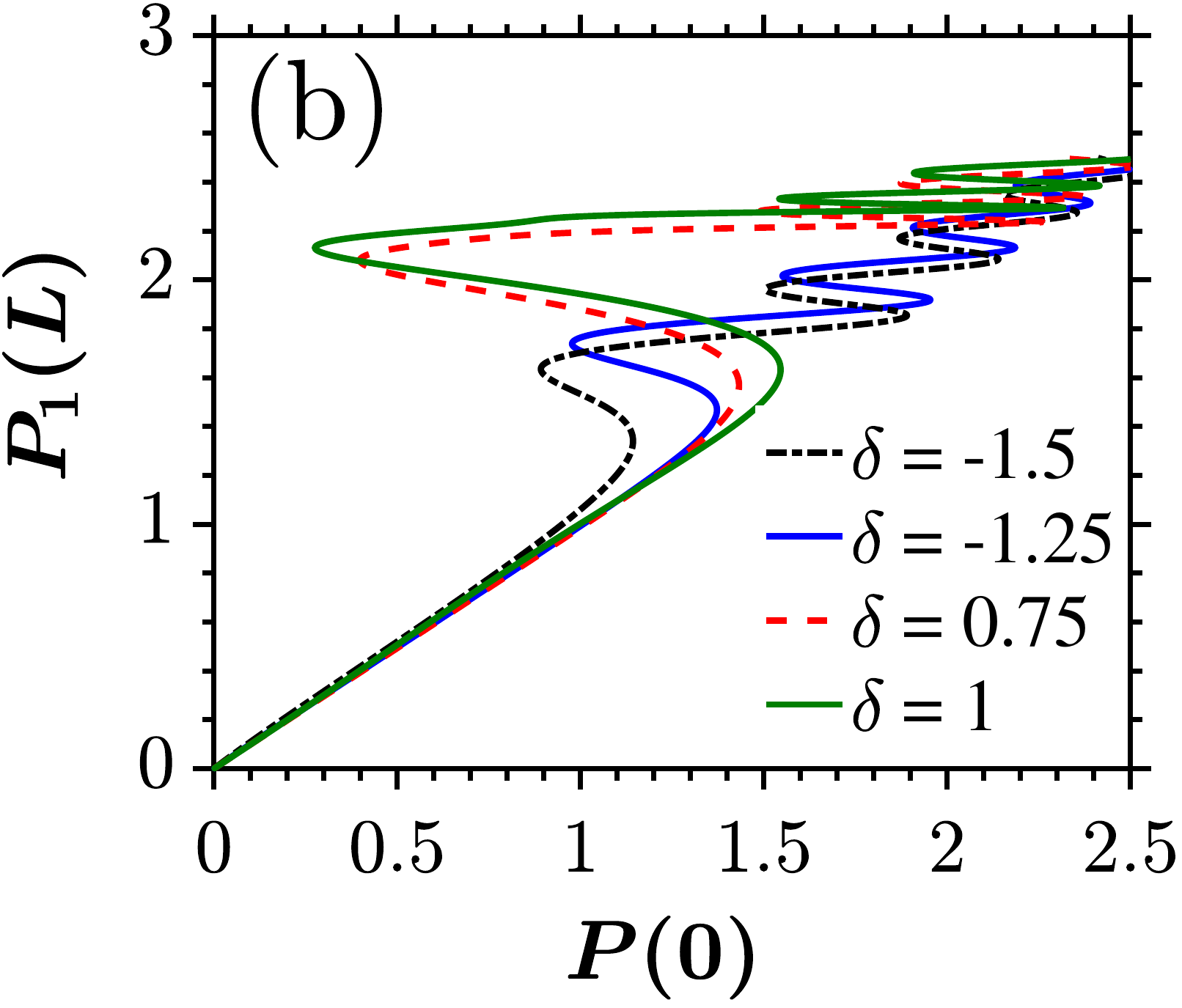}\\
	\includegraphics[width=0.5\columnwidth]{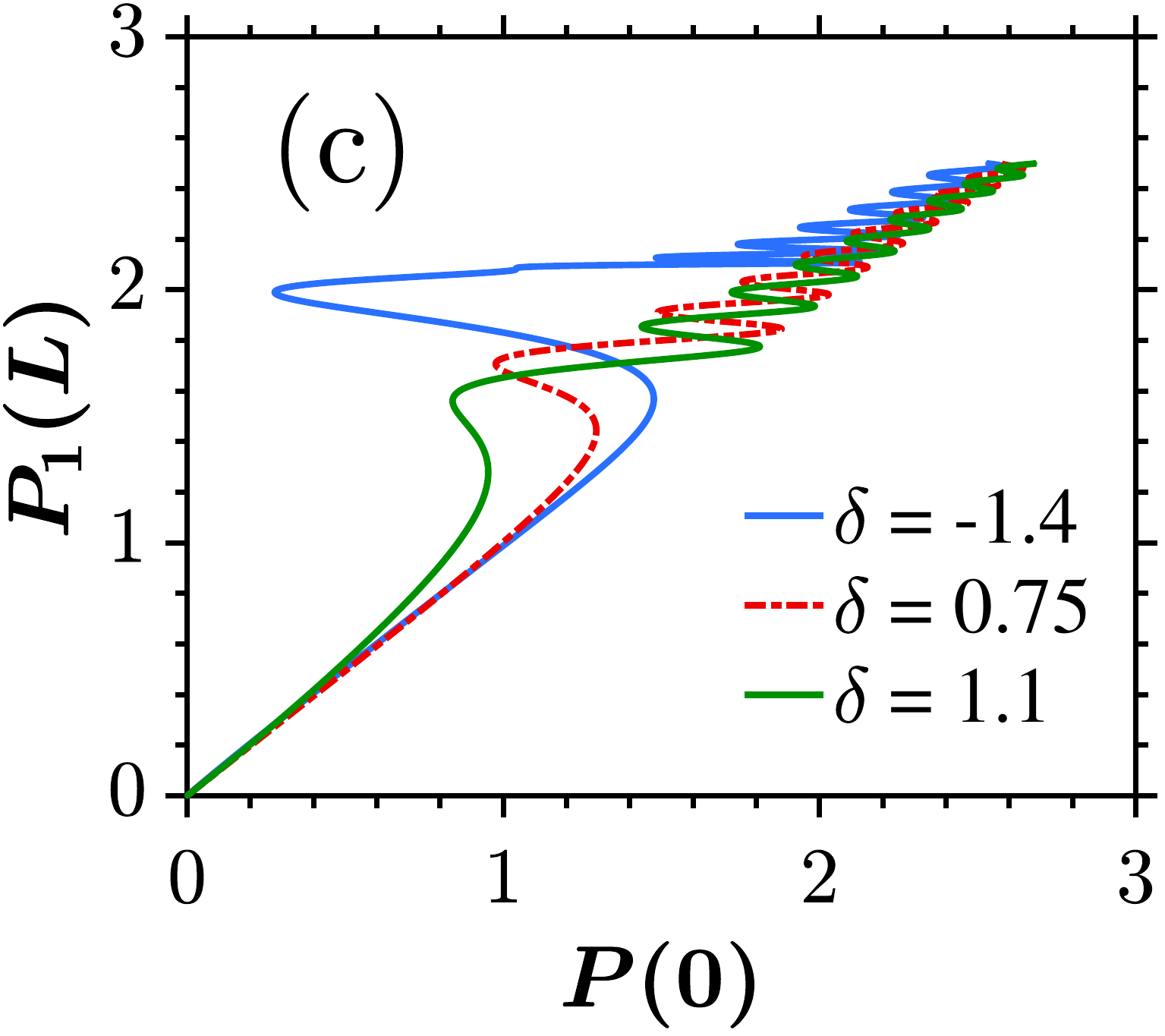}
	\caption{ (Color online) Effect of varying the detuning parameter $\delta$  on the input-output characteristics of a broken $\mathcal{PT}$-symmetric  FBG at $g=5$. Here the output and input powers are represented by $P_{1}(L)$ and $P(0)$ respectively. Figure (a) is plotted in the presence of cubic nonlinearity alone ($\gamma=1.5$, $\Gamma=$ $\sigma=0$). Figure (b) represents the simulated results in the presence of cubic-quintic nonlinearities ($\gamma=$ $\Gamma=1$, $\sigma=0$). Figure (c) is plotted in the presence of cubic-quintic-septic nonlinearities ($\gamma=$ $\Gamma=1$, $\sigma=0.6$). }
	\label{Fig2}
\end{figure}

\begin{figure}[ht]
	\centering
	\includegraphics[width=0.5\columnwidth]{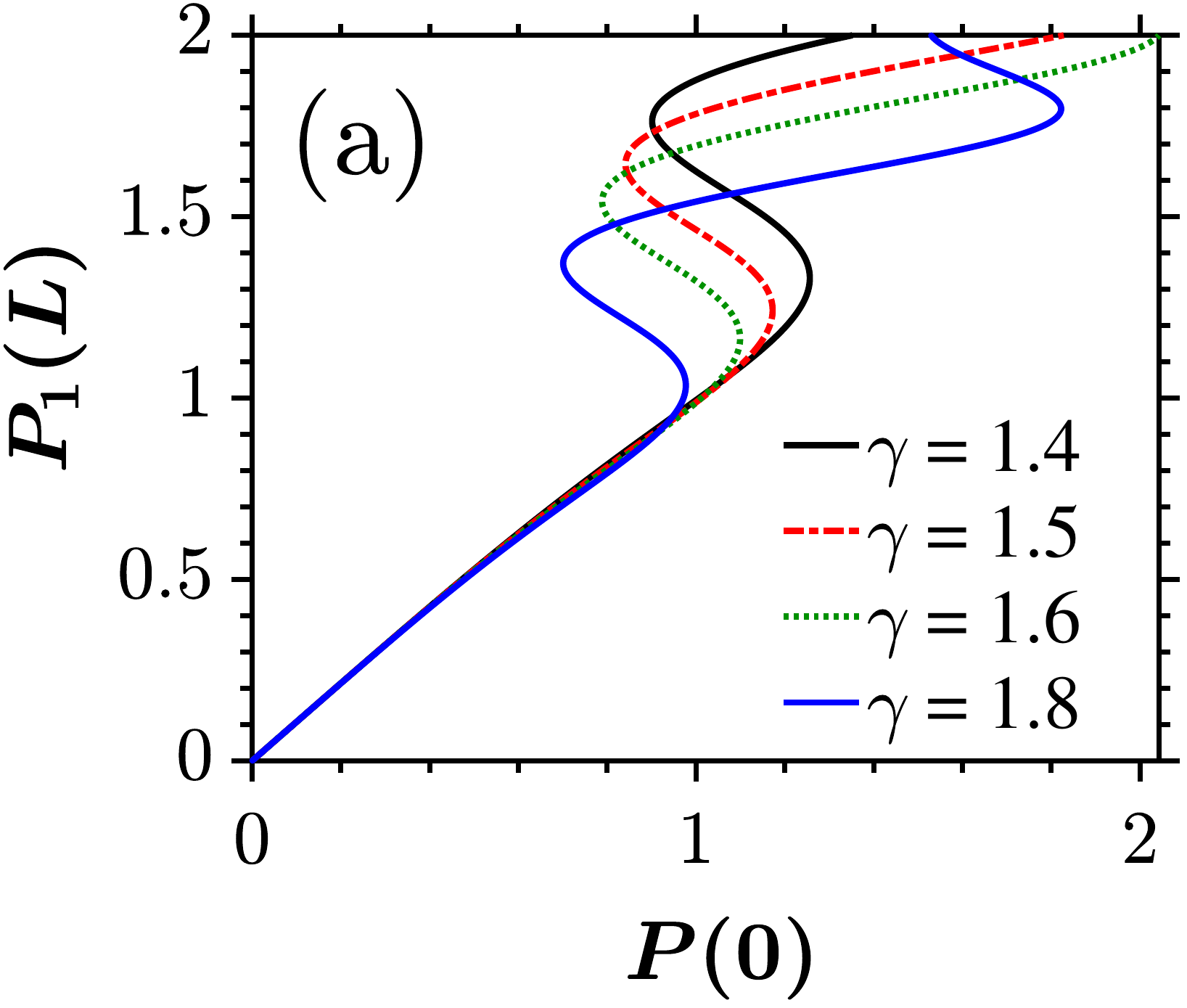}\includegraphics[width=0.5\columnwidth]{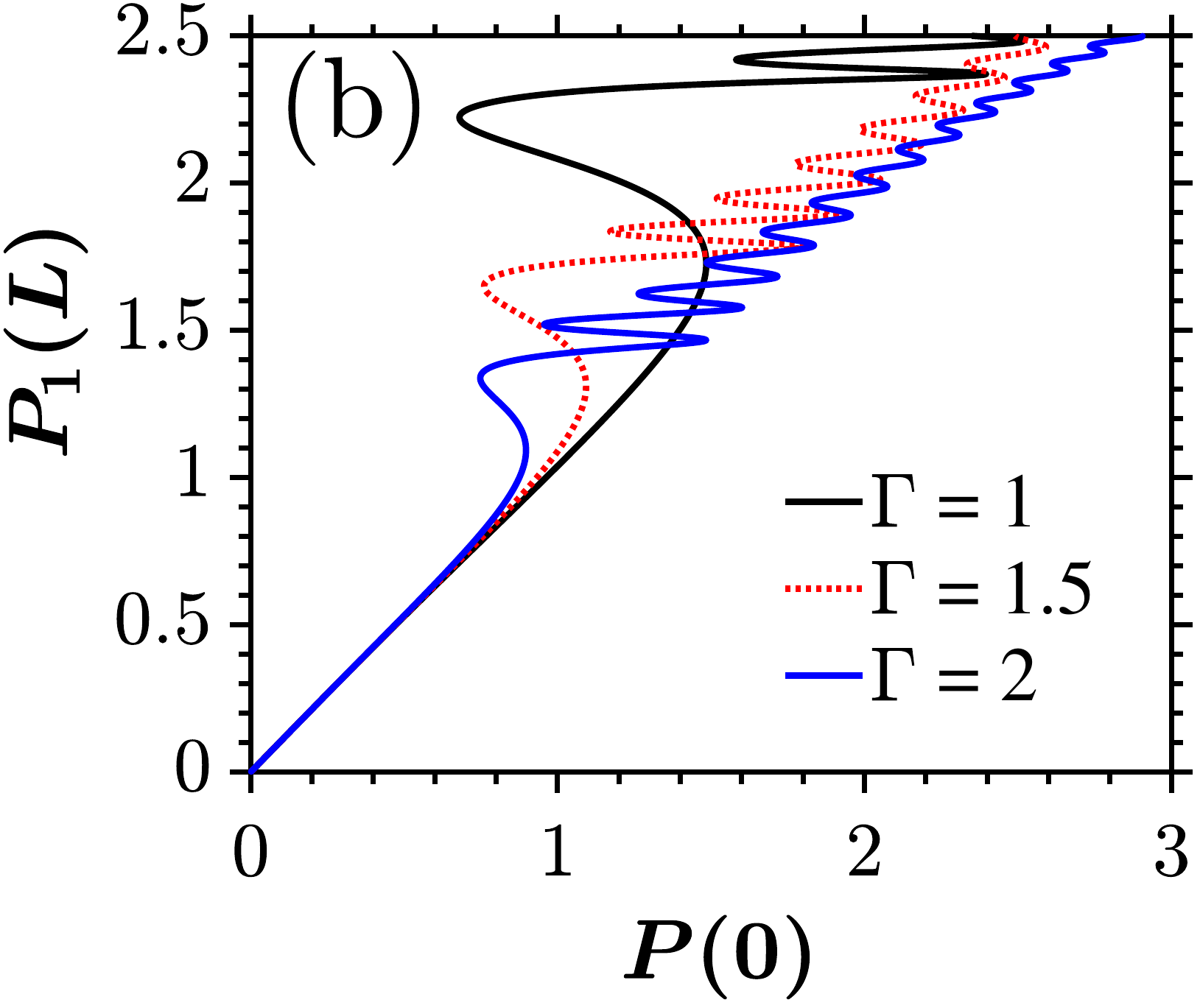}
	\includegraphics[width=0.5\columnwidth]{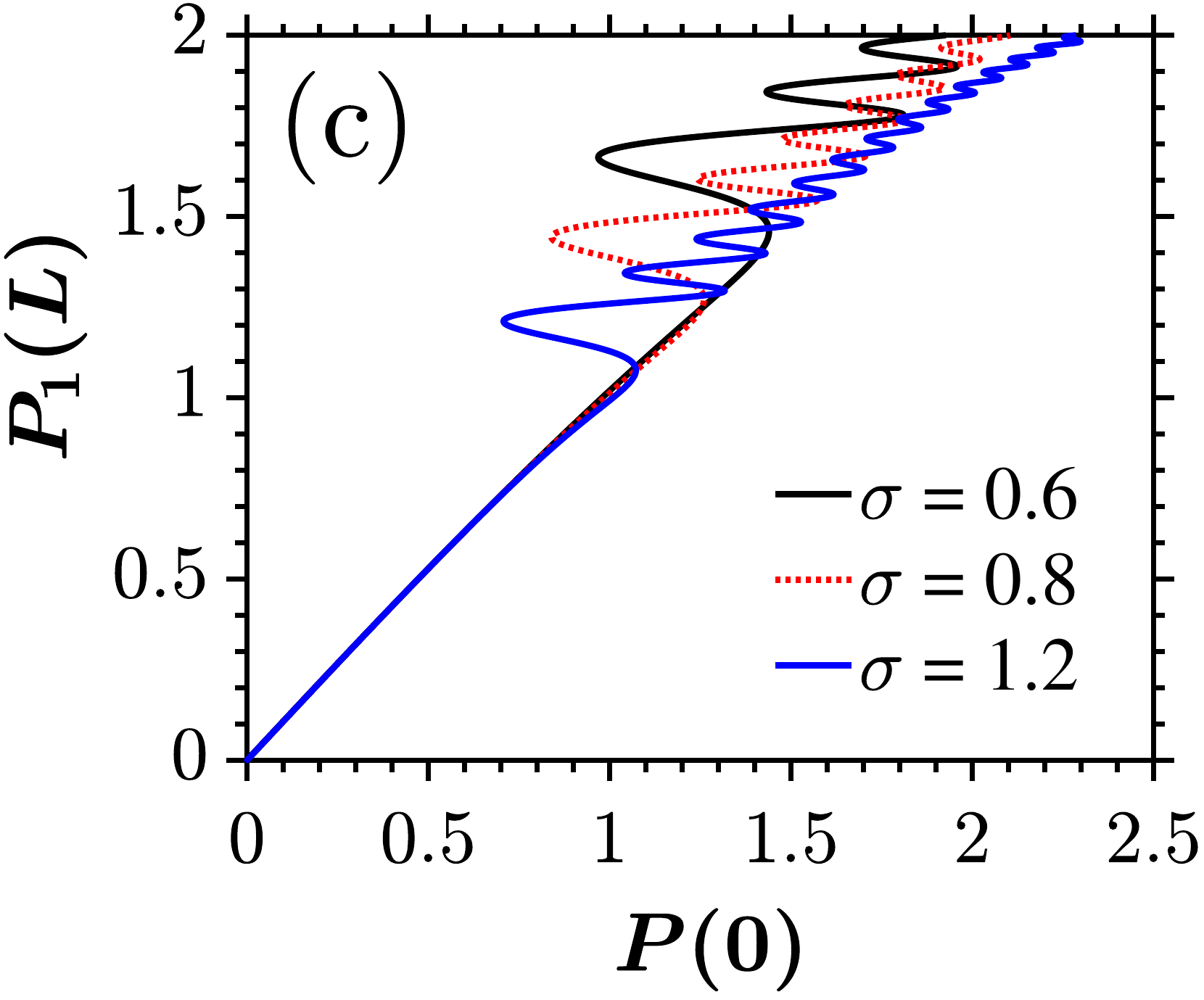}
	\caption{ (Color online) Role of nonlinear coefficients on the optical bistability (multistability) of a broken $\mathcal{PT}$-symmetric  FBG at $g=5$ and $\delta=0$. Here the output and input powers are represented by $P_{1}(L)$ and $P(0)$ respectively. Figure (a) is plotted for different values of cubic nonlinear coefficient ($\gamma$) at $\Gamma=\sigma=0$. Figure (b) represents the simulated results for different values of quintic nonlinear coefficient ($\Gamma$) at $\gamma=1.5$ and $\sigma=0$. Figure (c) is plotted for different values of septic nonlinear coefficient ($\sigma$) at $\gamma=1.5$ and $\Gamma=1$.}
	\label{Fig3}
\end{figure}

At the final stage of drafting this manuscript, we found an article which claims that  the OB/OM curves cannot occur in the broken $\mathcal{PT}$-symmetric regime \cite{komissarova2019pt}. Nevertheless, it has been recently revealed that the broken $\mathcal{PT}$-symmetric regime supports a new type of gap solitons which was further termed as dark gap solitons along with exhibiting novel optical bistability \cite{govindarjan2019}. Hence to analyze the issue further, we look at the switching characteristics of the $\mathcal{PT}$-symmetric FBGs with higher order nonlinearities for the  left light incidence. To demonstrate the role of $g$ in the broken regime, the detuning parameter is fixed at $\delta=0$ and in the
first case we neglect the higher order nonlinearities for the sake
of simplicity so that only the cubic nonlinearity has a significant role.
When $g=5.2$ and $\gamma=1.5$, the
system admits a desirable bistable curve with the switch up intensity around
$0.7443$ and the corresponding switch down intensity is measured to be $0.5924$ (see Fig. \ref{Fig1}(a)).  If the value of $g$ is reduced with
no changes to the other settings, a significant amount of increase
in the switch up intensity is observed.  In the broken (cubic) $\mathcal{PT}$-symmetric regime, if the value of $g$ is far from the singularity condition, the gain of the system is enhanced rather than absorption within the medium and it must be stressed that this is the much anticipated outcome in the context of any $\mathcal{PT}$- symmetric optical system. Also, from Fig. \ref{Fig1}(a), we observe that the value of gain/loss coefficient plays a vital role in deciding the range of intensities over which output state should remain either in the upper branch or the lower branch of the bistability curve. When $g$ is closer to $k$ ($g = 4.7$)  the output state remains in the lower branch for a larger range of input intensity. However, if the gain/loss coefficient value is far away from $k$ ($g = 5.2$), it switches to the upper stable state at relatively lower intensities. With a further reduction in
the value of gain/loss parameter ($g$), the system admits multistability at the same input power. This gives an overview of the
optimum choice of $g$ parameter to get the desired bi- or multi-stability. 

We further portray the effect of variation of parameter $g$ when the quintic nonlinearity
is added to the system. Unlike the previous case, the system
exhibits more than two stable states even at higher values of gain/loss
coefficient. A decrease in $g$ at a given set of values of nonlinearities $\gamma=1$
and $\Gamma=1$ brings about notable changes in the switch-up and
switch down power levels. The number of stable states in the input-output characteristics of the device decreases with an increase in the value of $g$. Thus $g$ serves as an additional degree of freedom to control the number of stable states for a given set of device parameters. A phenomenal outcome of the system is that it admits a
ramp like first stable output state for certain values of $g$ as seen
in Fig. \ref{Fig1}(b). This is yet another critical outcome of our investigation. These kinds of ramp like bistable states are already observed in plasmon resonance structures \cite{daneshfar2017switching,dai2015low}, graphene based structures \cite{sharif2016experimental,naseri2018terahertz}, silicon waveguide resonators \cite{rukhlenko2009} and plexcitonic systems \cite{naseri2018optical}. But this is the very first time such a OB/OM is observed in a $\mathcal{PT}$-symmetric FBG device, thanks to the judicious balance between the gain and loss. Analyzing Fig. \ref{Fig1}(b) further, one can observe that the range of input intensities over which this
ramp behavior is observed decreases when $g$ is reduced. When operated
at $g=4.7$, $5.65$, and $6.85$ the system concedes a step like response after a sharp transition from its initial value (see Fig. \ref{Fig1}(c)). It is important to note that such a mix of stable states resembling a ramp  and step has previously been observed in a complex nano-dimer with a semiconductor quantum dot \cite{naseri2018optical}. These kinds of bistable states with flat slopes can be explored
in the construction of signal regenerators  \cite{zang2012analysis}. Also, the difference between the two power
levels keeps mounting with increase in $g$ which inherently means
that the value of $g$ plays a central role in dictating the width of
the hysteresis loop. This implies that
the same device can be exploited for building applications like switching
with low hysteresis width or all optical memories with larger widths,
simply by tuning the value of $g$ which is very much feasible compared
to other device parameters. The effect of $g$ on the switch-up and
switch-down powers persists even at higher values of quintic nonlinearity but
the only difference being the increase in the number of stable states
as an effect of dominant self-defocusing nonlinearity over the focusing
one. 

We next consider a condition where all the nonlinearities are taken
into account ($\gamma=\Gamma=1$, $\sigma=0.6$) 
and $g$ is varied to find its impact on the switching operation. Similar to the previous case, the system exhibits multiple stable branches for any value of $g$ in this regime. One can easily
visualize that the $g$ parameter has a central role in deciding the intensity at which the system switches between the first stable state and the second one shown in Fig. \ref{Fig1}(d). At higher values of gain/loss parameter $(g=5.2)$, the second stable state is preceded by ramp like stable state. On the other hand, if the value of $g$ is fixed at $4.6$ and the intensity is tuned from zero, there is a sharp increase in the output intensity from zero to an intensity slightly greater than unity. Following the sharp transmission, the output intensity is steady over a large range of input intensities.
In the plot we observe that multiple stable branches start to  emanate at higher intensities. The width of each stable branch is lower than the previous one and the first stable branch is characterized by larger hysteresis width. The bistability  plots in this regime thus give a conclusive evidence that the value of the switching intensity is inversely proportional to the value of $g$. It is clear that
at two different sets of values of $g$ there is a drastic change in
the behavior of the system which confirms the fact that the system is quite sensitive
to smaller variations of gain and loss.

It is noteworthy to mention that the low intensity switching phenomenon for the right incidence can happen even in the broken $\mathcal{PT}$-symmetric regime. To do so, we simulate the system with the same set of parameters as in Fig. \ref{Fig1} for the right incidence. The hysteresis curve in Fig. \ref{Fig1_rr}(a) looks more or similar to the one that we observed in Fig. \ref{Fig1}(a). Nevertheless, the OB phenomenon is observed at very low input intensities. The switch up intensities for different values of $g = 4.7$, $5$, and $5.2$ are measured to be $0.23, 0.152$, and $0.08$, respectively. With the addition of quintic nonlinearity into the system, the device can exhibit both ramp and step like first stable states (see Figs. \ref{Fig1_rr}(b) and (c)), resembling the curves obtained  in Figs. \ref{Fig1}(b) and (c), respectively. For $P_0 < 1$, there is no formation of multistable states as seen in Figs. \ref{Fig1}(b) and (c), whereas we obtain multiple stable states with very low switching intensities in Figs. \ref{Fig1_rr}(b) and (c) for the same value of $P_0$. The same explanation holds good with the inclusion of septic nonlinearity too as observed in Fig. \ref{Fig1_rr}(d). The study of OB/OM curves in the broken $\mathcal{PT}$-symmetric regime for the right incidence thus opens a new avenue for fabricating all-optical switches and memory devices which require ultra low switching intensities with different launching conditions.

\subsection{Effect of variation of detuning parameter}
In the previous sections, the effect of variation of parameter $g$ under
various nonlinear regimes and the effect of nonlinearity at fixed $g$
were examined by setting the detuning parameter value to zero which
implies that the signal wavelength must be synchronized with the Bragg
wavelength. But there is a strong correlation between the switching
intensities (both up and down) and finite detuning values \cite{radic1995theory,zang2012analysis}.
To illustrate the effect of detuning parameter, the value of $g$ is fixed at $g=5$. Similar to the last section,
first we study the effect of detuning in the absence of higher order
nonlinearities. If the device is operated at $\delta=-0.25$, we get a wide bistable curve with switch-up power of $1.371$
and switch-down power of $0.9397$ and if the same system is operated at $\delta=0$ the switch up intensity reduces to $1.173$ as seen in Fig. \ref{Fig2}(a). This confirms
that operating in the negative detuning regime increases the
hysteresis width as well as the intensity required to switch between the two stable states. If one intends to reduce
the threshold, the device should be operated closer to the band gap or in the positive detuning regime. 
For instance, if the detuning value is assigned to be $\delta=0.25$,
the intensity to switch between first stable state and second stable
state reduces to $0.9741$. This intensity value further
reduces to $0.7939$ for $\delta=0.5$
with a simultaneous reduction in the hysteresis width. Thus negative detuning regime favors the device's preference to remain in the first stable branch for a larger range of input intensities 
whereas the positive detuning tend to keep the output intensity in the upper stable branch for a larger range of input intensities.

To portray the effect of detuning in the presence of both cubic and
quintic nonlinearities, we set $\gamma=\Gamma=1$ and $g=5$. The positive detuning parameter increases the hysteresis width and
reduces the number of stable states in the presence of quintic nonlinearity. The plots depicted in Fig. \ref{Fig2}(b) give a conclusive evidence that an increase in positive detuning
inflates the difference between switch-up and down intensities which implies
that the intensity inside the device is sufficient enough to keep
the output state dormant post the switching from its previous state.
Hence it is preferable to use this kind of multistable states in the construction
of optical memories rather than switches. If we look at the same system
working in the negative detuning regime the switch up intensity (between the first two stable states) keeps on deflating
and therefore if one intends to construct switches with a cubic-quintic
FBG in the broken regime, it is preferable to have signal wavelength longer
than the Bragg wavelength provided that it lies within the stop band. Also the number of stable states increases if the device is operated in the negative detuning regime whereas it decreases when operated in the positive detuning regime. Physically, the memory operation can be accomplished by varying the holding-beam input power \cite{ogasawara1986static}. As  the input intensity varies, the output intensity can stay in one of the stable branches and not in the unstable branch. The set can be accomplished by raising the input intensity beyond the switch-up threshold, whereas the reset operation can be effected by reducing the input intensity beyond switch-down intensity. So the switch-up and down intensities can serve as read and write bias pulse for the memory operation \cite{karimi2012all}. The memory holding width can be altered by changing the magnitude of the detuning parameter. 

In the presence of second focusing (septic) nonlinearity, the negative
detuning regime shows the growth in the dormant stable states whereas
in the positive detuning regime the switch down intensity
decreases in addition to the shrinkage in the hysteresis loop
as seen in Fig. \ref{Fig2}(c). 
Operating at longer wavelengths has a marginal impact in the
reduction of switch-up intensity, quite similar to those occurring at the shorter wavelengths in the cubic-quintic case with the only difference being the number of stable states above the dormant states is comparably larger and are desirable for multilevel
signal processing applications. Though we present only a few applications
here, the system's ability to retain its memory of the past state
for longer period can be subjected to detailed investigation in future
to build new all-optical devices.

\subsection{Impact  of nonlinear  parameters}
The nonlinear parameter purely depends on the type of glass material
used. From the application perspective, researchers are left with
many materials offering a wide range of nonlinearities from very high
to low values \cite{karimi2012all}.
The nonlinearity plays a crucial role in deciding the number of stable
states and the intensity required to switch between the stable branches. To illustrate this, we first consider a simple
case with only cubic nonlinearity. The first bistable curve starts
to emerge at $\gamma=1.4$ in our numerical simulations for $g=5$ and the intensity for up switching
is $1.256$ which further reduces to $1.099$ with an increase in $\gamma$
($1.6$). Any further increase in the value of $\gamma$ gives rise to
multistable states as a consequence of increase in the effective feedback
to the system as seen in Fig. \ref{Fig3}(a). Earlier in the unbroken regime, we stated a thumb rule to reduce the switching intensity which demands the nonlinear coefficient to be high. From our simulations, we confirm that the rule holds good in the broken regime too.

With the addition of quintic nonlinearity to the above system, it 
admits multistable states even at lower input intensities as seen in Fig. \ref{Fig3}(b). To comprehend the role
of quintic nonlinearity, we numerically vary $\Gamma$ at a fixed value of gain/loss coefficient
($g=5$). At $\Gamma=1$, when the input intensity is slowly
varied from zero, there is a linear increase of output intensity below
$1.479$ above which it switches to the second stable state. The output
intensity again starts to vary linearly in the second stable branch
for values between $1.479$ and $2.396$. Above this intensity, the system
switches to the third stable branch. If the intensity is decreased,
the switch-down intensity required to swap from third branch to the
second is $1.582$ and to switch back to its initial state, the input
intensity must be reduced further to $0.6803$. When $\Gamma$ is increased
further, the switching intensity between various stable branches tapers
off and new stable states appear within the same input intensity ($P_0 = 2$). For instance, when $\Gamma=1.5$ the switch up intensity to switch between the first and second stable state is measured to be $1.093$. The switch up intensity reduces below unity when the system is operated at $\Gamma=2$.

Finally, we present the effect of septic nonlinearity in the presence
of both cubic and quintic nonlinearities with simulation parameters as $g=5$,
$\gamma=1.5$, $\Gamma=1$ and septic nonlinearity
($\sigma$) is varied. A straight forward evidence one can
get from Fig. \ref{Fig3}(c)  is that inclusion of  the septic nonlinearity cuts down the
required intensity to switch between the stable states and the width of the hysteresis is further reduced when compared to the system
with same simulation parameters in its absence. The septic nonlinearity
also boosts the number of stable states similar to the other nonlinear
effects discussed already. When $\sigma=0.8$, it supports more than
five stable branches with each branch possessing a width narrower
than the previous one. When $\sigma=0.6$, a series of multistable states
appear and the intensities to jump from the preceding state are $1.44$ (1
to 2), $1.809$ (2 to 3) and the respective switch down intensities are
given by $0.9697$ and $1.433$ as shown in Fig. \ref{Fig3}(c).  The width of the upper stable branches drastically reduces. But with suitable adjustment in other device parameters, it is possible to increase the visibility of the upper stable states and hence such novel bistable states can lead to the efficient all-optical signal processing by controlling (output) light  with (input) light. Also, a possible experimental realization of the kind of structure envisaged in this paper is to identify a suitable material (preferably chalcogenide glass) which can allow the fabrication of alternate regions of gain (actively doped by erbium) and loss (no dopant by considering the intrinsic loss or dopant with high absorption by chromium element) into it and thereby serving as a $\mathcal{PT}$-symmetric periodic structure.

\section{Existence of Gap solitons}
\label{Sec:5}
\begin{figure}
	\centering
\includegraphics[width=1\columnwidth]{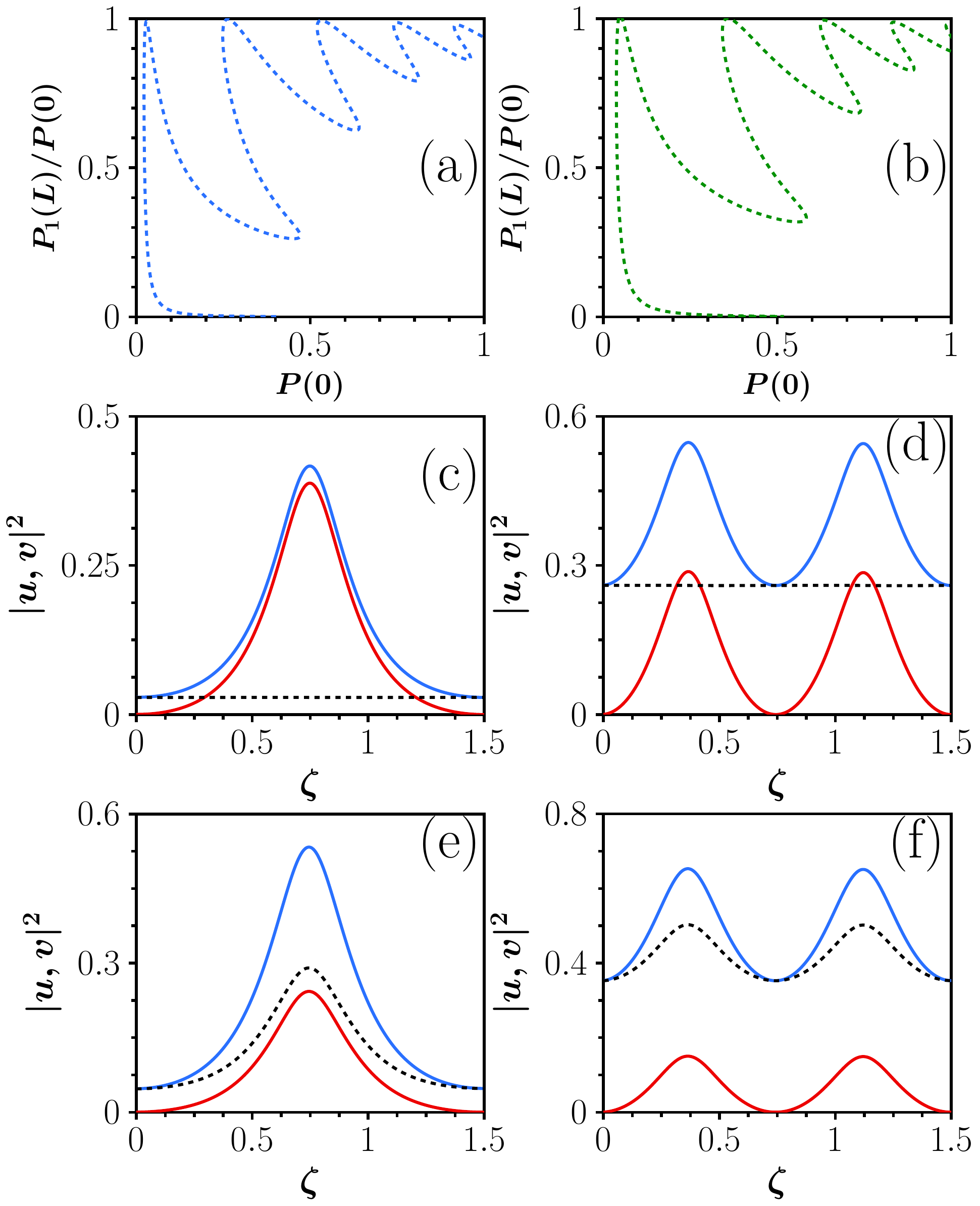}
	\caption{Top panels (a) and (b) illustrate the transmission characteristics of conventional and $\mathcal{PT}$-symmetric FBGs, respectively. Center and bottom panels delineate the stationary gap soliton formation at the different resonance peaks for conventional and $\mathcal{PT}$ symmetric systems, respectively when the parameters are kept as $L=1.5$, $k=3$, $g=2$, $\delta = 1$, $\gamma=4$, $\Gamma=1$ and $\sigma=2$. Here the blue solid lined curves indicate the forward field and red lined curves refer to backward field intensities. The total field intensities ($|u|^2-|v|^2$) are drawn by black dashed lines.
	}
	\label{G:1}
\end{figure}
Soliton formation is a universal phenomenon that can occur in any nonlinear optical structures through a delicate interplay between the group-velocity dispersion and the nonlinearity of the structure. The former tends to disperse the energy of the propagating pulse whereas the later is supposed to counter-balance the effect caused by the former by concentrating the energy of the pulse \cite{hasegawa1995}. Many types of solitons are reported theoretically and experimentally in the regular optical fibers which include dissipative \cite{zhao2010dissipative}, vector \cite{tang2008observation}, polarization domain wall \cite{zhang2009observation}, dispersion managed, Raman and paired solitons \cite{hasegawa1995}, as well as multisolitons \cite{chen2014formation}, which rely on the above mentioned phenomenon. Among them, gap solitons are a special type of solitons formed in the periodic structures which posses photonic band gap as a consequence of periodic variation in the linear dielectric constant. This periodic variation can often be engineered at ease \cite{winful2000raman} and hence gap solitons are well suited for applications such as optical buffering \cite{mok2006dispersionless}, optical delay lines \cite{d2004bright}, distributed feedback pulse generator \cite{winful1991modulational}, transmission filters \cite{kozhekin1995self}, and logic gates \cite{taverner1998all}.

The phenomenon of optical bistability in the feedback structures goes hand in hand with the formation of gap soliton \cite{winful1991modulational}. 
To understand the dynamics of gap solitons, it is necessary to look back at the light transmission ($T=\left|\cfrac{u(L)}{u(0)}\right|^2$) characteristics of the device. 
In the distributed feedback periodic structures, the light propagation is portrayed by the existence of stop bands and pass bands. In a linear FBG,  a wave is reflected if its wavelength falls within this forbidden band. Then again, a wave whose wavelength falls outside this stop band can traverse through the structure unhampered \cite{agrawal2001applications}. As pointed out by Winful \emph{et al.} \cite{winful1991modulational} that the wave amplitude falls off exponentially  along the propagation direction in this case. However, in the presence of nonlinearities, the light is totally transmitted for certain intensities due to the formation of \emph{localized modes} or \emph{spatial nonlinear resonances} within the stop band. These spatial resonances are coined as gap solitons simply for two reasons. First, they have the trademark shape of $sech^{2}$ solitons. Secondly they reside within this photonic band gap. The formation of such gap solitons in the presence of Kerr effect in the conventional periodic structure has already been investigated by many authors \cite{winful1991modulational, d2004bright, litchinitser2007optical}. Recently, the formation of gap solitons has been explored in $\mathcal{PT}$-symmetric periodic structures and some noteworthy properties were highlighted. In particular, it is shown that these phenomenological $\mathcal{PT}$-symmetric structures can support the interesting formation of dark gap solitons \cite{govindarjan2019}. In this section, our aim is to show that the gap solitons can persist even in the presence of higher order nonlinearities in such $\mathcal{PT}$-symmetric systems. The first nonlinear resonance is observed at very low input intensities for a conventional FBG with $L=1.5,k=3,\gamma=4,\Gamma=1,\sigma=2$ and $g =0$ (see Fig. \ref{G:1}(a)). The plot of forward field intensities against the propagation distance at the first peaks reveals that a bright gap soliton like entity corresponding to the resonance value appears. We can observe that the difference between the forward and backward field intensity is marginal in Fig. \ref{G:1}(c). At sufficiently larger input intensities, a second order soliton like entity is formed at the transmission resonance. The peak power of the forward field distribution curve is enhanced whereas the peak power of the backward field distribution is reduced at the successive transmission peaks as seen in Fig  \ref{G:1}(d). In both cases the total power remains constant throughout the propagation length.

\begin{figure}
	\centering
	\includegraphics[scale=0.25]{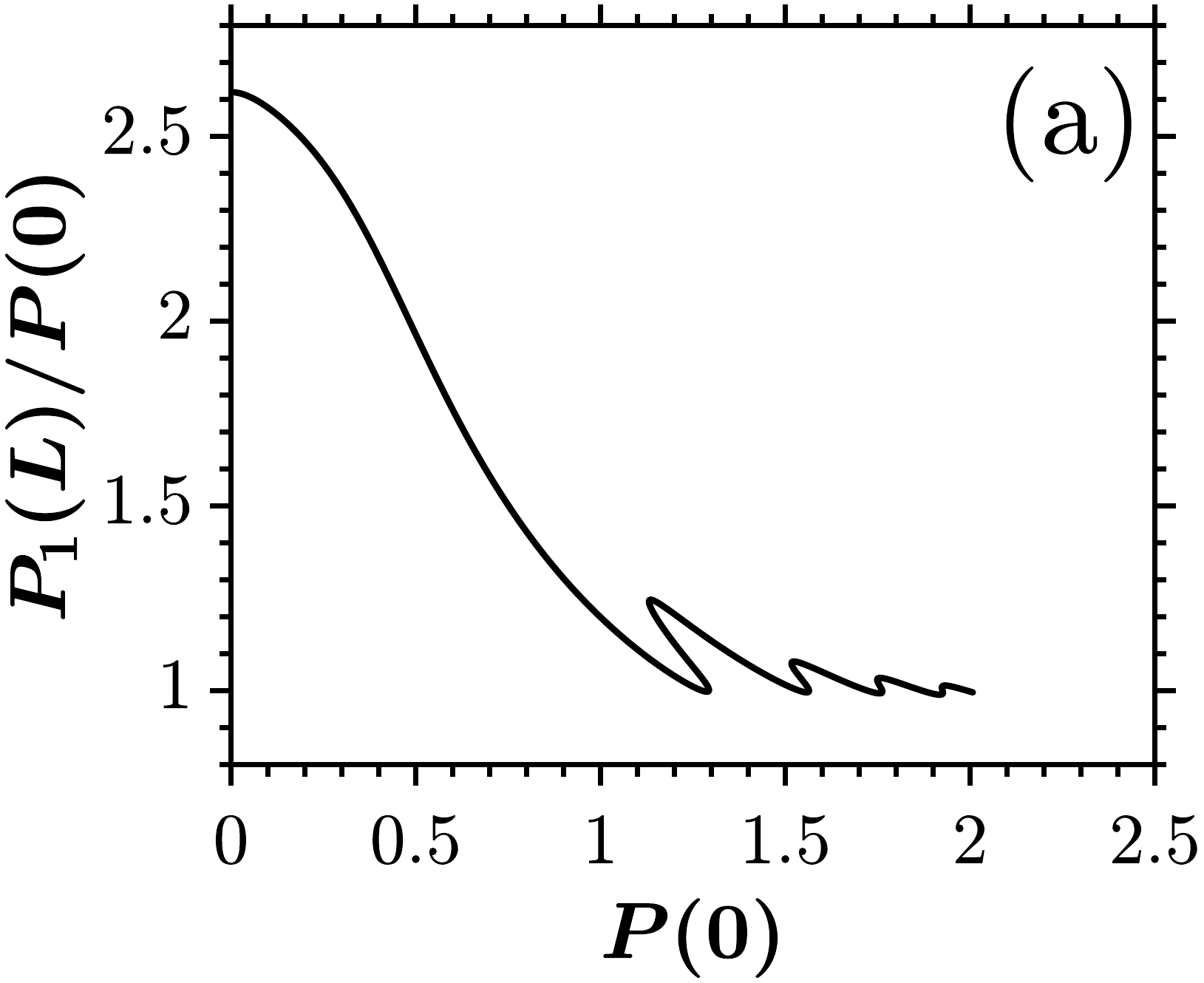}\includegraphics[scale=0.25]{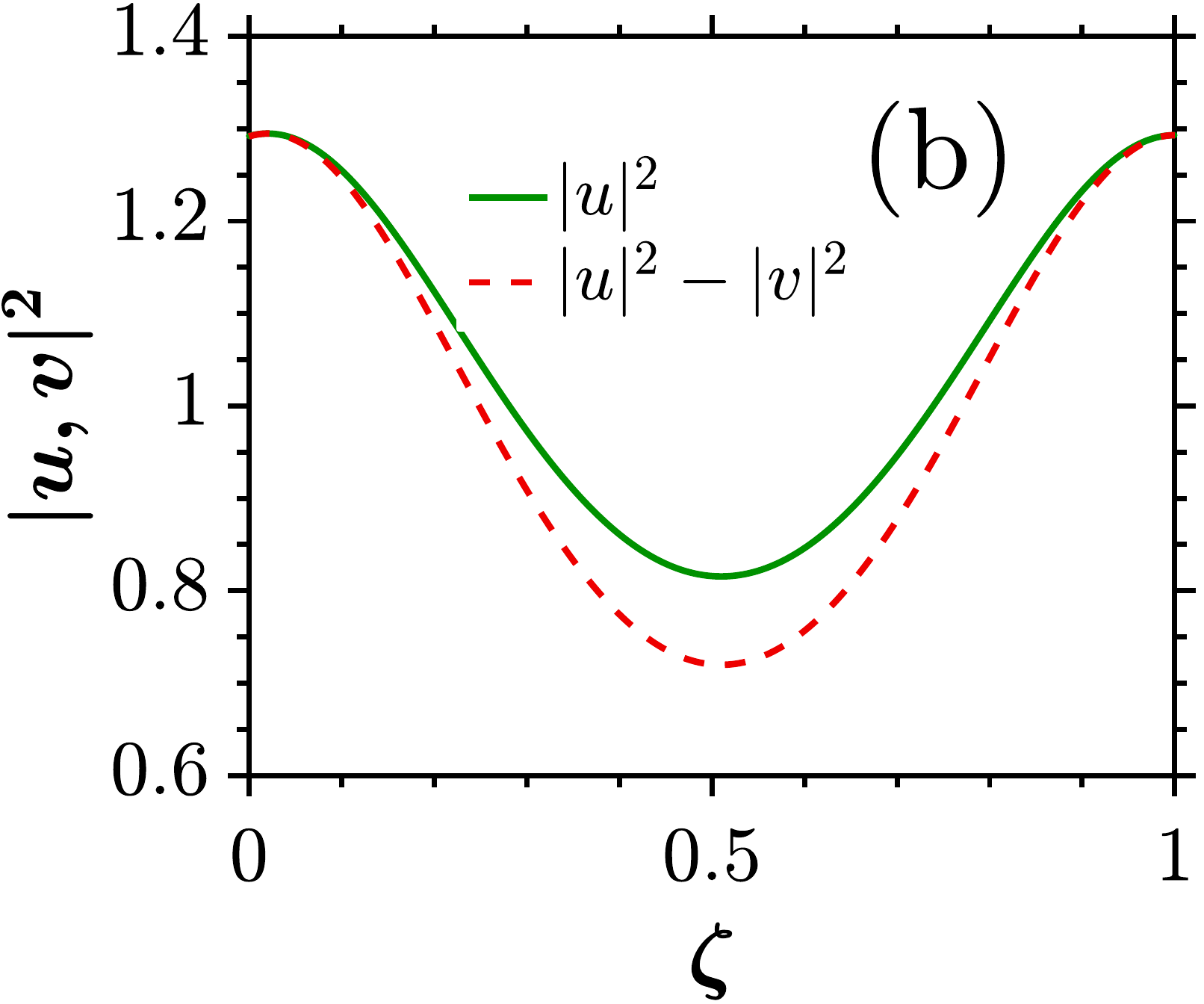}\\\includegraphics[scale=0.25]{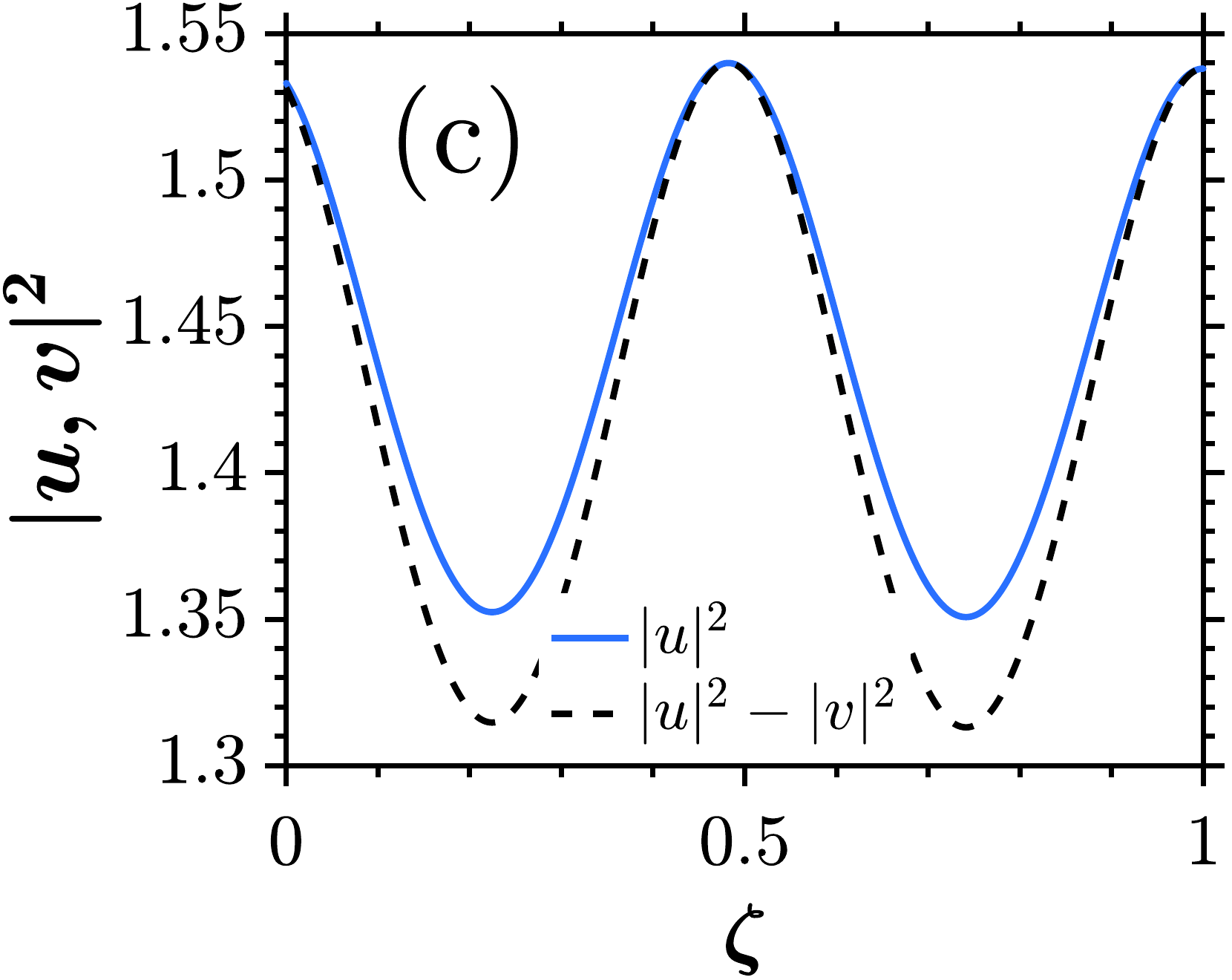}\includegraphics[scale=0.25]{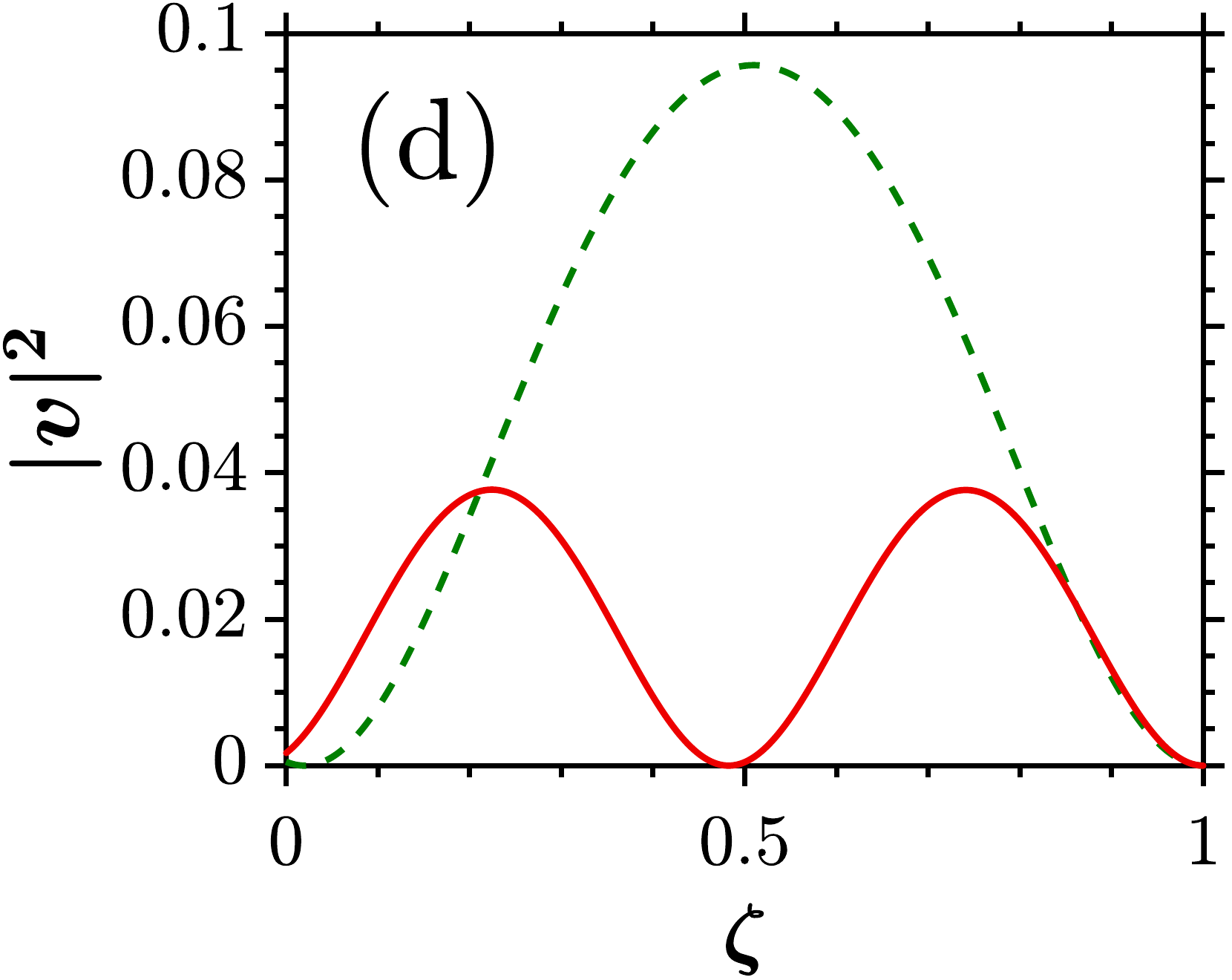}
	\caption { (color online) Plots of (a) transmission (b) and (c) dark-gap solitons formation in the broken $\mathcal{PT}$-symmetric regime and (d) illustrates the bright-gap soliton like entity in the broken regime. In fig (d) the dotted line corresponds to backward field intensity at the first transmission resonance and the solid line indicate the same at the second transmission resonance. The system parameters are chosen as $L=1$, $\kappa=2$, $g=3$, $\delta=0$ and the values of nonlinear parameters are taken to be $\gamma$ = $\Gamma$ = $\sigma$ = 1.}
	\label{B:1}
\end{figure}

When $\mathcal{PT}$-symmetry is included to the system ($g=1$), these nonlinear resonances can still exist in Fig. \ref{G:1}(b)). The plot also depicts that the $\mathcal{PT}$-symmetry plays a significant role in altering the peak intensities of these nonlinear resonances. The first and second transmission peaks occur at slightly larger input intensities when compared to the conventional case in Fig. \ref{G:1}(c). Similar to the conventional case, the value of the forward field peak is enhanced and the backward field peak is reduced at the second transmission peak as seen in Fig. \ref{G:1}(f). The plot of total intensity (see Figs. \ref{G:1}(e) and (f))  against the propagation distance also shows a bright soliton like entity unlike the conventional case. Motivated by the formation of unique gap soliton obtained in the unbroken regime, we next intend to examine whether these kinds of nonlinear resonances can occur in the broken regime too in the presence of higher order nonlinearities. Numerical simulations turn out a surprising outcome of  localized modes at the band gap which resembles a dark soliton in the broken regime as seen in Fig. \ref{B:1}. The $\mathcal{PT}$-symmetry dictates the value of the dip of the dark soliton like entity as in the case of forward field intensity distribution and total power distribution as seen in Figs . \ref{B:1}(b) and (c). On the other hand, one can observe that the simultaneous existence of bright soliton like entity in the plot of backward field distribution against the propagation distance (see Fig. \ref{B:1}(d)), which further can be regarded as the unique outcome of $\mathcal{PT}$-symmetry in such periodic structures.

\section{Nonlinear Reflection spectrum for constant pump power}
\label{Sec:6}
In the previous sections, we elaborated the switching exhibited by the $\mathcal{PT}$-symmetric system under different conditions. But in all the systems discussed previously, switching is achieved via optical bi- (multi-) stability under continuous variation in the pump power (P(0)). We can also find that the FBG exhibits another type of switching mechanism in the presence of constant pump power ($P(0)$) as a function of detuning parameter ($\delta$) in the literature in both linear \cite{erdogan1997fiber} as well as nonlinear regimes \cite{zang2012analysis,karimi2012all}. These studies are restricted to only conventional FBGs without gain and loss. Hence we are interested in studying this kind of  switching behavior in the presence of $\mathcal{PT}$-symmetry. To do so, we fix $L=0.5$, $k=2$ and vary the other parameters for the unbroken $\mathcal{PT}$-symmetric regime, whereas we fix $L=1$ and $k=2$ for the broken $\mathcal{PT}$-symmetric regime.
\begin{figure}[t]
	\centering
	\includegraphics[width=0.5\columnwidth]{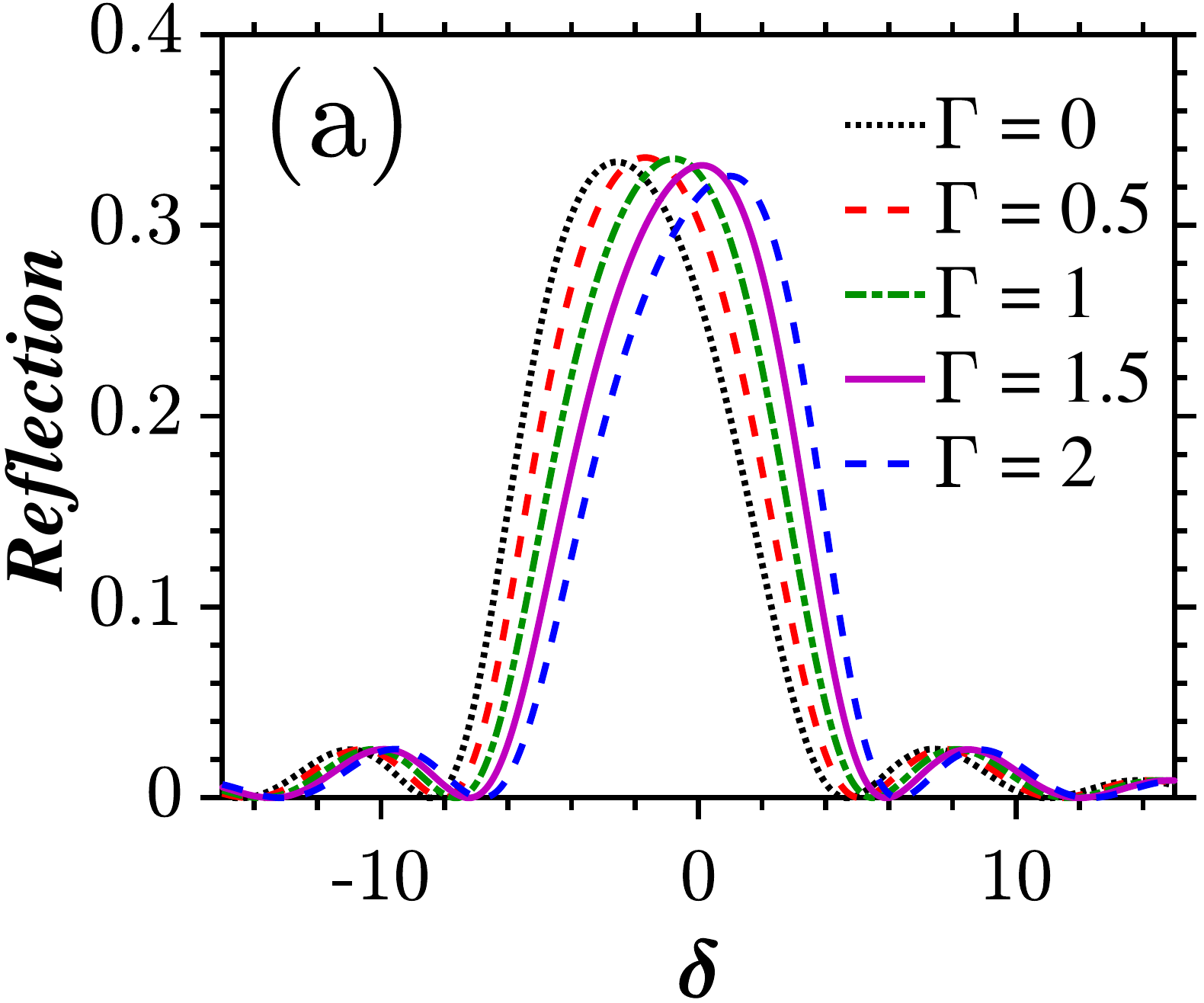}\includegraphics[width=0.5\columnwidth]{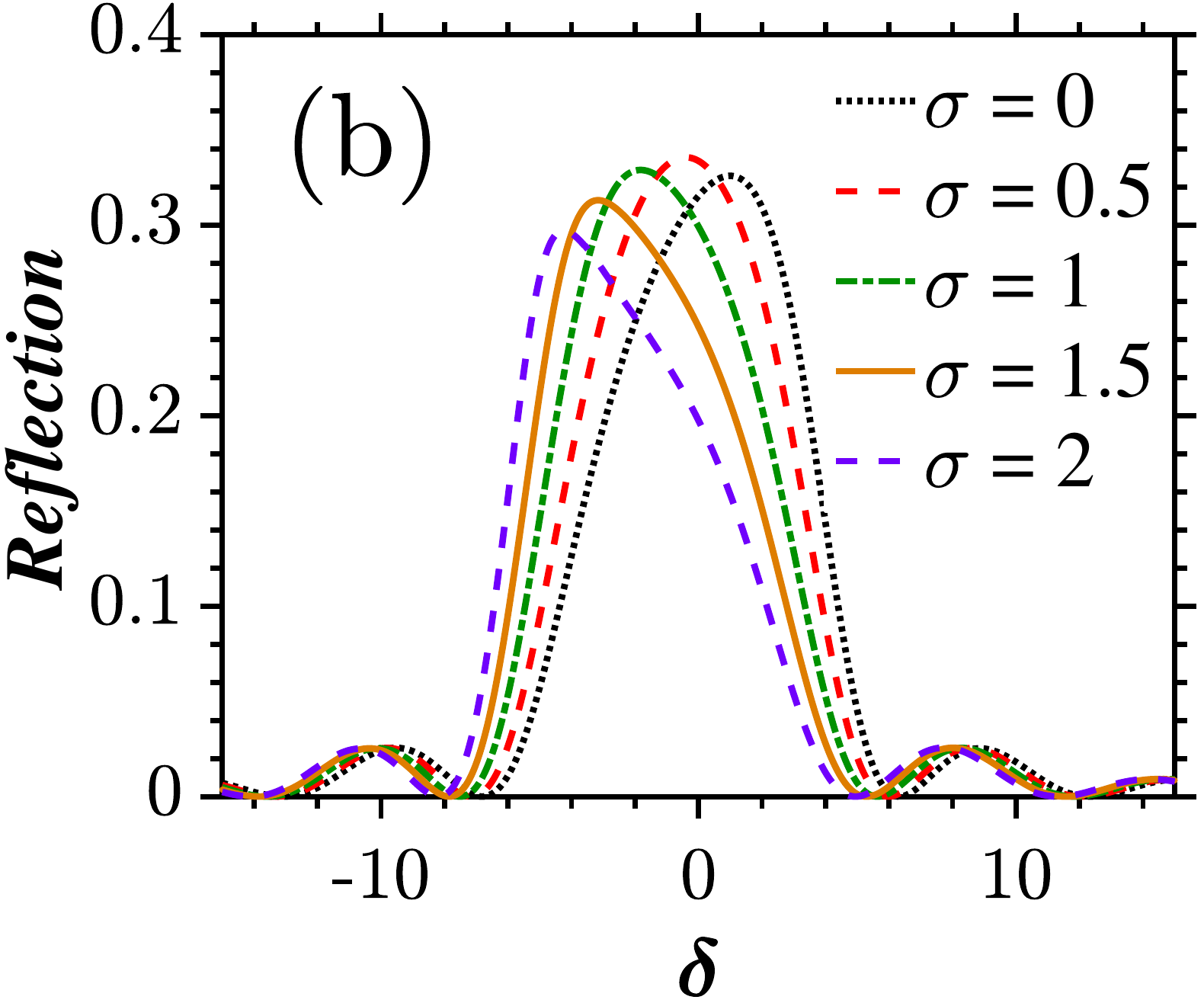}
	\caption{ (Color online) Role of nonlinear coefficients on the reflection characteristics under constant pump power of an unbroken $\mathcal{PT}$-symmetric  FBG at $g=0.5$.  Figure (a) represents the simulated results for different values of quintic nonlinear coefficient ($\Gamma$) at $\gamma=2$ and $\sigma=0$. Figure (b) is plotted for different values of septic nonlinear coefficient ($\sigma$) at $\gamma$ = $\Gamma=2$.}
	\label{Fig_de_1}
\end{figure}
\subsection{Effect of variations in the nonlinearity in the unbroken $\mathcal{PT}$-symmetric regime}
 In the absence of any nonlinearity the reflection spectrum is centered at $\delta=0$. The spectrum is shifted towards longer wavelength when a cubic nonlinearity is added to the system.  With further increase in the cubic nonlinearity parameter ($\gamma$), the spectrum is blue shifted  \cite{govindarjan2019}. Since quintic nonlinearity is a self-defocusing nonlinearity the spectrum is shifted towards shorter wavelength with increase in $\Gamma$. In addition to shifting, the amount of light which is reflected by the system is reduced slightly with increase in $\Gamma$ as seen in Fig. \ref{Fig_de_1}(a). The same kind of phenomenon is observed in the presence of an additional self-focusing nonlinearity (septic) but the shifting of the spectrum is towards longer wavelength similar to cubic nonlinearity case (see Fig. \ref{Fig_de_1}(b)). Thus we can conclude that the shifting of the spectrum towards longer or shorter wavelength is dependent upon the nature of nonlinearities. Note that the effect of nonlinearity on the side lobes of the spectra (spectrum outside the band edges) is minimal.

\subsection{Effect of variations in the gain/loss in the unbroken $\mathcal{PT}$-symmetric regime}
\begin{figure}[t]
	\centering
\includegraphics[width=0.5\columnwidth]{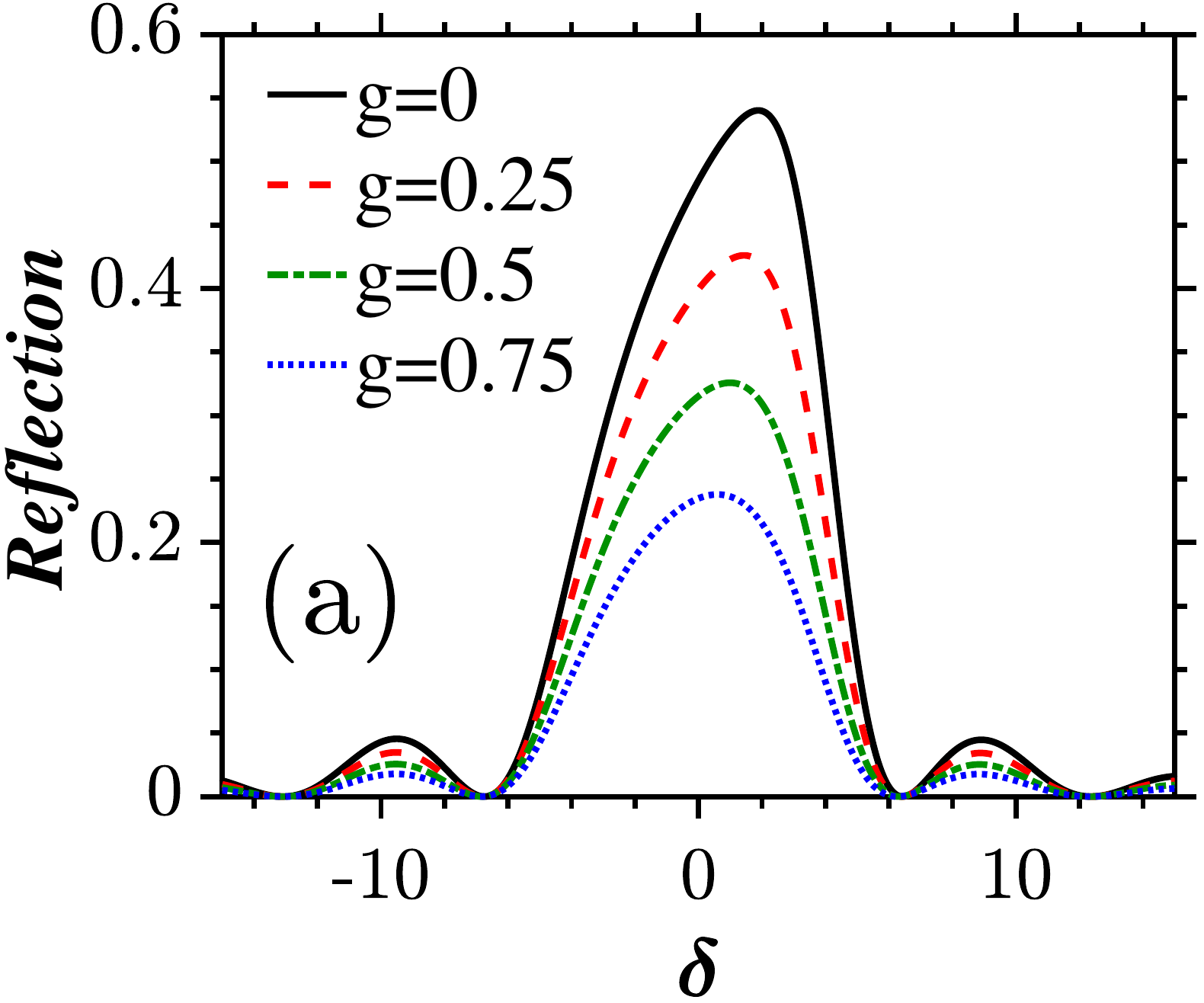}\includegraphics[width=0.5\columnwidth]{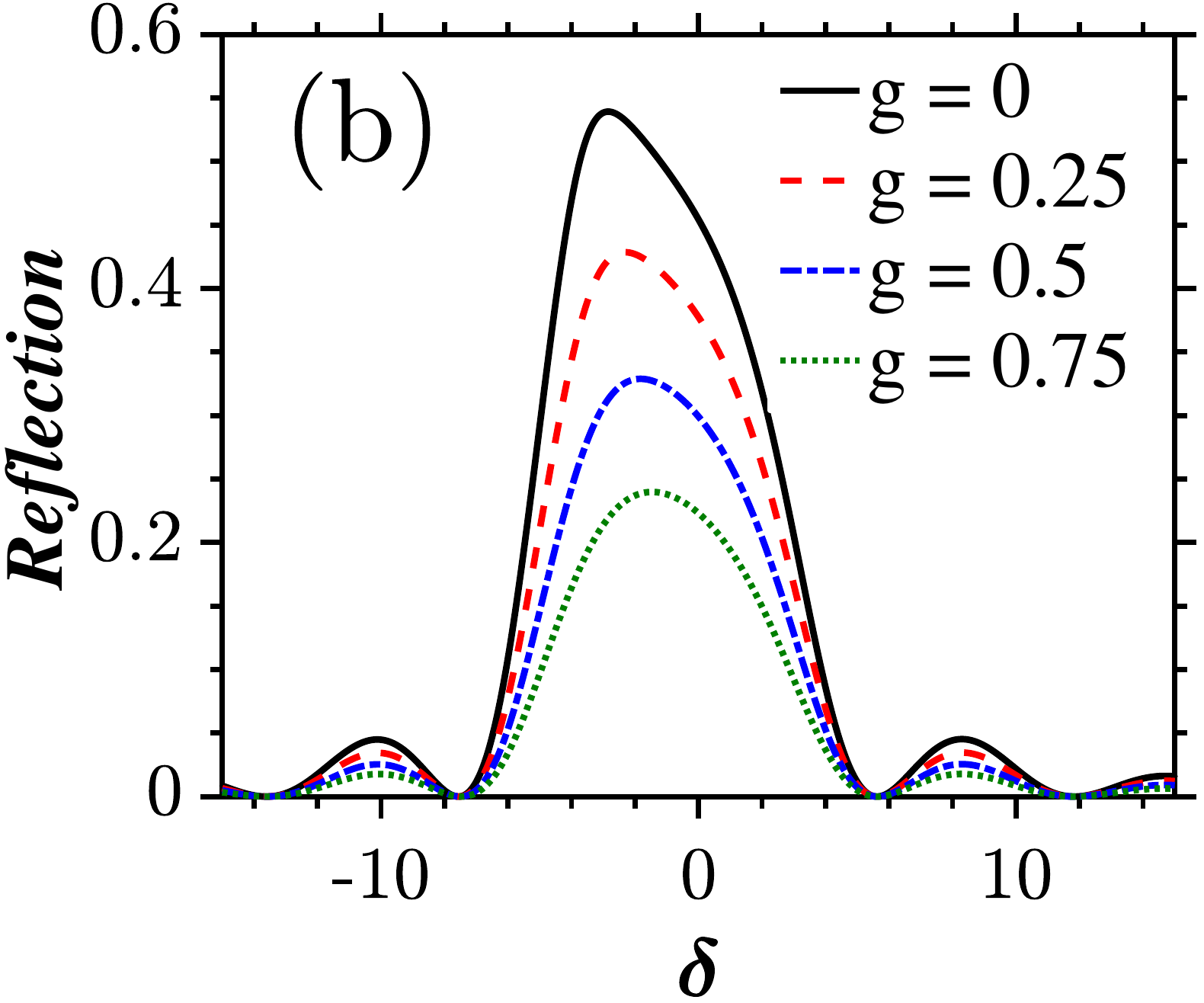}
	\caption{ (Color online) Effect of variation in the gain/loss parameter ($g$) on the reflection characteristics under constant pump power of an unbroken $\mathcal{PT}$-symmetric  FBG.  Figure (a) represents the simulated results for different values of $g$ in the presence of cubic-quintic nonlinearities ($\gamma$ = $\Gamma=2$, $\sigma=0$). Figure (b) is plotted for different values of $g$ in the presence of cubic-quintic-septic nonlinearities ($\sigma$) at $\gamma$ = $\Gamma=2$.}
	\label{Fig_de_2}
\end{figure}

 In the absence of any $\mathcal{PT}$-symmetry and nonlinearities the reflection is maximum around the Bragg wavelength. The wavelength at which the peak reflectivity occurs is varied by the presence of nonlinearities \cite{govindarjan2019}. In Figs. \ref{Fig_de_2}(a) and (b) we can observe that the reflection peak is slightly off-centered as a consequence of higher order nonlinearity in the system. Any variation in the gain/loss does not shift the spectrum as in the case of nonlinearity. But, the actual effect of variation in the parameter $g$ depends on the launching conditions. The band gap is more or less symmetric about the center wavelength in the presence of cubic nonlinearity alone ($\gamma=1, \Gamma= \sigma=0$). For left incidence, the reflectivity is suppressed with increase in the gain/loss. When we look into the system with higher order nonlinearities, the same effect persists. Thus we can conclude that irrespective of the type of nonlinearity, the amount of light reflected is reduced by the gain/loss parameter $g$ for the left incidence. 

\subsection{Unique spectra in the broken $\mathcal{PT}$-symmetric regime}
\begin{figure}[t]
	\centering
	\includegraphics[width=0.5\columnwidth]{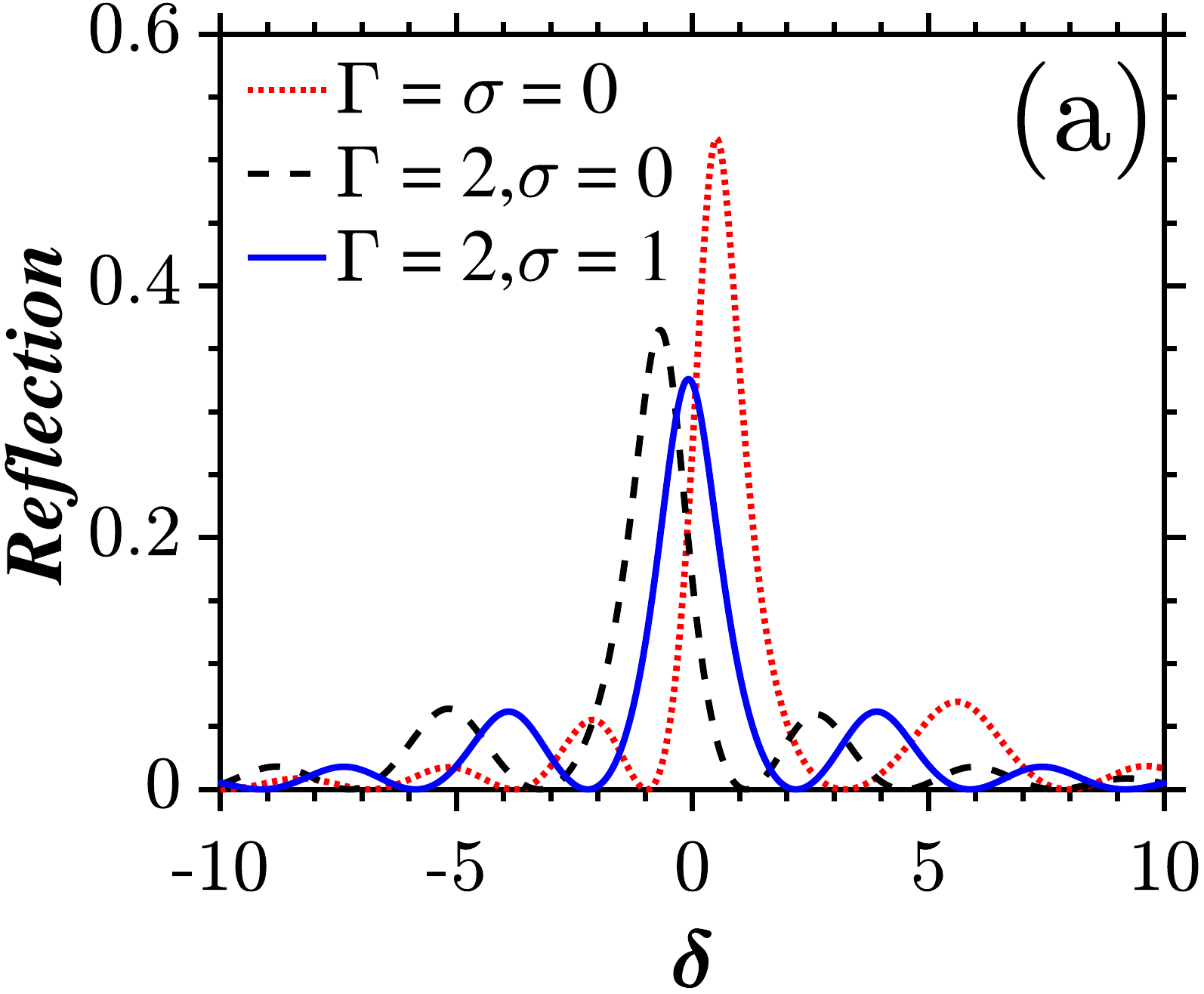}\includegraphics[width=0.5\columnwidth]{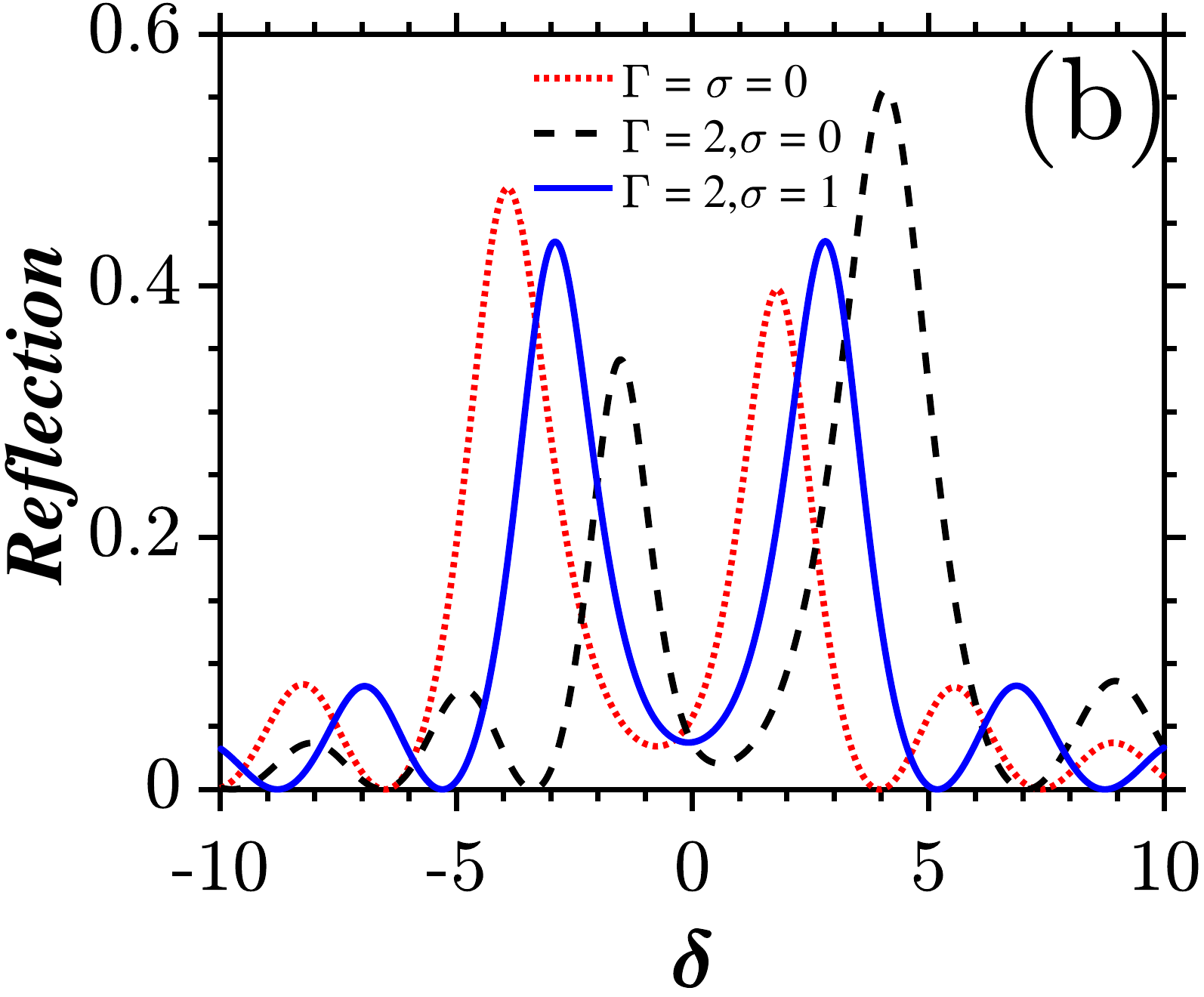}
	\includegraphics[width=0.5\columnwidth]{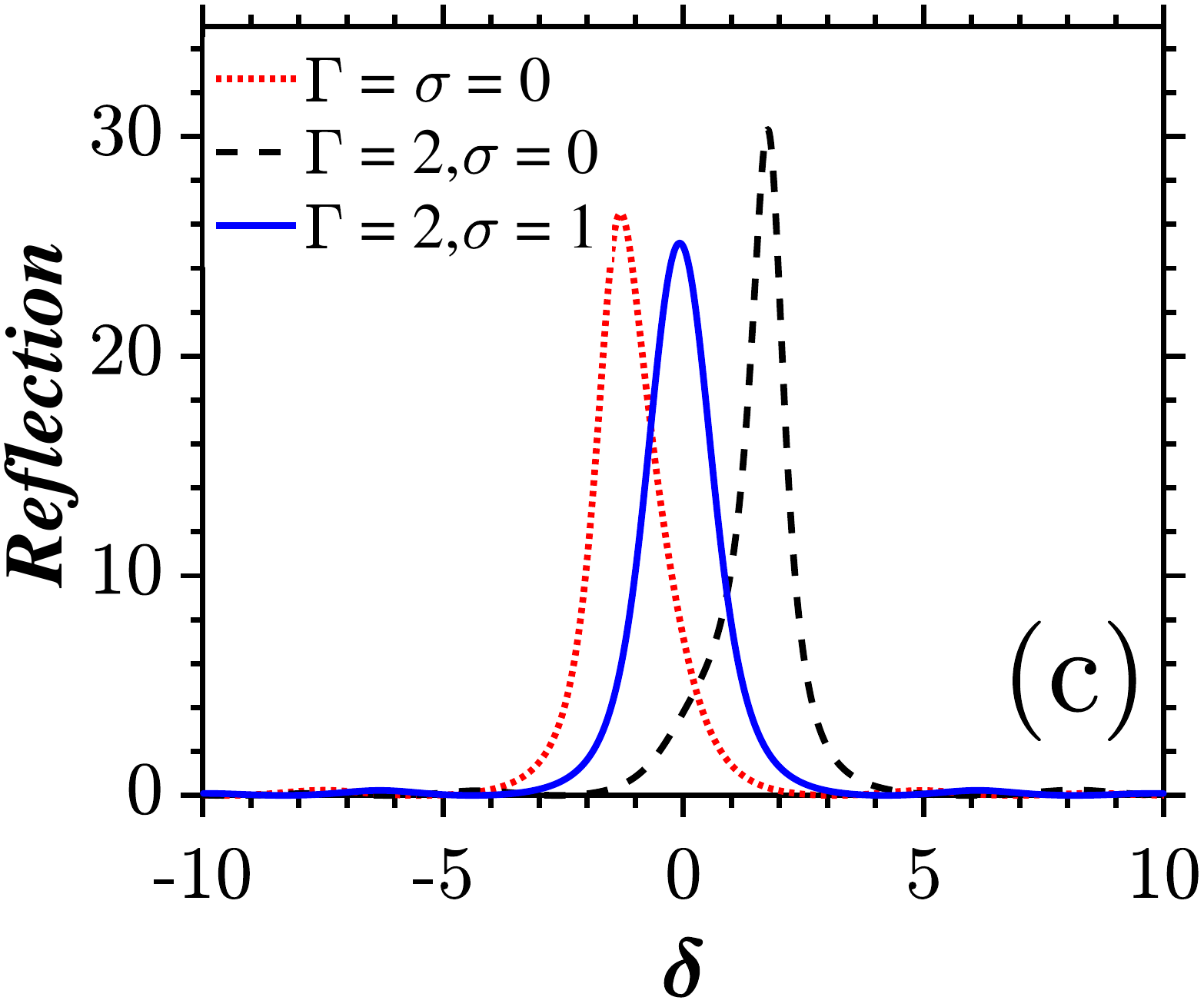}
	\caption{ (Color online)  Reflection characteristics under constant pump power of a broken $\mathcal{PT}$-symmetric FBG. Figures (a), (b), and (c) are plotted at $g=3, 4, 5$ respectively. Throughout the figure the value of cubic nonlinear coefficient is set to $\gamma=1$. The dotted lines (red) represents the cubic nonlinear regime ($\Gamma= \sigma= 0$). The  dashed lines (black) indicate the quintic nonlinear regime ($\Gamma=2, \sigma=0$). The solid lines (blue) represents the septic nonlinear regime ($\Gamma=2$, $\sigma=1$).}
	\label{Fig_de_3}
\end{figure}

In the unbroken $\mathcal{PT}$-symmetric regime, the reflection characteristics are found to be more or less flat within the stop band. But in the broken $\mathcal{PT}$-symmetric regime, the peak of the reflection occurs at a single wavelength instead of the entire stop band as in the case of Fig. \ref{Fig_de_3}(a) which is simulated at $g=3$. Any additional nonlinearities in the form of quintic nonlinearity or septic nonlinearity shifts the peak of the reflection towards longer wavelengths. When $g=4$,  the reflection spectrum begins to divide into two and thus we can observe two separate
peaks on either side of $\delta=0$ (see Fig. \ref{Fig_de_3}(b)). The first peak corresponds to negative detuning values of $\delta$ and the
second peak corresponds to positive detuning values. The reflected power is not the same in these peaks. More power is accumulated at the peak lying on the shorter wavelength side compared to its counterpart. But closer to $\delta=0$ the reflection is minimum. This kind of behavior is unique because in both conventional FBG \cite{erdogan1997fiber} and unbroken $\mathcal{PT}$-symmetry the reflection is maximum within the stop band and at the band edges the reflection is minimum. Even more interestingly, when $g=5$ the reflection is completely prohibited in the side lobes unlike the other cases and the reflection is maximum at a single wavelength rather than band of wavelengths. Otherwise this can be called as a lasing like behavior, with large reflected intensity at the lasing wavelength. Moreover, the presence of quintic nonlinearity red shifts the spectrum instead of a blue shift in both Figs. \ref{Fig_de_3}(b) and \ref{Fig_de_3}(c) unlike the previous cases.

\section{Conclusion}
\label{Sec:7}
In this paper we have presented a detailed study on the optical bi- and
multi-stability phenomena as well as nonlinear reflection spectra in a highly nonlinear fiber Bragg grating
influenced by $\mathcal{PT}$-symmetry conditions.  We have reported unique behaviors such as ramp and step like stable states in the broken $\mathcal{PT}$-symmetric regime which were previously believed to exist only in complex optical structures involving plasmons, graphene etc. We affirm that these states are feasible in a simple FBG device with minimal effort by having a balance between gain and loss of the system. We have also found novel optical bistabilities in the broken $\mathcal{PT}$-symmetric regime
which can pave a new way to the low power switches via reversal of the direction of light incidence. This confirms that
FBG offers a fertile ground to unearth unique nonlinear functionalities
in the $\mathcal{PT}$- symmetry broken regime in addition 
to the unbroken regime. We have also depicted the existence of both dark and bright soliton like entities at the transmission resonances.  We would like to leave an endnote that all the numerical experiments presented here deserve further investigations through practical observations, and this could open the door for new generation of multi-functional
optical devices including optical switches and memories. To be regarded as the next generation multi-functional devices, any system
must address some of the important criteria like size miniature, reduction in cost, low
power consumption, design flexibility and so on and our ramifications rely on the last two aspects i.e., optical switches with low switching intensities and
flexibility to set-up the desired application simply by tuning one of the control parameters.
In addition our system can play a key role to set-up a new generation of all-optical regenerators employing PAM scheme in the near future as
our system admits a large number of stable states with low switching intensities which is
an ideal requirement for any all-optical systems.

\section*{Acknowledgments}
SVR is indebted to the financial assistantship provided
by Anna University through Anna Centenary Research Fellowship (CFR/ACRF-2018/AR1/24).
AG thanks the Department of Science and Technology (DST)
and Science and Engineering Research Board (SERB), Government of India,
for providing a National Postdoctoral Fellowship (Grant No. PDF/2016/002933). ML is supported by DST-SERB through a Distinguished Fellowship (Grant No. SB/DF/04/2017).


\begin{thebibliography}{61}
	\providecommand{\url}[1]{\texttt{#1}}
	\providecommand{\urlprefix}{URL }
	\providecommand{\eprint}[2][]{\url{#2}}
	
	\bibitem{jensen}
	S.~Jenson, The nonlinear coherent coupler, IEEE J. Quantum Electron.
	\textbf{18}, 1580 (1982).
	
	\bibitem{govindaraji2015}
	A.~Govindaraji, A.~Mahalingam, and A.~Uthayakumar, Numerical investigation of
	dark soliton switching in asymmetric nonlinear fiber couplers, Appl. Phys. B
	\textbf{120}, 341 (2015).
	
	\bibitem{zang2012analysis}
	Z.~Zang and Y.~Zhang, Analysis of optical switching in a $yb^{3+}$-doped fiber
	{B}ragg grating by using self-phase modulation and cross-phase modulation,
	Appl. Opt. \textbf{51}, 3424 (2012).
	
	\bibitem{sethi2014all}
	P.~Sethi and S.~Roy, All-optical ultrafast switching in 2$\times$ 2 silicon
	microring resonators and its application to reconfigurable demux/mux and
	reversible logic gates, J. Light. Technol. \textbf{32}, 2173 (2014).
	
	\bibitem{erdogan1997fiber}
	T.~Erdogan, Fiber grating spectra, J. Light. Technol. \textbf{15}, 1277 (1997).
	
	\bibitem{ramaswami2009optical}
	R.~Ramaswami, K.~Sivarajan, and G.~Sasaki, \emph{Optical networks: {A}
		practical perspective}  (Morgan Kaufmann 2009).
	
	\bibitem{hill1997fiber}
	K.~O. Hill and G.~Meltz, Fiber {B}ragg grating technology fundamentals and
	overview, J. Light. Technol. \textbf{15}, 1263 (1997).
	
	\bibitem{radic1995theory}
	S.~Radic, N.~George, and G.~P. Agrawal, Theory of low-threshold optical
	switching in nonlinear phase-shifted periodic structures, J. Opt. Soc. Am. B
	\textbf{12}, 671 (1995).
	
	\bibitem{jeong2006all}
	Y.~D. Jeong, J.~S. Cho, Y.~H. Won, H.~J. Lee, and H.~Yoo, All-optical flip-flop
	based on the bistability of injection locked {F}abry--{P}\'{e}rot laser
	diode, Opt. Express \textbf{14}, 4058 (2006).
	
	\bibitem{li2014bistability}
	Q.~Li, H.~Yuan, and X.~Tang, Bistability and all-optical flip--flop with active
	microring resonator, Appl. Opt. \textbf{53}, 3049 (2014).
	
	\bibitem{litchinitser2007optical}
	N.~M. Litchinitser, I.~R. Gabitov, and A.~I. Maimistov, Optical bistability in
	a nonlinear optical coupler with a negative index channel, Phys. Rev. Lett.
	\textbf{99}, 113902 (2007).
	
	\bibitem{ping2005bistability}
	Yosia, S.~Ping, and L.~Chao, Bistability threshold inside hysteresis loop of
	nonlinear fiber {B}ragg gratings, Opt. Express \textbf{13}, 5127 (2005).
	
	\bibitem{yousefi2015all}
	E.~Yousefi, M.~Hatami, and A.~T. Jahromi, All-optical ternary signal processing
	using uniform nonlinear chalcogenide fiber {B}ragg gratings, J. Opt. Soc. Am.
	B \textbf{32}, 1471 (2015).
	
	\bibitem{gibbs2012optical}
	H.~Gibbs, \emph{Optical bistability: {C}ontrolling light with light}  (Elsevier
	2012).
	
	\bibitem{winful1979theory}
	H.~G. Winful, J.~Marburger, and E.~Garmire, Theory of bistability in nonlinear
	distributed feedback structures, Appl. Phys. Lett. \textbf{35}, 379 (1979).
	
	\bibitem{shi1995optical}
	C.-X. Shi, Optical bistability in reflective fiber gratings, IEEE J. Quantum
	Electron. \textbf{31}, 2037 (1995).
	
	\bibitem{harbold2002highly}
	J.~Harbold, F.~Ilday, F.~Wise, J.~Sanghera, V.~Nguyen, L.~Shaw, and
	I.~Aggarwal, Highly nonlinear {A}s--{S}--{S}e glasses for all-optical
	switching, Opt. Lett. \textbf{27}, 119 (2002).
	
	\bibitem{chen2006measurement}
	Y.-F. Chen, K.~Beckwitt, F.~W. Wise, B.~G. Aitken, J.~S. Sanghera, and I.~D.
	Aggarwal, Measurement of fifth-and seventh-order nonlinearities of glasses,
	J. Opt. Soc. Am. B \textbf{23}, 347 (2006).
	
	\bibitem{karimi2012all}
	M.~Karimi, M.~Lafouti, A.~A. Amidiyan, and J.~Sabbaghzadeh, All-optical
	flip-flop based on nonlinear effects in fiber {B}ragg gratings, Appl. Opt.
	\textbf{51}, 21 (2012).
	
	\bibitem{porsezian2005modulational}
	K.~Porsezian, K.~Senthilnathan, and S.~Devipriya, Modulational instability in
	fiber {B}ragg grating with non-kerr nonlinearity, IEEE J. Quantum Electron.
	\textbf{41}, 789 (2005).
	
	\bibitem{triki2016}
	H.~Triki, K.~Porsezian, A.~Choudhuri, and P.~T. Dinda, Chirped solitary pulses
	for a nonic nonlinear {S}chr{\"o}dinger equation on a continuous-wave
	background, Phys. Rev. A \textbf{93}, 063810 (2016).
	
	\bibitem{triki2017}
	H.~Triki, K.~Porsezian, P.~T. Dinda, and P.~Grelu, Dark spatial solitary waves
	in a cubic-quintic-septimal nonlinear medium, Phys. Rev. A \textbf{95},
	023837 (2017).
	
	\bibitem{broderick1998nonlinear}
	N.~Broderick, D.~Taverner, and D.~J. Richardson, Nonlinear switching in fibre
	{B}ragg gratings, Opt. Express \textbf{3}, 447 (1998).
	
	\bibitem{yosia2007double}
	Y.~Yosia and S.~Ping, Double optical bistability and its application in
	nonlinear chalcogenide-fiber {B}ragg gratings, Physica B \textbf{394}, 293
	(2007).
	
	\bibitem{ping2005nonlinear}
	S.~Ping, L.~Chao \emph{et~al.}, Nonlinear switching and pulse propagation in
	phase-shifted cubic quintic grating, IEEE Photonics Technol. Lett.
	\textbf{17}, 2670 (2005).
	
	\bibitem{lupu2013switching}
	A.~Lupu, H.~Benisty, and A.~Degiron, Switching using $\mathcal{PT}$-symmetry in
	plasmonic systems: {P}ositive role of the losses, Opt. Express \textbf{21},
	21651 (2013).
	
	\bibitem{kottos2010optical}
	T.~Kottos, Optical physics: {B}roken symmetry makes light work, Nat. Phys.
	\textbf{6}, 166 (2010).
	
	\bibitem{el2007theory}
	R.~El-Ganainy, K.~Makris, D.~Christodoulides, and Z.~H. Musslimani, Theory of
	coupled optical $\mathcal{PT}$-symmetric structures, Opt. Lett. \textbf{32},
	2632 (2007).
	
	\bibitem{govindarajan2018tailoring}
	A.~Govindarajan, A.~K. Sarma, and M.~Lakshmanan, Tailoring
	$\mathcal{PT}$-symmetric soliton switch, Opt. Lett. \textbf{44}, 663 (2019).
	
	\bibitem{karthi2}
	S.~Karthiga, V.~K. Chandrasekar, M.~Senthilvelan, and M.~Lakshmanan,
	Controlling of blow-up responses by nonlinear $\mathcal{PT}$-symmetric
	coupling, Phys. Rev. A \textbf{95}, 033829 (2017).
	
	\bibitem{feng2013experimental}
	L.~Feng, Y.-L. Xu, W.~S. Fegadolli, M.-H. Lu, J.~E. Oliveira, V.~R. Almeida,
	Y.-F. Chen, and A.~Scherer, Experimental demonstration of a unidirectional
	reflectionless parity-time metamaterial at optical frequencies, Nat. Mater.
	\textbf{12}, 108 (2013).
	
	\bibitem{longhi2014pt}
	S.~Longhi and L.~Feng, $\mathcal{PT}$-symmetric microring laser-absorber, Opt.
	Lett. \textbf{39}, 5026 (2014).
	
	\bibitem{huang2014type}
	C.~Huang, R.~Zhang, J.~Han, J.~Zheng, and J.~Xu, Type-ii perfect absorption and
	amplification modes with controllable bandwidth in combined
	$\mathcal{PT}$-symmetric and conventional {B}ragg-grating structures, Phys.
	Rev. A \textbf{89}, 023842 (2014).
	
	\bibitem{phang2013ultrafast}
	S.~Phang, A.~Vukovic, H.~Susanto, T.~M. Benson, and P.~Sewell, Ultrafast
	optical switching using parity--time symmetric {B}ragg gratings, J. Opt. Soc.
	Am. B \textbf{30}, 2984 (2013).
	
	\bibitem{feng2014single}
	L.~Feng, Z.~J. Wong, R.-M. Ma, Y.~Wang, and X.~Zhang, Single-mode laser by
	parity-time symmetry breaking, Science \textbf{346}, 972 (2014).
	
	\bibitem{longhi2010pt}
	S.~Longhi, $\mathcal{PT}$-symmetric laser absorber, Phys. Rev. A \textbf{82},
	031801(R) (2010).
	
	\bibitem{bender1998real}
	C.~M. Bender and S.~Boettcher, Real spectra in non-hermitian hamiltonians
	having $\mathcal{PT}$- symmetry, Phys. Rev. Lett. \textbf{80}, 5243 (1998).
	
	\bibitem{lin2011unidirectional}
	Z.~Lin, H.~Ramezani, T.~Eichelkraut, T.~Kottos, H.~Cao, and D.~N.
	Christodoulides, Unidirectional invisibility induced by
	$\mathcal{PT}$-symmetric periodic structures, Phys. Rev. Lett. \textbf{106},
	213901 (2011).
	
	\bibitem{razzari2012optics}
	L.~Razzari and R.~Morandotti, Optics: Gain and loss mixed in the same cauldron,
	Nature \textbf{488}, 163 (2012).
	
	\bibitem{ruter2010observation}
	C.~E. R{\"u}ter, K.~G. Makris, R.~El-Ganainy, D.~N. Christodoulides, M.~Segev,
	and D.~Kip, Observation of parity--time symmetry in optics, Nat. Phys.
	\textbf{6}, 192 (2010).
	
	\bibitem{baum2015parity}
	B.~Baum, H.~Alaeian, and J.~Dionne, A parity-time symmetric coherent plasmonic
	absorber-amplifier, J. Appl. Phys \textbf{117}, 063106 (2015).
	
	\bibitem{chang2014parity}
	L.~Chang, X.~Jiang, S.~Hua, C.~Yang, J.~Wen, L.~Jiang, G.~Li, G.~Wang, and
	M.~Xiao, Parity-time symmetry and variable optical isolation in
	active-passive-coupled microresonators, Nat. Photonics \textbf{8}, 524
	(2014).
	
	\bibitem{phang2015versatile}
	S.~Phang, A.~Vukovic, T.~M. Benson, H.~Susanto, and P.~Sewell, A versatile
	all-optical parity-time signal processing device using a {B}ragg grating
	induced using positive and negative Kerr-nonlinearity, Opt. Quant. Electron.
	\textbf{47}, 37 (2015).
	
	\bibitem{phang2014impact}
	S.~Phang, A.~Vukovic, H.~Susanto, T.~M. Benson, and P.~Sewell, Impact of
	dispersive and saturable gain/loss on bistability of nonlinear parity--time
	{B}ragg gratings, Opt. Lett. \textbf{39}, 2603 (2014).
	
	\bibitem{1555-6611-25-1-015102}
	J.~Liu, X.-T. Xie, C.-J. Shan, T.-K. Liu, R.-K. Lee, and Y.~Wu, Optical
	bistability in nonlinear periodical structures with
	{$\mathcal{P}\mathcal{T}$}-symmetric potential, Laser Phys. \textbf{25},
	015102 (2015).
	
	\bibitem{sarma2014modulation}
	A.~K. Sarma, Modulation instability in nonlinear complex parity-time symmetric
	periodic structures, J. Opt. Soc. Am. B \textbf{31}, 1861 (2014).
	
	\bibitem{miri2012bragg}
	M.-A. Miri, A.~B. Aceves, T.~Kottos, V.~Kovanis, and D.~N. Christodoulides,
	{B}ragg solitons in nonlinear $\mathcal{PT}$-symmetric periodic potentials,
	Phys. Rev. A \textbf{86}, 033801 (2012).
	
	\bibitem{komissarova2019pt}
	M.~Komissarova, V.~Marchenko, and P.~Y. Shestakov, $\mathcal{PT}$-symmetric
	periodic structures with the modulation of the Kerr nonlinearity, Phys. Rev.
	E \textbf{99}, 042205 (2019).
	
	\bibitem{agrawal2001applications}
	G.~Agrawal, \emph{Applications of nonlinear fiber optics}  (Elsevier 2001).
	
	\bibitem{zhou2014optical}
	X.-Y. Zhou, B.-J. Wu, Q.-Y. Wan, F.~Wen, and K.~Qiu, Optical multi-stability in
	fiber {B}ragg gratings with application to all-optical regeneration of
	multi-level pulse-amplitude-modulation signals, in \emph{Asia Communications
		and Photonics Conference}  (Optical Society of America 2014), pp. ATh3A--119.
	
	\bibitem{radic1994optical}
	S.~Radic, N.~George, and G.~P. Agrawal, Optical switching in
	$\lambda$/4-shifted nonlinear periodic structures, Opt. Lett. \textbf{19},
	1789 (1994).
	
	\bibitem{govindarjan2019}
	A.~Govindarajan and M.~Lakshmanan, Dark gap solitons in broken
	$\mathcal{PT}$-symmetry (unpublished).
	
	\bibitem{daneshfar2017switching}
	N.~Daneshfar and T.~Naseri, Switching between optical bistability and
	multistability in plasmonic multilayer nanoparticles, J. Appl. Phys.
	\textbf{121}, 023111 (2017).
	
	\bibitem{dai2015low}
	X.~Dai, L.~Jiang, and Y.~Xiang, Low threshold optical bistability at terahertz
	frequencies with graphene surface plasmons, Sci. Rep. \textbf{5}, 12271
	(2015).
	
	\bibitem{sharif2016experimental}
	M.~A. Sharif, M.~M. Ara, B.~Ghafary, S.~Salmani, and S.~Mohajer, Experimental
	observation of low threshold optical bistability in exfoliated graphene with
	low oxidation degree, Opt. Mater. \textbf{53}, 80 (2016).
	
	\bibitem{naseri2018terahertz}
	T.~Naseri, N.~Daneshfar, M.~Moradi-Dangi, and F.~Eynipour-Malaee, Terahertz
	optical bistability of graphene-coated cylindrical core--shell nanoparticles,
	J. Theo. Appl. Phys. \textbf{12}, 257 (2018).
	
	\bibitem{rukhlenko2009}
	I.~D. Rukhlenko, M.~Premaratne, and G.~P. Agrawal, Analytical study of optical
	bistability in silicon-waveguide resonators, Opt. Express \textbf{17}, 22124
	(2009).
	
	\bibitem{naseri2018optical}
	T.~Naseri and N.~Daneshfar, Optical bistability of a plexcitonic system
	consisting of a quantum dot near a metallic nanorod, J. Theo. Appl. Phys.
	\textbf{12}, 183 (2018).
	
	\bibitem{ogasawara1986static}
	N.~Ogasawara and R.~Ito, Static and dynamic properties of nonlinear
	semiconductor laser amplifiers, Jpn. J. Appl. Phys. \textbf{25}, L739 (1986).
	
	\bibitem{hasegawa1995}
	A.~Hasegawa and Y.~Kodama, \emph{Solitons in optical communications}, 7
	(Oxford University Press, USA 1995).
	
	\bibitem{zhao2010dissipative}
	L.~Zhao, D.~Tang, X.~Wu, and H.~Zhang, Dissipative soliton generation in
	yb-fiber laser with an invisible intracavity bandpass filter, Opt. Lett.
	\textbf{35}, 2756 (2010).
	
	\bibitem{tang2008observation}
	D.~Tang, H.~Zhang, L.~Zhao, and X.~Wu, Observation of high-order
	polarization-locked vector solitons in a fiber laser, Phys. Rev. Lett.
	\textbf{101}, 153904 (2008).
	
	\bibitem{zhang2009observation}
	H.~Zhang, D.~Y. Tang, L.~Zhao, and X.~Wu, Observation of polarization domain
	wall solitons in weakly birefringent cavity fiber lasers, Phys. Rev. B
	\textbf{80}, 052302 (2009).
	
	\bibitem{chen2014formation}
	Y.~Chen, M.~Wu, P.~Tang, S.~Chen, J.~Du, G.~Jiang, Y.~Li, C.~Zhao, H.~Zhang,
	and S.~Wen, The formation of various multi-soliton patterns and noise-like
	pulse in a fiber laser passively mode-locked by a topological insulator based
	saturable absorber, Laser Phys. Lett. \textbf{11}, 055101 (2014).
	
	\bibitem{winful2000raman}
	H.~G. Winful and V.~Perlin, Raman gap solitons, Phys. Rev. Lett.
	\textbf{84}, 3586 (2000).
	
	\bibitem{mok2006dispersionless}
	J.~T. Mok, C.~M. De~Sterke, I.~C. Littler, and B.~J. Eggleton, Dispersionless
	slow light using gap solitons, Nat. Phys. \textbf{2}, 775 (2006).
	
	\bibitem{d2004bright}
	G.~D'Aguanno, N.~Mattiucci, M.~Scalora, and M.~J. Bloemer, Bright and dark gap
	solitons in a negative index {F}abry--{P}\'{e}rot etalon, Phys. Rev. Lett.
	\textbf{93}, 213902 (2004).

\bibitem{winful1991modulational}
H.~G. Winful, R.~Zamir, and S.~Feldman, Modulational instability in nonlinear
periodic structures: Implications for ‘‘gap solitons’’, Appl. Phys.
Lett. \textbf{58}, 1001 (1991).
	
	\bibitem{kozhekin1995self}
	A.~Kozhekin and G.~Kurizki, Self-induced transparency in bragg reflectors: gap
	solitons near absorption resonances, Phys. Rev. Lett. \textbf{74}, 5020
	(1995).
	
	\bibitem{taverner1998all}
	D.~Taverner, N.~Broderick, D.~Richardson, M.~Ibsen, and R.~Laming, All-optical
	and gate based on coupled gap-soliton formation in a fiber bragg grating,
	Opt. Lett. \textbf{23}, 259 (1998).
	
	
\end{thebibliography}
\end{document}